%% file: DBatNLO.tex
\documentclass[a4paper,fleqn,10pt]{article}
\pdfoutput=1
\usepackage{amsmath}
\usepackage{amssymb}
\usepackage{graphicx}
\usepackage[merge,numbers,compress]{natbib}
\usepackage[T1]{fontenc}
\usepackage{booktabs}
\usepackage{xcolor}
\usepackage{xspace}
\usepackage{dcolumn}
\usepackage{placeins}
\usepackage{todonotes}
\usepackage{siunitx}

\usepackage[colorlinks=true,citecolor=blue!50!black,linkcolor=black]{hyperref}
\usepackage{caption}
\usepackage
[subrefformat=parens,position=top,skip=-15pt,margin=15pt,justification=justified,singlelinecheck=false]
{subcaption}
\usepackage{authblk}

\usepackage{mciteplus}
\usepackage{geometry}
\usepackage{sectsty}
\usepackage{multirow}

\numberwithin{equation}{section}
\input{macros}

\hypersetup{
  pdfauthor={Enrico Bothmann, Davide Napoletano, Marek Sch\"onherr, Steffen Schumann, Simon Luca Villani},
  pdftitle={Higher-order EW corrections in ZZ and ZZj production at the LHC}
}

\preprint{IPPP/21/51\\MCNET-21-16}

\usepackage[symbol]{footmisc}

\renewcommand{\thefootnote}{\fnsymbol{footnote}}

\author[1]{Enrico Bothmann\footnote[1]{E-mail: \texttt{enrico.bothmann@uni-goettingen.de}}}
\author[2]{Davide Napoletano\footnote[2]{E-mail: \texttt{davide.napoletano@unimib.it}}}
\author[3]{Marek Sch\"onherr\footnote[3]{E-mail: \texttt{marek.schoenherr@durham.ac.uk}}}
\author[1]{Steffen Schumann\footnote[4]{E-mail: \texttt{steffen.schumann@phys.uni-goettingen.de}}}
\author[1]{Simon Luca Villani\footnote[5]{E-mail: \texttt{simonluca.villani@uni-goettingen.de}}}

\affil[1]{Institut f{\"u}r Theoretische Physik,
  Georg-August-Universit{\"a}t G{\"o}ttingen, 37077 G{\"o}ttingen, Germany}

\affil[2]{Universit\`a degli Studi di Milano-Bicocca \& INFN, Piazza della Scienza 3, Milano 20126, Italy}

\affil[3]{Institute for Particle Physics Phenomenology, Department of Physics, Durham University,
  Durham DH1 3LE, United Kingdom}

\title{Higher-order EW corrections in $ZZ$ and $ZZj$ production at the LHC}

\begin{document}
\maketitle

\renewcommand{\thefootnote}{\fnsymbol{footnote}}

\begin{abstract}
  We consider the production of a pair of $Z$ bosons at the LHC and study
  the inclusion of EW corrections in theoretical predictions at fixed order
  and based on multijet-merged parton-shower simulations. To this end we
  present exact NLO EW results for $pp\to e^+e^-\mu^+\mu^-$, and,
  for the first time, for $pp\to e^+e^-\mu^+\mu^-j$, and compare them to the
  EW virtual and NLL Sudakov approximation.
  We then match the exact NLO EW result to the resummed
  Sudakov logarithms to achieve an improved \NLOEWNLLEWsudexp result.

  Further, we discuss the inclusion of the above EW corrections in
  \MEPSatNLO event simulations in the framework of the \Sherpa event generator.
  We present detailed phenomenological predictions for inclusive $ZZ$ and $ZZj$
  production taking into account the dominant EW corrections through the EW
  virtual approximation, as well as through (exponentiated) EW Sudakov
  logarithms.
\end{abstract}

\clearpage

\vspace{10pt}
\noindent\rule{\textwidth}{1pt}
\tableofcontents
\noindent\rule{\textwidth}{1pt}
\vspace{10pt}

\section{Introduction}
With the successful completion of the Large Hadron Collider (LHC) Run~1 and Run~2
data-taking campaigns and the upcoming Run~3 at even higher luminosities, we have
entered the precision era of high-energy hadron-collider physics. Precision
measurements are meanwhile routinely done and are used to further scrutinise the
Standard Model (SM) of particle physics and to search for faint hints of
New Physics.

For the success of the LHC precision-measurement program it is vital that tools
used to quantify theoretical expectations consistently include higher-order perturbative
corrections. While typically QCD effects are dominant, there is a growing need to
also account for electroweak (EW) effects, in particular for remote regions of
phase space, where these can be significantly enhanced. Such quantitative SM
predictions are essential for the interpretation of actual measurements. They also
serve as important inputs to the determination of parton distribution functions (PDFs),
see for instance~\cite{Ball:2021leu}.
The list of processes for which precise predictions are needed steadily gets longer and
the inherent complexity of the corresponding calculations continues to rise, \emph{e.g.}\
due to the presence of intermediate resonances or the need to account for QCD jets
accompanying the desired signal. For an extensive recent review on EW corrections
for collider physics see~\cite{Denner:2019vbn}.

Similar to the case of next-to-leading order (NLO) QCD corrections, the computation
of exact NLO EW contributions has in recent years been largely automated.
Central for these developments are dedicated codes providing one-loop amplitudes,
such as \Gosam~\cite{Cullen:2014yla}, \MadLoop~\cite{Hirschi:2011pa,Frederix:2018nkq},
\MCFM~\cite{Campbell:2016dks}, \NLOX~\cite{Honeywell:2018fcl},
\OpenLoops~\cite{Buccioni:2019sur}, or \Recola~\cite{Actis:2016mpe}. These get
supplemented by implementations of infrared subtraction schemes such as
Catani--Seymour~\cite{Catani:1996vz,Catani:2002hc} or Frixione--Kunszt--Signer~\cite{Frixione:1995ms}
subtraction, suitably generalised to the case of QED infrared singularities~\cite{Dittmaier:1999mb,Dittmaier:2008md},
see for instance~\cite{Gleisberg:2007md,Frederix:2010cj,Gehrmann:2010ry,Schonherr:2017qcj,Frederix:2018nkq}.
With these technologies NLO EW accurate predictions have been achieved for a variety
of higher-multiplicity processes already, see for
instance~\cite{Kallweit:2014xda,Denner:2016jyo,Granata:2017iod,Chiesa:2017gqx,Greiner:2017mft,Schonherr:2018jva,Denner:2019tmn,Dittmaier:2019twg,Pagani:2020mov,Pagani:2021iwa}.
Furthermore, with the given methods also full NLO SM accurate predictions can be computed, including
all possible EW, QCD and mixed contributions.
For example Refs.~\cite{Frederix:2016ost} and~\cite{Reyer:2019obz} presented corresponding
results for two- and three-jet production in hadronic collisions, respectively.
In Refs.~\cite{Biedermann:2017bss,Denner:2020zit,Denner:2021hsa} full NLO SM results for
the class of Vector-Boson-Scattering processes have been obtained.
Many of the quoted calculations have been compiled in generator frameworks such as
\mgfive~\cite{Alwall:2014hca}, \PB~\cite{Alioli:2010xd}, or \Sherpa~\cite{Gleisberg:2008ta,Bothmann:2019yzt},
which provide backbone infrastructures such as the overall process organisation,
phase-space integration,
and interfaces to simulations of other physics aspects, including parton showers or models for
non-perturbative phenomena.

However, when aiming for full particle-level simulations as accomplished by
general-purpose Monte Carlo (MC) generators~\cite{Buckley:2011ms}, the consistent inclusion
of EW corrections to the hard-scattering process poses a severe theoretical challenge,
in particular in the context of multijet-merged calculations.
It requires using an interleaved shower evolution for QCD and QED emissions. As in
the pure QCD case, a detailed matching of QED real-emission matrix elements with
corresponding shower expressions is needed. Furthermore, the assignment of parton-shower
starting conditions and the determination of
parton-shower emission histories is significantly complicated by (mixed) EW
contributions.

As an alternative to the complete set of NLO EW corrections, methods restricted to the
leading effects due to EW loops are available. In particular at energy scales $Q$ large
compared to the masses of the EW gauge bosons, contributions from virtual $W$- and $Z$-boson
exchange and corresponding collinear real emissions dominate. The leading contributions are
Sudakov-type logarithms of the form~\cite{Sudakov:1954sw,Ciafaloni:1998xg}
\begin{equation}
  \frac{\alpha}{4\pi \sin^2\theta_W}\log^2\left(\frac{Q^2}{M^2_W}\right)\quad\text{and}\quad
  \frac{\alpha}{4\pi \sin^2\theta_W}\log\left(\frac{Q^2}{M^2_W}\right)\,.
\end{equation}
The one-loop EW Sudakov approximation, dubbed \EWsud\ here, has been developed for general processes
in~\cite{Denner:2000jv,Denner:2001gw}. An implementation for multi-parton
LO channels, in particular $Z+n$ jets ($n=1,2,3$), is available within the \Alpgen
generator~\cite{Chiesa:2013yma}. The \MCFM\ program currently provides EW Sudakov corrections
in combination with NLO QCD accuracy for Drell--Yan, top-quark pair, and dijet
production~\cite{Campbell:2016dks}. A corresponding automated implementation in the
\Sherpa framework, applicable for LO and NLO contributions, has been presented
in~\cite{Bothmann:2020sxm}. Quite recently Ref.~\cite{Pagani:2021vyk} reported on an
implementation in the \mgfive framework.

Another available approximation, dubbed \EWvirt, was devised in~\cite{Kallweit:2015dum}.
It comprises exact renormalised NLO EW virtual corrections and integrated
approximate real-emission subtraction terms, thereby neglecting in particular hard
real-emission contributions. However, both methods qualify for a rather straightforward
inclusion of the dominant EW corrections in state-of-the-art matrix-element plus
parton-shower simulations, subject of this study.

We here aim for a detailed comparison of the \EWvirt\ and \EWsud\ approximations for the production of a pair of $Z$
bosons under LHC conditions and consider their inclusion in \MEPSatNLO\ multijet-merging shower
simulations in the \Sherpa framework. To this end we consider the channels $pp\to e^+e^-\mu^+\mu^-$
and $pp\to e^+e^-\mu^+\mu^-j$. We present the first evaluation of the full set of NLO EW corrections
and corresponding differential distributions for the $e^+e^-\mu^+\mu^-j$ channel that we use to
quantify the quality of the two approximations. Furthermore, we describe the
generalisation of the \Sherpa implementation of the EW Sudakov approximation to the case of
parton-shower evolved processes and NLO multijet merging in particular. Our final deliverable
are NLO QCD accurate shower simulations for the $e^+e^-\mu^+\mu^- + 0, 1$\,jet channels supplemented
by \EWvirt\ and \EWsud\ corrections.

In general, the class of diboson-production processes is of utmost importance at the LHC.
They constitute important signal and background contributions to Higgs-boson production. Furthermore,
they can be used to probe the self-interactions of weak gauge bosons
and thus provide insights into the mechanisms of electroweak symmetry breaking. In particular
for the $ZZ$ channel, NNLO QCD predictions have been presented in
\cite{Cascioli:2014yka,Grazzini:2015hta,Heinrich:2017bvg,Kallweit:2018nyv}. NLO EW corrections
for the four-lepton final state have first been presented in~\cite{Biedermann:2016yvs,Biedermann:2016lvg}.
Their combination with NLO and NNLO QCD corrections was discussed in Refs.~\cite{Kallweit:2017khh,Chiesa:2018lcs}
and \cite{Grazzini:2019jkl}, respectively. The matching of NLO QCD calculations with parton showers was first
presented in~\cite{Melia:2011tj,Frederix:2011ss}. Ref.~\cite{Chiesa:2020ttl} presented the matching of NLO
QCD and NLO EW corrections to a QCD+QED parton shower in the \PB\ framework. In
Ref.~\cite{Alioli:2021egp} the NNLO QCD calculation matched to parton showers
was presented in the \textsc{GENEVA}
framework~\cite{Alioli:2012fc,Alioli:2013hqa,Alioli:2015toa,Alioli:2016wqt},
followed by Ref.~\cite{Buonocore:2021fnj}
showing results in the MiNNLO$_\text{PS}$~\cite{Monni:2019whf}
method, both including the loop-induced gluon-fusion contribution. For gluon-initiated
four-lepton production the NLO QCD corrections are known and have been studied
extensively for example in Refs.~\cite{Caola:2015psa,Caola:2016trd,Grazzini:2018owa,Grazzini:2021iae,Buonocore:2021fnj}.
The matching of these loop-induced processes to the parton shower in the \Powheg framework
has been presented in~\cite{Alioli:2016xab,Alioli:2021wpn}. In these studies the
virtual correction was considered in the massless limit~\cite{vonmanteuffel2015twoloop}
and supplemented by approximate finite top-mass effects. Only recently the calculation
with the full top-mass dependence has become available~\cite{Agarwal:2020dye}.

Recent measurements of four-lepton production by the ATLAS and CMS experiments
at the LHC have for example been presented
in~\cite{Aaboud:2017rwm,Aaboud:2019lxo,Aaboud:2019lgy,ATLAS:2019qet,ATLAS:2021kog,Khachatryan:2016txa,Sirunyan:2017zjc,Sirunyan:2018vkx,CMS:2021pqj}.
In general, when compared to data, higher-order QCD calculations often show deviations of up to \SI{20}{\%} in the tail of transverse
momentum and mass distributions, which can be accounted for by higher-order corrections
in the EW sector \cite{Monni:2019whf}, further motivating works like the one
presented here.
Thus, the ideal
set-up would provide not only fixed-order NLO EW effects, but also resum the Sudakov
logarithmic contributions discussed above and match them to the fixed order result.
In this paper we will provide exactly such a resummation-improved
calculation, \NLOEWNLLEWsudexp, matching next-to-leading logarithmic
exponentiated Sudakov corrections to the exact NLO EW expression
for the off-shell $ZZ$ and $ZZj$ production processes.

The paper is organised as follows: In Sec.~\ref{sec:fomatching} we briefly review the general
structure of NLO EW corrections, and in particular in the high-energy limit, and present our
approach to match exact $\order{\alpha}$ EW corrections with the resummation of Sudakov
logarithms, \NLOEWNLLEWsudexp. In Sec.~\ref{sec:setup} we recapitulate the theoretical framework
for merging (N)LO QCD matrix elements of variable parton multiplicity dressed with truncated QCD parton showers
as used in \Sherpa. We then detail the available methods for including EW corrections in
such multijet-merging calculations, \emph{i.e.}\ the \EWvirt\ and \EWsud\ approximations.
Sec.~\ref{sec:applZZ} is devoted to the application and validation of the calculational methods
to $e^+e^-\mu^+\mu^-$ and $e^+e^-\mu^+\mu^-j$ production in proton--proton collisions. This includes
novel \NLOEWNLLEWsudexp\ predictions for the zero- and one-jet processes, as well as results
from NLO QCD multijet merging supplemented with EW corrections in the \EWvirt\ and \EWsud\ schemes.
We conclude and give an outlook in Sec.~\ref{sec:conclusions}.

\section{NLO EW and resummed EW Sudakov corrections}
\label{sec:fomatching}

The ${\cal{O}}(\alpha)$ EW corrections to a given tree-level process comprise
virtual one-loop contributions, \emph{i.e.}\ the interference of
loop-diagrams with tree-level amplitudes, as well as a real-emission part.
Due to the non-vanishing mass of the $W$ and $Z$ bosons, as well as their
straight-forward experimental identifiability, these are typically discarded
as real-emission corrections. At NLO EW, the differential cross section in
the fiducial region of an observable that is non-trivial at Born-level
can be written as a correction, \deltaEW, to its LO expression
given by the Born matrix element $\mr{B}$ in the Born phase space $\Phi$, as
\begin{equation}\label{eq:fo-nloew}
  \begin{split}
    \done\sigma^\text{NLO EW}
    =&\;
      \done\Phi\;\mr{B}(\Phi)\;\left(1+\deltaEW(\Phi)\mhhl\right)
    \qquad\text{with}\qquad
    \deltaEW(\Phi)
    =
      \frac{\tilde{\mr{V}}^\text{EW}(\Phi)}{\mr{B}(\Phi)}
      +\done\Phi_1\,\frac{\mr{R}^\text{EW}(\Phi\!\cdot\!\Phi_1)}{\mr{B}(\Phi)}\;.
  \end{split}
\end{equation}
$\tilde{\mr{V}}^\text{EW}$ includes the virtual correction
$\mr{V}^\text{EW}$ as well as the collinear
mass-factorisation terms. The phase space associated with the real-emission
correction $\mr{R}^\text{EW}$ refers to one additional particle compared to the Born process.

\subsection{EW corrections in the Sudakov limit}
\label{sec:fomatching:Ewsud}

In the high-energy limit, \emph{i.e.}\ when all invariants of
a process are large compared to the EW scale, the so-called
EW Sudakov regime, the exact NLO EW corrections are dominated
by the exchange of virtual electroweak gauge bosons and the
running of the EW input parameters \cite{Denner:2000jv}.
Thus, using the above building blocks, the NLO EW corrections
can be approximated by \cite{Kallweit:2015dum}
\begin{equation}\label{eq:fo-nllew-virt}
  \begin{split}
    \done\sigma^\text{LO${}+{}$\EWvirt}
    =&\;
      \done\Phi\;\mr{B}(\Phi)\;\left(1+\deltaEWvirt(\Phi)\mhhl\right)
    \qquad\text{with}\qquad
    \deltaEWvirt(\Phi)
    =
      \frac{\mr{V}^\text{EW}(\Phi)+\mr{I}^\text{EW}(\Phi)}{\mr{B}(\Phi)}
    \;.
  \end{split}
\end{equation}
In this so-called EW virtual approximation, \EWvirt, the exact NLO EW
virtual correction $\mr{V}^\text{EW}$ is supplemented with
integrated approximate real-emission corrections $\mr{I}^\text{EW}$,
corresponding to the EW Catani--Seymour subtraction $\mb{I}$-operator
\cite{Schonherr:2017qcj}.
The such constructed correction is infrared finite and contains
both all EW next-to-leading logarithms (NLL) of the high-energy limit,
as well as important finite corrections throughout phase space and
EW scheme-dependent renormalisation terms that reduce the
scheme dependence at higher orders in $\alpha$.
On the other hand, as exact NLO EW one-loop matrix elements
need to be evaluated, this type of correction is computationally
rather expensive. In practice the \EWvirt\ approximation is therefore
applicable for rather low-multiplicity processes.

Alternatively, without the need to know the
exact one-loop amplitude, an equally NLL
accurate universal Sudakov
factor can be constructed~\cite{Denner:2000jv,Denner:2001gw}:
\begin{equation}\label{eq:fo-nllew-sud}
  \begin{split}
    \done\sigma^\text{LO${}+{}$\EWsud}
    =&\;
      \done\Phi\;\mr{B}(\Phi)\;\left(1+\deltaEWsud(\Phi)\mhhl\right)
    \quad\;\text{with}\quad\;
    \deltaEWsud(\Phi)
    =
      \frac{\mr{V}_\text{NLL}^\text{EW}(\Phi)+\mr{I}_\text{NLL}^\text{EW}(\Phi)}{\mr{B}(\Phi)}
    =
      \KEWsud(\Phi)\;.
  \end{split}
\end{equation}
Therein, $\mr{V}_\text{NLL}^\text{EW}$ is based mainly on the soft and
collinear limits of the dominating one-loop virtual electroweak
gauge-boson exchange diagrams.
Also included in $\deltaEWsud$
are the integrated real-photon emission corrections $\mr{I}_\text{NLL}^\text{EW}$
to the same accuracy.
This results in residual corrections proportional to logarithms
of the large kinematic invariants that are universal. The above
defined approximate correction thus
directly translates into the relative correction $\KEWsud$
to the LO cross section.\footnote{
  Please note, in variance with $K_\text{NLL}$ given in \cite{Bothmann:2020sxm},
  $\KEWsud$ used here is defined such that $1+\KEWsud=K_\text{NLL}$.}
A general and automated implementation of such NLL EW Sudakov corrections
in the \Sherpa framework has been presented in~\cite{Bothmann:2020sxm}.
In the course of the work presented in this paper, several improvements to
this implementation have been accomplished, related to the presence of
intermediate resonances and multiple energy scales in the pairwise invariants,
see App.~\ref{sec:app:ewsud} for details.

Further, EW corrections in the Sudakov limit can be resummed through
exponentiation~\cite{Fadin:1999bq,Melles:2001ye,Melles:2001dh,
  Chiu:2007yn,Chiu:2007dg,Chiu:2008vv,Chiu:2009mg,Chiu:2009ft,Fuhrer:2010eu}.
This provides a more accurate description at very high energies, when Sudakov
logarithms become large, as it takes account of these effects to all orders.
While the relative correction of the \EWvirt\ approximation is generally
not suitable for exponentiation as it contains a number of non-universal
finite terms, the pure NLL correction of the \EWsud\ approximation can be
directly used as the basis of the resummation.
Thus, the resulting cross section including NLL accurate resummed EW Sudakov
corrections reads
\begin{equation}\label{eq:fo-nllew-exp}
  \begin{split}
    \done\sigma^\text{LO${}+{}$\EWsudexp}
    =&\;
      \done\Phi\;\mr{B}(\Phi)\;\exp\left(\deltaEWsud(\Phi)\mhhl\right)\;.
  \end{split}
\end{equation}
All of the above approximations can be supplemented with a soft-photon
resummation in the YFS scheme~\cite{Yennie:1961ad,Schonherr:2008av} without
losing the formal NLL accuracy in the Sudakov regime, \emph{see}\ Sec.~\ref{sec:setup:ew}.
In consequence, the restoration of the differential description of
real-photon emissions will typically improve the agreement with
the exact NLO EW calculation.

\subsection{Matched \texorpdfstring{\NLOEWNLLEWsudexp}{NLO EW + NLL Sudakov} corrections}
\label{sec:fomatching:match}

The resummed NLL EW Sudakov corrections can be matched
to the exact NLO EW result to achieve the optimal description
for inclusive observables \emph{and} the high-energy tails of kinematic
distributions. We choose a matching scheme in which we replace
the $\order{\alpha}$ coefficient in the expansion of the exponential
with the exact NLO EW expression, \emph{i.e.}\
\begin{equation}\label{eq:nloew-nllsud}
  \begin{split}
    \done\sigma^{\text{\NLOEWNLLEWsudexp}}
    =&\;
      \done\Phi\;\mr{B}(\Phi)\;
      \left[
        \exp\left(\deltaEWsud(\Phi)\mhhl\right)
        -\deltaEWsud(\Phi)
        +\deltaEW(\Phi)\mhhl
      \right]\;.
  \end{split}
\end{equation}
In this way, at high energies when $\deltaEW\approx
\deltaEWsud$, we obtain the resummed result as expected.
On the other hand, when the Sudakov logarithms are small,
\emph{i.e.}\ $\exp\left(\deltaEWsud\right)\approx 1 +
\deltaEWsud$, the fixed-order result is recovered.

It is worth stressing that this matching formula, when expanded
to $\order{\alpha^2}$, coincides with the approach used in
\cite{Lindert:2017olm} to estimate the approximate NNLO EW
corrections, and its third-order coefficient serving as an estimate for
its uncertainty.
Here we will use the complete matched result to obtain
an improved central value that incorporates the dominant terms
to all orders, and estimate the uncertainty by a change of
the EW renormalisation scheme. In terms of the uncertainty
estimation, this ansatz is more conservative, as scheme-compensating
terms are not included beyond NLO in the matched calculation.
However, such uncertainty estimate is expected to be more reliable
in the fixed-order regime.

\section{QCD Multijet merging with approximate EW corrections}
\label{sec:setup}

Multijet-merging calculations attempt to describe the high-energetic QCD radiation
accompanying a production process through exact matrix elements rather than solely
relying on the parton-shower approximation. However, as discussed in
Sec.~\ref{sec:fomatching}, the high-energy tails of physical distributions are also
rather susceptible to EW corrections and in particular Sudakov-suppression effects.
This motivates the development of calculational schemes to incorporate the leading EW
contributions in QCD merging approaches.

In this section we briefly review the merging formalism used in the \Sherpa framework. To
this end we revisit in particular the definitions of the \MEPSatLO and \MEPSatNLO
schemes. We then describe their respective combinations with approximate EW corrections, \emph{i.e.}\
the \EWvirt\ and \EWsud\ approaches. To account for higher-order QED corrections, \emph{i.e.}\
the emission of soft photons from the final-state leptons, we employ the
Yennie--Frautschi--Suura (YFS) approach to QED resummation~\cite{Yennie:1961ad}. Its combination
with the approximate EW corrections will also be addressed in what follows.

\subsection{QCD multijet merging at LO and NLO}
\label{sec:setup:merging}

QCD multijet merging aims for the consistent combination of processes with varying multiplicity
of associated jets that get evolved by a QCD parton-shower model. The generic multijet-merged cross
section can be written as
\begin{equation}\label{eq:meps-master}
  \begin{split}
    \done\sigma
    =&\;
      \sum\limits_{n=0}^{\nmax-1}
       \done\sigma_n^\text{excl}
      +\done\sigma_{\nmax}^\text{incl}\;.
  \end{split}
\end{equation}
Therein, each exclusive $n$-jet process---relative to some core process which may already contain
jets not counted in $n$---can be evaluated either at LO or at NLO in QCD. The notion of exclusive
and inclusive cross sections is closely linked to the shower evolution of the underlying
multi-parton ensembles and is realised through a jet-resolution parameter, \emph{i.e.}\
the merging scale \Qcut. By including higher-multiplicity matrix elements, multijet-merged
simulations account both for the bulk of the total production cross section \emph{as well as} rare
configurations with (multiple) associated high-energetic QCD jets.

\paragraph{M\protect\scalebox{0.8}{E}P\protect\scalebox{0.8}{S}@L\protect\scalebox{0.8}{O}.}
The \Sherpa leading-order merging prescription, \emph{i.e.}\ the \MEPSatLO method~\cite{Hoeche:2009rj},
is based on tree-level matrix elements dressed by a truncated-shower evolution using the Catani--Seymour
dipole shower~\cite{Schumann:2007mg}.
The exclusive $n$-jet cross section is thereby derived from the
simple combination of the LO $n$-parton matrix element $\mr{B}_n$ (including all symmetry, flux and PDF factors)
and the appropriate parton-shower functional $\Fcal_n$ restricted to emissions below the merging scale \Qcut,
\begin{equation}\label{eq:meps-lo-excl}
  \begin{split}
    \done\sigma_n^{\text{excl},\MEPSatLO}
    =\;&
      \done\Phi_n\;\mr{B}_n\left(\Phi_n\right)\;
      \Theta_n(\Qcut)\;
      \Fcal_n\left(\mu_Q^2;\Qcut\right)\;.
  \end{split}
\end{equation}
Here, $\Theta_n(\Qcut)$ implements cuts and constraints on the fiducial phase space for
the final state of the core process, and further requires at least $n$ resolved jets with
the merging parameter \Qcut\ playing the role of the jet-resolution scale. The
inverse of the parton shower must thereby be used
as a clustering and recombination algorithm
to produce a consistent result.
In general, the parton-shower functional \Fcal\ takes the form
\begin{equation}\label{eq:def:Fcal}
  \begin{split}
    \mathcal{F}_{n}\left(t\right)
     =&\;
       \Delta_n\left(t_c,t\right)\,
       + \int_{t_c}^t
         \done\Phi_1^\prime\,\mathrm{K}_n\left(\Phi_1^\prime\right)\,
         \Delta_{n}\left(t^\prime,t\right)\;
         \Fcal_{n+1}\left(t^\prime\right) \;,
  \end{split}
\end{equation}
with $t$ the shower-evolution variable and $t_c$ its IR cutoff, $\mr{K}_n$ the splitting kernel,
and  $\Delta_n\left(t^\prime,t\right)
= \exp(-\int_{t^\prime}^{t}\done\Phi_1\,\mr{K}_n(\Phi_1))$
the corresponding Sudakov form factor.
As mentioned above,
in the context of the exclusive processes entering the multijet-merging,
the shower emissions are restricted to occur below \Qcut\ to prevent
filling the phase space governed by higher-multiplicity matrix elements twice.
This constraint is implemented as follows:
\begin{equation}
  \begin{split}
    \mathcal{F}_{n}\left(t;\Qcut\right)
    =&\;
      \Delta_n\left(t_c,t\right)\,
      + \int_{t_c}^t
        \done\Phi_1^\prime\,\mathrm{K}_n\left(\Phi_1^\prime\right)\,
        \Theta\left(\Qcut -Q_{n+1}(t^\prime)\right)\,
        \Delta_{n}\left(t^\prime,t\right)\,
        \Fcal_{n+1}\left(t^\prime\right)\;,
  \end{split}
\end{equation}
with $Q_{n+1}$ being the smallest reconstructed emission scale of
the newly formed $(n+1)$-parton ensemble.
In consequence, its unitary is broken, giving a Sudakov
suppression that turns the inclusive Born expression into
the description of an exclusive $n$-jet cross section down to
a jet resolution of $\Qcut$.
Hence, $\mu_Q^2$ in Eq.~\eqref{eq:meps-lo-excl} defines
the parton-shower starting scale.
In contract,
the highest multiplicity $n=\nmax$ needs to
be treated inclusively as indicated in Eq.~\eqref{eq:meps-master}.
Accordingly, \Qcut\ is in this case event-wise replaced
with the lowest reconstructed emission scale $Q_{n_\mathrm{max}}\geq \Qcut$, and
the parton shower $\Fcal_{\nmax}=\Fcal_{\nmax}\left(t;Q_{n_{\mathrm{max}}}\right)$ is allowed to fill the complete phase space
below it.

\paragraph{M\protect\scalebox{0.8}{E}P\protect\scalebox{0.8}{S}@N\protect\scalebox{0.8}{LO}.}
The described merging method can be extended to NLO accuracy in the description
of hard-jet production, by replacing the LO matrix elements with their respective NLO
counterpart. In the resulting \MEPSatNLO approach the exclusive $n$-jet contribution is given
by the \SMCatNLO expression~\cite{Hoeche:2011fd,Hoeche:2012yf,Gehrmann:2012yg,Hoeche:2014lxa,Hoeche:2014rya},
\begin{equation}\label{eq:meps-nlo-excl}
  \begin{split}
    \done \sigma_{n}^{\text{excl},\MEPSatNLO}
    =&\;
      \done\Phi_n\;\Bbar_n\left(\Phi_n\right)\;
      \Theta_n(\Qcut)\;
      \Fbarcal_n\left(\mu_Q^2;\Qcut\right)\\
    &\;{}+
     \done\Phi_{n+1}\;\mr{H}_n\left(\Phi_{n+1}\right)\;
     \Theta_n(\Qcut)\;
     \Theta\left(\Qcut - Q_{n+1}\right)\,
     \mathcal{F}_{n+1}\left(\mu_Q^2;\Qcut\right)\;.
  \end{split}
\end{equation}
The first line describes the so-called standard, or $\mm{S}$, events with Born kinematics. Their
weight is given by the familiar \Bbar-function, including among other terms the renormalised NLO QCD
virtual corrections. These $\mm{S}$-events are matched to a modified parton shower
\begin{equation}
  \begin{split}
    \Fbarcal_n\left(t;\Qcut\right)
    =&\;
      \overline{\Delta}_n\left(t_c,t\right)\,
      + \int_{t_c}^t
        \done\Phi_1^\prime\,\overline{\mr{K}}_n\left(\Phi_1^\prime\right)\,
        \Theta\left(\Qcut -Q_{n+1}(t^\prime)\right)\,
        \overline{\Delta}_{n}\left(t^\prime,t\right)
        \mathcal{F}_{n+1}\left(t^\prime\right)\;.
  \end{split}
\end{equation}
The modified Sudakov $\overline{\Delta}_n$ is determined by the modified
splitting kernels $\overline{\mr{K}}_n$, which differ from the standard parton-shower
kernels $\mr{K}_n$ in that they reproduce the exact soft colour- and collinear
spin-correlations of NLO QCD $n$-jet matrix elements. Further secondary
radiation is then generated through the standard shower \Fcal. In the second line of
Eq.~\eqref{eq:meps-nlo-excl}, $\mr{H}_n$ is defined
as $\mr{H}_n=\mr{R}_n-\mr{D}_n$, \emph{i.e.}\ as the
difference of the exact NLO QCD real-emission correction
$\mr{R}_n=\mr{B}_{n+1}$ and its soft-collinear approximation
as generated in the modified shower \Fbarcal, given by
$\mr{D}_n=\mr{B}_n\overline{\mr{K}}_n\,\Theta(\mu_Q^2-t)$. In consequence,
the corresponding hard-correction, or $\mm{H}$, events lift the emission
pattern to the exact NLO QCD expression. $\mm{H}$ events are dressed by
applying the standard shower \Fcal\ to their $n+1$ parton configuration.%
\footnote{
  It is interesting to note that \cite{Danziger:2021xvr} adds an additional Sudakov factor
  with respect to the $n$-parton configuration to the $\mm{H}$-events.
  This is done to, among other objectives, reduce the impact of negative weights.
  These additional Sudakov factors, however, are outside the
  accuracy targeted in this article and are thus not included
  in the presented argument.
}

Although a highest-multiplicity treatment in NLO multijet merging follows
along the same lines as in the LO case,
it is in practice never used. In typical applications yet higher-multiplicity
matrix elements can be calculated at LO.
Thus, the highest multiplicity, \nmax, will always be larger than the highest
multiplicity calculated at NLO, \nmaxnlo.
We multiply these additional LO processes at multiplicities
$n=\nmaxnlo+l$ with an additional $k$-factor
$k_n$ which supplies corrections beyond their perturbative orders, such
that the dependency on the merging scale \Qcut~\cite{Gutschow:2018tuk} is minimised
and the overall NLO accuracy is not affected.
The corresponding exclusive $n$-jet cross section is then given by
\begin{equation}\label{eq:menlops-excl}
  \begin{split}
    \done\sigma_{n>\nmaxnlo}^{\text{excl},\MEPSatNLO}
    =\;&
      \done\Phi_n\;
      k_{\nmaxnlo}\left(\Phi_{\nmaxnlo}\left(\Phi_n\right),\Phi_{\nmaxnlo+1}\left(\Phi_n\right)\right)\;
      \mr{B}_n\left(\Phi_n\right)\;
      \Theta_n(\Qcut)\;
      \Fcal_n\left(\mu_Q^2;\Qcut\right)\;.
  \end{split}
\end{equation}
With the kinematic mapping $\Phi_m(\Phi_{m+l})$ and
$\Phi_{m+1}(\Phi_{m+l})$ taken from the identified cluster
history, the local $k$-factor is defined as
\begin{equation}\label{eq:kfac}
  \begin{split}
    k_{m}\left(\Phi_m,\Phi_{m+1}\right)
    =&\;
      \frac{\Bbar_m\left(\Phi_m\right)}
           {\mr{B}_m\left(\Phi_m\right)}
      \left(1 - \frac{\mr{H}_m\left(\Phi_{m+1}\right)}
                     {\mr{B}_{m+1}\left(\Phi_{m+1}\right)}\right)
      +\frac{\mr{H}_m\left(\Phi_{m+1}\right)}
            {\mr{B}_{m+1}\left(\Phi_{m+1}\right)} \;.
  \end{split}
\end{equation}
This $k$-factor is constructed such that the NLO merged expression
with $\nmax=\nmaxnlo=m$ coincides with the merged expression with
$\nmaxnlo=m$ and $\nmax=m+1$.

\paragraph{M\protect\scalebox{0.8}{E}P\protect\scalebox{0.8}{S}@L\protect\scalebox{0.8}{OOP}$^2$.}
Finally, in this paper we are also considering loop-induced contributions
to four-lepton production in association with zero and one additional jet
and merge them into one inclusive sample.
To this end, we use the methods described in Refs.~\cite{Cascioli:2013gfa,Goncalves:2016bkl}.
In essence, the merging itself, dubbed \MEPSatLOOP, follows
the same principles as \MEPSatLO, see Eq.~\eqref{eq:meps-lo-excl}.
The only difference is that the Born matrix element $\mr{B}_n$
of each multiplicity is replaced with its loop-squared
counterpart.

\subsection{Incorporating EW corrections}
\label{sec:setup:ew}
The expressions for the QCD contributions to multijet-merged samples can
be combined with EW corrections, \deltanEW{n}. However, in order not to interfere
with the ordering of the QCD emissions in the matrix elements and parton
showers this requires well-defined approximations. We here consider the two
cases introduced in Sec.\ \ref{sec:fomatching:Ewsud},
the high-energy limit treated either in the \EWvirt\ or \EWsud\ approximation,
supplemented by final-state \emph{resolved} photon radiation off the charged leptons,
treated in the YFS resummation approach.
With such factorised technique we can account for EW corrections on top of QCD ones,
which would otherwise not be viable if we were to include the full set of NLO EW
contributions.

First we deal with the region dominated by virtual gauge-boson
exchange and/or the emission of a soft or collinear \emph{unresolved}
photon. In this scenario, \deltanEW{n}\ is local in the corresponding $n$-particle
phase space, affecting the weight of Born-, $\mm{S}$-, and $\mm{H}$-type events,
such that EW corrections are easily incorporated via
\begin{equation}\label{eq:setup:ew:deltanEW}
  \begin{split}
    \mr{B}_n\to\mr{B}_n\,\left(1+\deltanBEW{n}\mhhl\right)
  \end{split}
\end{equation}
for LO QCD contributions, and
\begin{equation}\label{eq:setup:ew:deltanEWSH}
  \begin{split}
    \Bbar_n\to\Bbar_n\,\left(1+\deltanSEW{n}\mhhl\right)
    \qquad\text{and}\qquad
    \mr{H}_n\to\mr{H}_n\,\left(1+\deltanHEW{n}\mhhl\right)
  \end{split}
\end{equation}
for NLO QCD $\mm{S}$- and $\mm{H}$-type contributions.\footnote{Of course,
the Born matrix element $\mr{B}_n$ appearing within $\Bbar_n$ does not
receive a second separate EW correction.}
The case where LO matrix elements are merged on top of NLO
matrix elements deserves special attention.
The key object in its construction is the local $k$-factor
$k_{n}$ which in turn contains lower-multiplicity Born-,
$\mm{S}$-, and $\mm{H}$ contributions, see Eq.~\eqref{eq:kfac}.
For both possibilities to construct the correction
factors \deltanEW{n}\ that we discuss in the following,
we also detail
how EW corrections are included in $k_n$ and how this
impacts the achieved accuracy.

Please note, for loop-induced processes we set all $\deltanEW{n}$ to
zero and only effect the YFS soft-photon resummation.

\paragraph{\EWvirt\ approximation.}
A first way to define the \deltanEW{n}\ is by using the exact
NLO EW virtual corrections, supplemented with real
corrections approximated in the soft-collinear limit
integrated over the extra-emission phase space.
This corresponds to the EW virtual approximation \EWvirt\
introduced in Sec.\ \ref{sec:fomatching:Ewsud}, with
\begin{equation}\label{eq:deltaEWvirt}
  \deltanSEWvirt{n}\left(\Phi_n\right) =
  \frac{\mathrm{V}_n^{\mathrm{EW}}\left(\Phi_n\right)
   + \mathrm{I}_n^{\mathrm{EW}}\left(\Phi_n\right)}
  {\mathrm{B}_n\left(\Phi_n\right)}
  \;,\qquad
  \deltanHEWvirt{n}\left(\Phi_{n+1}\right) =
  0\;,\qquad\text{and}\qquad
  \deltanBEWvirt{n}\left(\Phi_{n}\right)=0
  \;.
\end{equation}
As remarked in Sec.\ \ref{sec:fomatching:Ewsud},
the \EWvirt\ is in practice computationally expensive, and therefore
the approximation is
restricted to low-multiplicity processes.
These are typically the ones that QCD
one-loop matrix elements are computed for, \emph{i.e.}\ $n\leq\nmaxnlo$.
For the same reason the EW corrections for $\mm{H}$-event topologies
are neglected.
The higher-multiplicity LO processes receive an approximate
correction through the local $k$-factor $k_m$ with $m = \nmaxnlo$
defined in Eq.~\eqref{eq:kfac}.
With the definitions of Eq.~\eqref{eq:deltaEWvirt}, this $k$-factor
now takes the form
\begin{equation}\label{eq:kfacewvirt}
  \begin{split}
    k_{\text{virt},m}^\text{EW}\left(\Phi_m,\Phi_{m+1}\right)
    =&\;
      \frac{\Bbar_m\left(\Phi_m\right)\left(1+\deltanSEWvirt{m}(\Phi_m)\mhhl\right)}
      {\vphantom{\Bbar_m\left(\Phi_m\right)\left(1+\deltanSEWvirt{m}(\Phi_m)\mhhl\right)}
      \mr{B}_m\left(\Phi_m\right)}
      \left(1 - \frac{\mr{H}_m\left(\Phi_{m+1}\right)}
                     {\mr{B}_{m+1}\left(\Phi_{m+1}\right)}\right)
      +\frac{\mr{H}_m\left(\Phi_{m+1}\right)}
            {\mr{B}_{m+1}\left(\Phi_{m+1}\right)} \;.
  \end{split}
\end{equation}
This has the effect that in $(\nmaxnlo+l)$-jet events the underlying
\nmaxnlo-jet topology receives the correct EW correction.
In particular, the such-constructed approximate EW correction is
insensitive to additional jet production just above \Qcut.
However, once one of the additional jets enters the EW Sudakov regime,
for energy scales much larger than \Qcut, the effected correction will be incomplete.

\noindent
An alternative variant of \EWvirt\ is given by
\begin{equation}\label{eq:deltaEWvirtplus}
  \deltanSEWvirt{n}\left(\Phi_n\right) =
  \frac{\mathrm{V}_n^{\mathrm{EW}}\left(\Phi_n\right)
   + \mathrm{I}_n^{\mathrm{EW}}\left(\Phi_n\right)}
  {\Bbar_n\left(\Phi_n\right)}\;.
\end{equation}
This differs from Eq.~\eqref{eq:deltaEWvirt} by terms of relative $\order{\alpha_s\alpha}$.\footnote{
  It is important to note that this definition represents
  the original formulation of the method \cite{Kallweit:2015dum},
  and allows the incorporation of additional subleading
  tree-level contributions  of relative $\order{\alpha_s^{-1}\alpha}$
  with respect to $\mr{B}_n$, due to its complete cancellation
  of $\Bbar_n$ in the $\Bbar_n\,(1+\deltanSEWvirt{n})$ construction.
  Unwanted higher-order terms are thus avoided.
}
While a superficial correspondence of Eq.~\eqref{eq:deltaEWvirt}
to the multiplicative and Eq.~\eqref{eq:deltaEWvirtplus} to the
additive scheme used to combine QCD and EW corrections in
fixed-order calculations is tantalising, it is important to note
that these schemes combine relative correction factors
on the level of observable histograms, whereas the present
formulation operates point-wise in phase space prior to adding
higher-order QCD corrections through the parton shower.
As a result, both formulations automatically induce
terms of $\order{\alpha_s\alpha}$ relative to the Born
expression, and typically their difference is found to
be small~\cite{Brauer:2020kfv}.

\paragraph{\EWsud\ approximation.}
Another way to define the corrections \deltanSEW{n}, \deltanHEW{n}\ and \deltanBEW{n}\
is given by the EW Sudakov approximation, \emph{see} Sec.\ \ref{sec:fomatching:Ewsud},
\begin{equation}\label{eq:deltaEWsud}
  \deltanBEWsud{n}\left(\Phi_n\right) =
  \deltanSEWsud{n}\left(\Phi_n\right) =
  \KnEWsud{n}\left(\Phi_n\right)
  \qquad\text{and}\qquad
  \deltanHEWsud{n}\left(\Phi_{n+1}\right) =
  \KnEWsud{n+1}\left(\Phi_{n+1}\right)
  \,.
\end{equation}
Therein, the approximate NLO correction $\delta^{\text{EW}}_{\text{sud}}$ contains
all contributions up to NLL in the high-energy limit of the one-loop
EW corrections to the $n$-jet process.
The evaluation of the Sudakov corrections is only slightly more involved than
the underlying LO matrix-element computations and can therefore be applied to
all contributions of all multiplicities used in a given calculation, \emph{i.e.}\ Born-, $\mm{S}$-
and $\mm{H}$-type events.
In particular, all higher-multiplicity LO
contributions receive their own multiplicity-dependent correction
factor \deltanBEWsud{n}.
The correction factors in the local $k$-factor $k_n$ of Eq.~\eqref{eq:kfacewvirt}
cancel, thus avoiding a double counting of the higher-order EW effects.
In consequence, and in contrast to the \EWvirt\ approach, the additional
LO multiplicities at $n>\nmaxnlo$ receive the complete EW Sudakov factor
when all jets are produced in the EW Sudakov regime.

\paragraph{Exponentiated \EWsud\ corrections.}
The all-orders resummation of the NLL EW Sudakov logarithms is achieved by
the replacement $1+\deltanEW{n}\to\exp{\left(\deltanEW{n}\right)}$ throughout.
Though, as discussed in Sec.\ \ref{sec:fomatching:Ewsud}, this does
not generally hold for the \EWvirt\ approximation,
where a naive exponentiation would also include non-universal
finite terms, thus introducing an error that depends on their relative size.
However, the \EWsud\ approximation is ideally suited for this task.
In particular, in the case of a LO QCD calculation
we modify Eq.~\eqref{eq:setup:ew:deltanEW} to
\begin{equation}
  \begin{split}
    \mr{B}_n\to\mr{B}_n\,\exp\left(\deltanBEWsud{n}\mhhl\right),
  \end{split}
\end{equation}
while in the NLO QCD case, in analogy to Sec.\ \ref{sec:fomatching:match}, Eq.~\eqref{eq:setup:ew:deltanEWSH} becomes
\begin{equation}
  \begin{split}
    \Bbar_n\to\Bbar_n\,\exp\left(\deltanSEWsud{n}\mhhl\right)
    \qquad\text{and}\qquad
    \mr{H}_n\to\mr{H}_n\,\exp\left(\deltanHEWsud{n}\mhhl\right)\;.
  \end{split}
\end{equation}

\paragraph{Matched NLO \EWvirt\ + NLL Sudakov.}
Due to the absence of a suitable
matching implementation that achieves full NLO EW accuracy for
inclusive observables in the present formalism, we match the
resummed \EWsud\ corrections to the NLO \EWvirt\ ones.
Although there is no improvement in the formal accuracy of such a
matching---both the \EWvirt\ and \EWsud\ approximations have formal
NLL accuracy in the EW Sudakov regime---such a matched calculation
benefits from the combination of the better handling of EW
renormalisation-scheme dependence and phenomenologically important
finite $\order{\alpha}$ terms included in the \EWvirt\ scheme
on the one hand side, and the improved all-orders structure of
the resummed \EWsud corrections on the other.
We thus set
\begin{equation}
  \begin{split}
    \mr{B}_n\to\mr{B}_n\,\exp\left(\deltanBEWsud{n}\mhhl\right),
  \end{split}
\end{equation}
while in the NLO QCD case, Eq.~\eqref{eq:setup:ew:deltanEWSH}
becomes
\begin{equation}\label{eq:setup:deltamatchSH}
  \begin{split}
    \Bbar_n\to\Bbar_n\,\left[\exp\left(\deltanSEWsud{n}\mhhl\right)-\deltanSEWsud{n}+\deltanSEWvirt{n}\right]
    \qquad\text{and}\qquad
    \mr{H}_n\to\mr{H}_n\,\exp\left(\deltanHEWsud{n}\mhhl\right)\;.
  \end{split}
\end{equation}
As is evident, neither the structure of the resummation nor that of
the \EWvirt approximation at $\order{\alpha}$ have been upset.
With the above choice and setting
$\deltanEWsud{n}=\deltanBEWsud{n}=\deltanSEWsud{n}=\deltanHEWsud{n-1}=\KnEWsud{n}$,
the local $k$-factor gets
modified to
\begin{equation}\label{eq:kfacmatch}
  \begin{split}
    k_{\text{matched},n}^\text{EW}\left(\Phi_n,\Phi_{n+1}\right)
    =&\;
      \frac{\Bbar_n\left(\Phi_n\right)
            \left[\exp\left(\deltanEWsud{n}(\Phi_n)\mhhl\right)
                  -\deltanEWsud{n}(\Phi_n)+\deltanSEWvirt{n}(\Phi_n)\right]}
           {\mr{B}_n\left(\Phi_n\right)
            \exp\left(\deltanEWsud{n}(\Phi_n)\mhhl\right)}
      \left(1 - \frac{\mr{H}_n\left(\Phi_{n+1}\right)}
                     {\mr{B}_{n+1}\left(\Phi_{n+1}\right)}\right)\\
    &{}
      +\frac{\mr{H}_n\left(\Phi_{n+1}\right)}
            {\mr{B}_{n+1}\left(\Phi_{n+1}\right)} \;.\\[-5mm]
  \end{split}
\end{equation}
As \deltanSEWvirt{n} and \deltanEWsud{n} have the same formal
accuracy, $k_{\text{matched},n}^\text{EW}$ contains no EW corrections
to NLL accuracy.
However, beyond the formal accuracy one can show that the inclusive
behaviour of the \EWvirt\ approximation is preserved with the
above definitions, see App.\ \ref{sec:app:match}.

With the formulae presented so far, an automatic implementation of the full
matched result is in principle a technical matter only, which, however, we
leave for future work.

\paragraph*{Soft-photon resummation.}
The inherent approximation of the above \EWvirt\ and \EWsud\
constructions can be partially unfolded again by
adding the effects of final-state photon radiation.
In particular, we use the soft-photon resummation in the YFS
scheme~\cite{Yennie:1961ad}. The \Sherpa implementation described in
\cite{Schonherr:2008av} is restricted to photon emission off
final-state leptons in order not to interfere with the strongly
ordered resummation of QCD radiation in the parton shower.
For both same-flavour lepton pairs, \emph{i.e.}\ $e^+e^-$ and $\mu^+\mu^-$,
it constructs a pseudo-resonant $Z$-boson decay and then
corrects the apparent LO decay width $\done\Gamma_0$
to the all-orders resummed decay rate
\begin{equation}\label{eq:setup:yfsmaster}
  \begin{split}
    \done\Gamma^\text{YFS}
    =\;&
      \done\Gamma_0\cdot e^{\alpha Y(\omega_\text{cut})}\cdot
      \sum\limits_{n_\gamma}\frac{1}{n_\gamma!}
      \left[
        \prod\limits_{i=1}^{n_\gamma}\done \Phi_{k_i}\cdot\alpha\,
        \tilde{S}(k_i)\,\Theta(k_i^0-\omega_\text{cut})
      \right]
      \cdot\mathcal{C}
      \,.
  \end{split}
\end{equation}
Therein, the YFS form factor $Y(\omega_\text{cut})$ resums
unresolved real and virtual soft-photon corrections.
Individual resolved photons with momenta $k_i$ are distributed according
to the eikonal $\tilde{S}(k_i)$ in their phase space $\Phi_{k_i}$.
The parameter $\omega_\text{cut}$ separates the explicitly-generated resolved
from the integrated-over unresolved real-photon emission phase-space
regions. As default value we use $\omega_\text{cut}=1\,\text{MeV}$.
The correction factor $\mathcal{C}$ contains exact higher-order
corrections which we incorporate up to NLO in QED.\footnote{
  Although NNLO QED + NLO EW corrections are available for
  $Z\to\ell\ell$ decays \cite{Krauss:2018djz}, their impact
  is numerically too insignificant to be included in this study.
}

It is important to note that the constructions used here,
\LOPlusEWVirtYFS\ and \LOEWSudYFS, do not achieve full
NLO EW accuracy.
In particular, there is an overlap in the virtual corrections
and the unresolved integrated real-emission corrections
generated in the YFS resummation and the \EWvirt\
and \EWsud\ approximations.
This overlap, however, is non-logarithmic in the high-energy
limit where our approximation is valid, and therefore does
not compromise its accuracy.
On the contrary, the addition of a detailed all-orders
description of final-state radiation allows for an accurate
calculation of realistic fiducial cross sections.

\section{$ZZ$ production in association with jets}\label{sec:applZZ}
As testbed for the validation and comparison of the calculational schemes
to include approximate EW NLO corrections in a multijet-merged computation,
we consider the hadronic production of a pair of $Z$-bosons in proton--proton collisions,
\emph{i.e.}\ the four-lepton final state $e^+e^-\mu^+\mu^-$. We first compile exact NLO
EW results for the $pp \to e^+e^-\mu^+\mu^-$ and $pp \to e^+e^-\mu^+\mu^-j$ processes. These
serve as benchmarks for the \EWvirt\ and \EWsud\ approximations.\footnote{
  The \EWvirt approximation for inclusive $ZZ$ production has been already
  investigated in \cite{Gutschow:2020cug} and an excellent reproduction
  of the exact NLO EW distributions was found.
} Together with the fixed-order benchmarks we give a detailed description of
the effects of matching the exponentiated \EWsud\ to either LO and NLO. We then move on
to study multijet merging based on the $ZZ+0,1j$ NLO QCD and $ZZ+2,3j$ LO QCD
matrix elements. A similar study for $WW$ production has been presented in
Ref.~\cite{Brauer:2020kfv}. However, there the EW Sudakov approximation was not
considered.

\subsection{Contributions at Born-level and NLO EW}
\label{sec:res_contribs}

We first review the contributions which the two processes,
\emph{i.e.}\ $pp\to e^+e^-\mu^+\mu^-$ both
inclusively and in association with at least one additional parton,
comprise at NLO EW.

\subsubsection*{Inclusive $e^+e^-\mu^+\mu^-$ production}
\begin{figure}[t!]
  \captionsetup[subfigure]{justification=centering}
  \begin{subfigure}{0.5\textwidth}
  \centering
    \includegraphics[height=\diagheight]{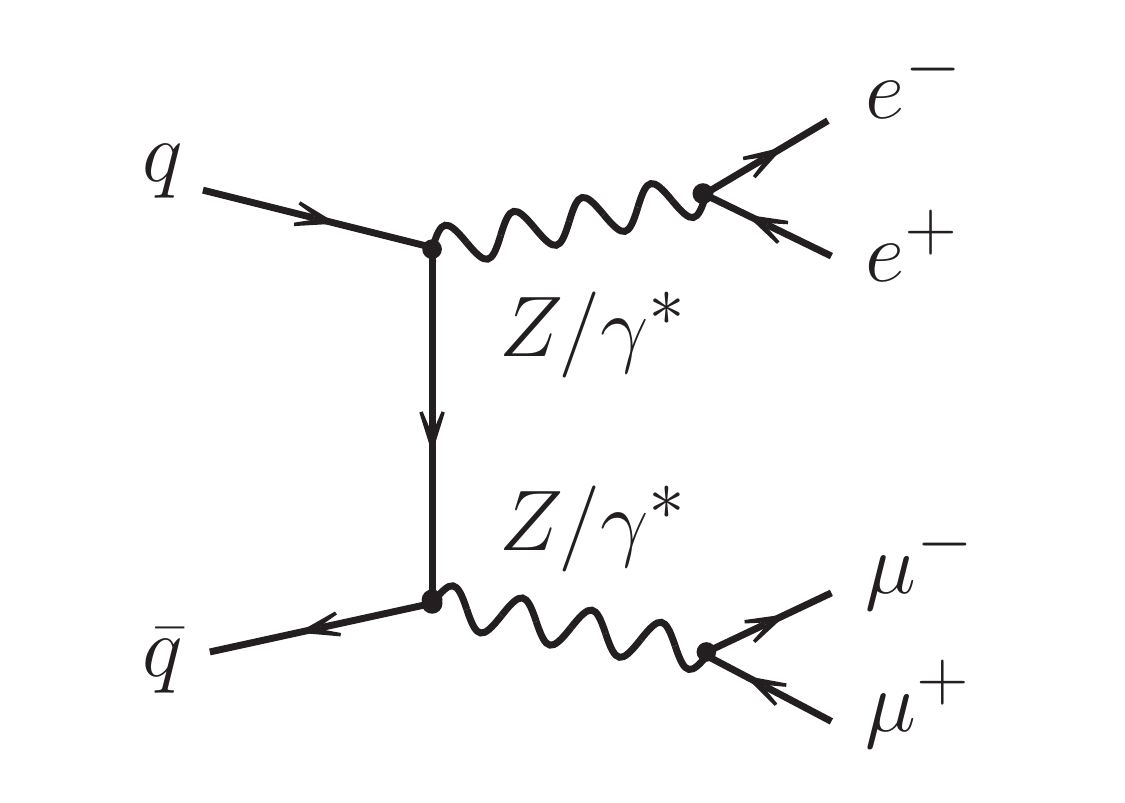}
    \hspace*{\diagsepB}
    \includegraphics[height=\diagheight]{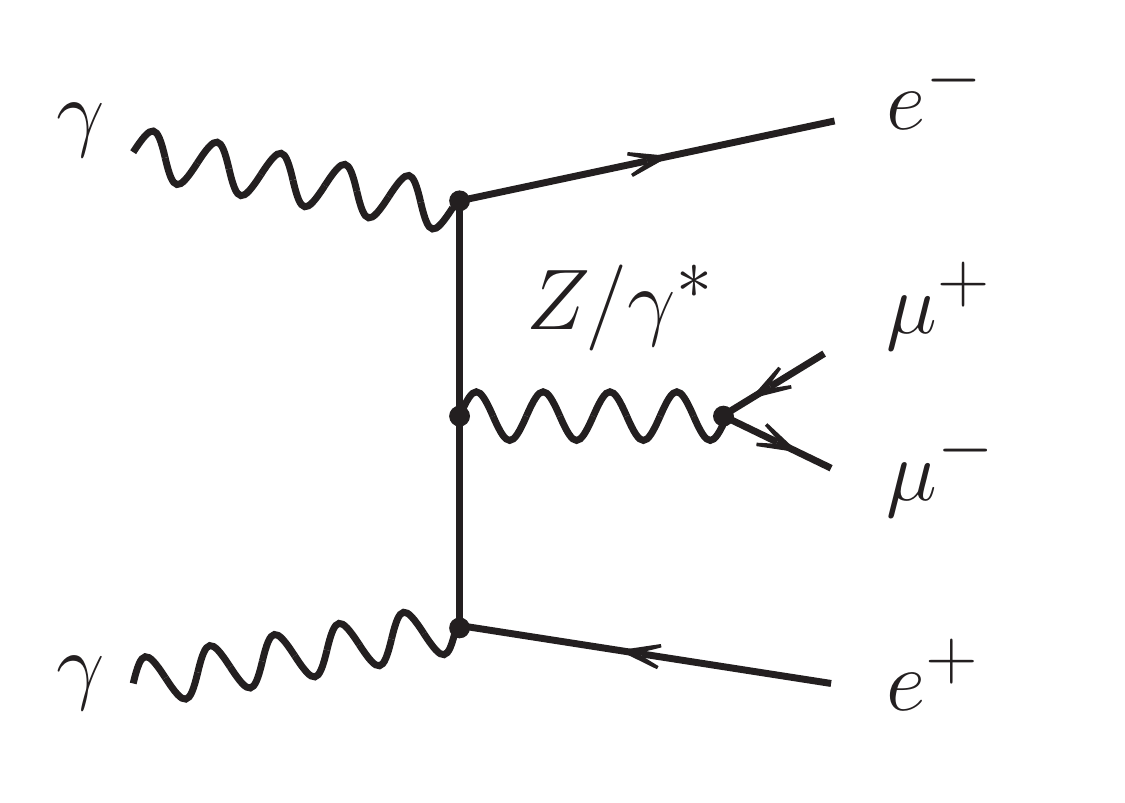}
    \caption{}
    \label{fig:diags:4l-B}
  \end{subfigure}
  \hfill
  \begin{subfigure}{0.5\textwidth}
    \includegraphics[height=\diagheight]{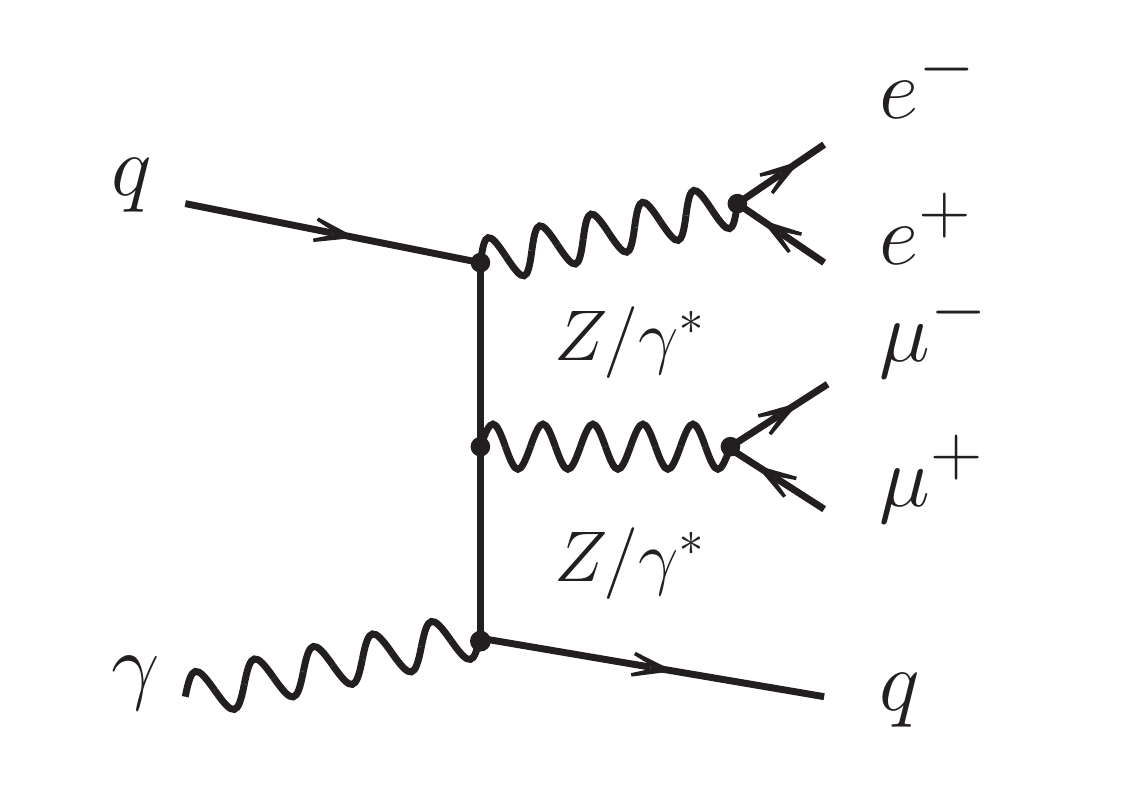}
    \hspace*{\diagsepB}
    \includegraphics[height=\diagheight]{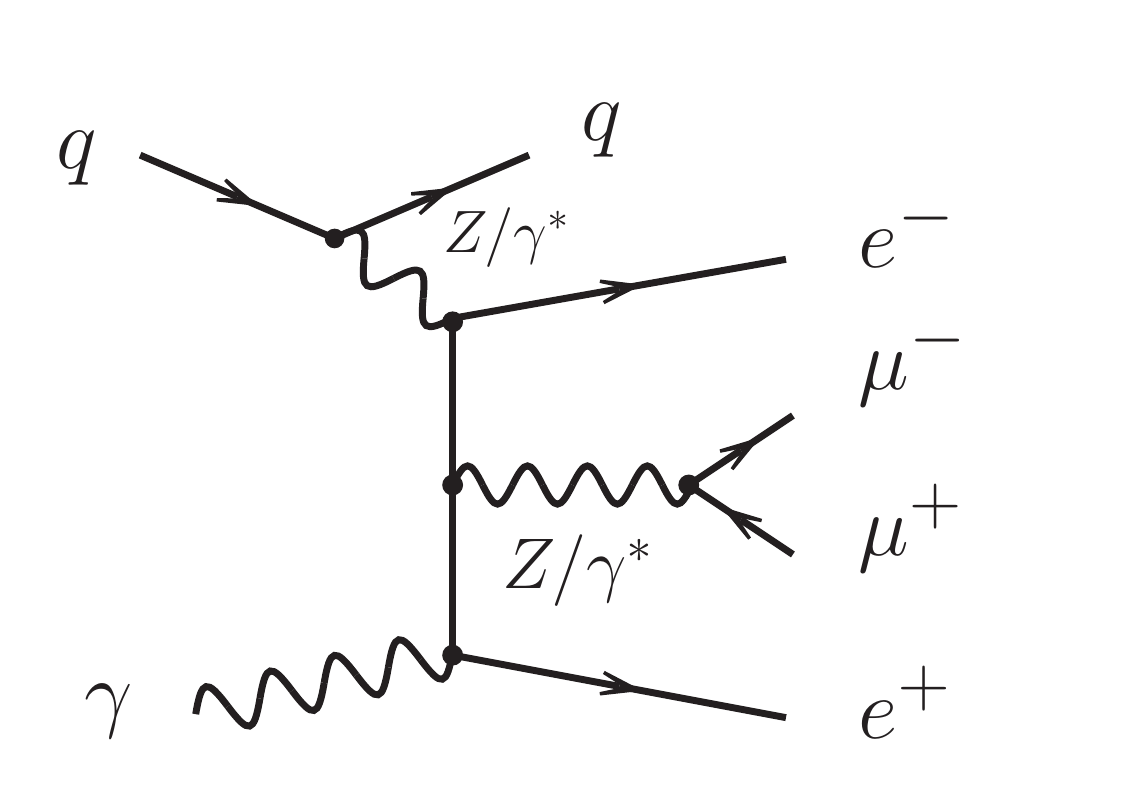}
  \caption{}
    \label{fig:diags:4l-R}
  \end{subfigure}\\[3mm]
  \caption{%
    Example Born-level (a) and real-emissions diagrams (b) contributing to
    $pp\to e^+e^-\mu^+\mu^-+X$ at $\order{\alpha^4}$ and $\order{\alpha^5}$, respectively.
  }
  \label{fig:diags:4l-BR}
\end{figure}

\begin{figure}[t!]
  \centering
  \includegraphics[height=\diagheight]{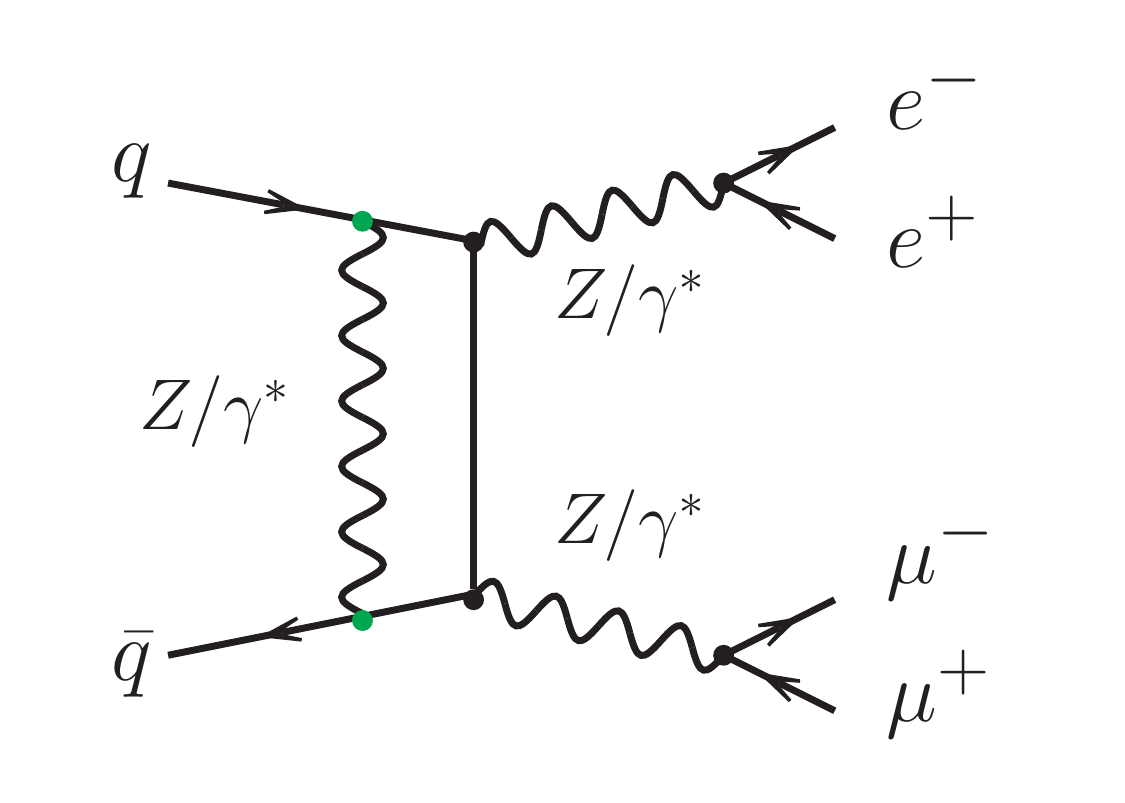}
  \hspace*{\diagsepB}
  \includegraphics[height=\diagheight]{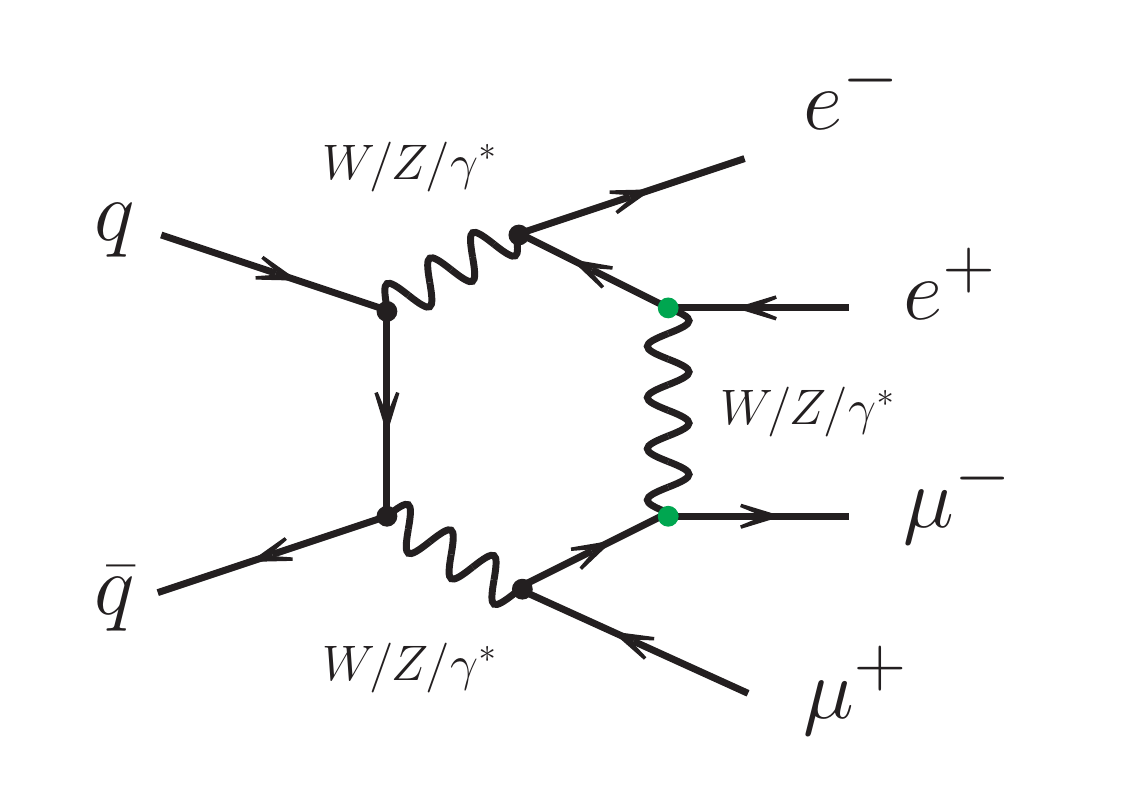}
  \hspace*{\diagsepB}
  \includegraphics[height=\diagheight]{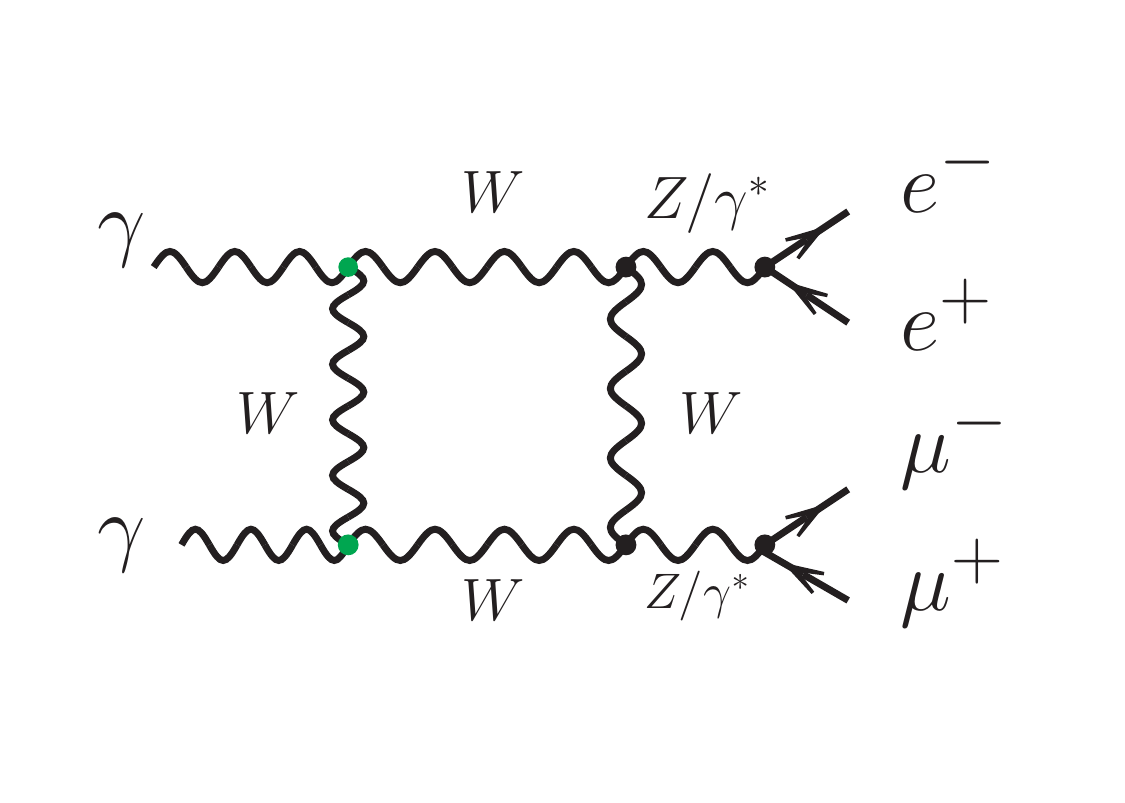}
  \hspace*{\diagsepB}
  \includegraphics[height=\diagheight]{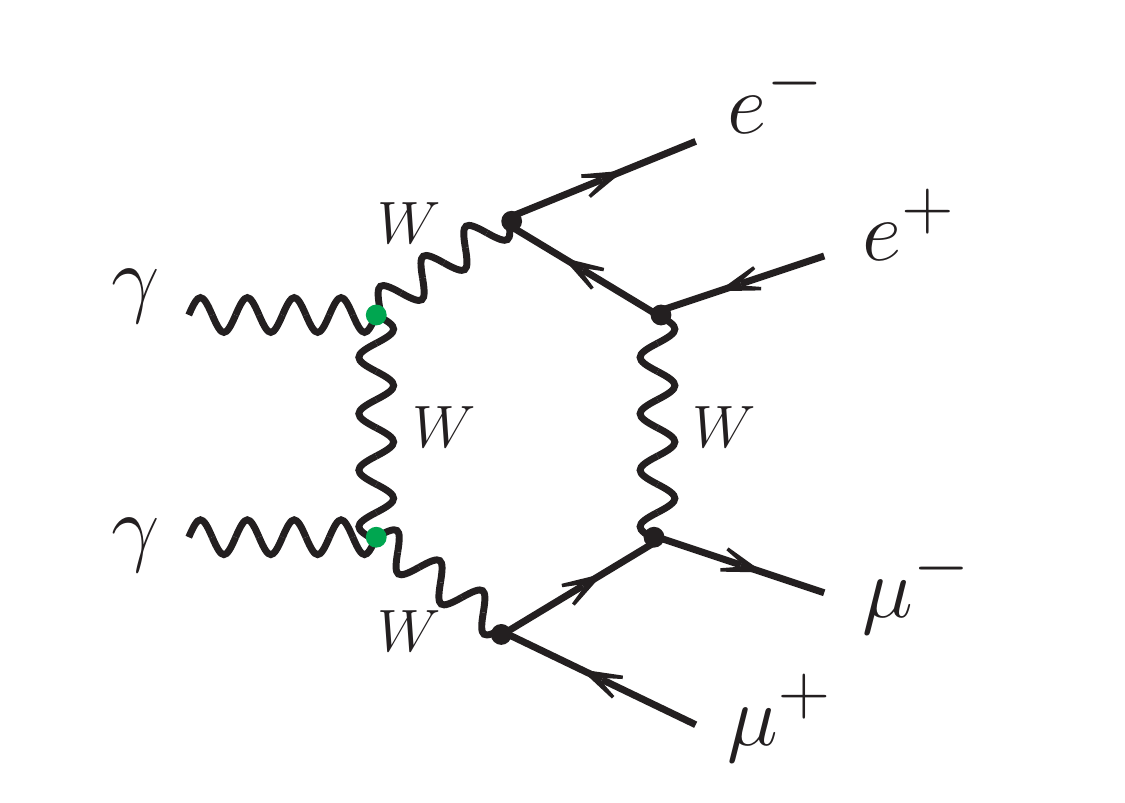}
  \caption{%
    Example one-loop diagrams contributing to $pp\to e^+e^-\mu^+\mu^-$
    at $\order{\alpha^5}$.
  }
  \label{fig:diags:4l-V}
\end{figure}

The partonic processes contributing to the
LO cross section are given by
\begin{equation}
  q\,\bar{q}\to e^+e^-\mu^+\mu^-\quad\text{and}\quad
  \gamma\,\gamma\to e^+e^-\mu^+\mu^-\quad \text{at}\quad \order{\alpha^4}\;.\nnb
\end{equation}
Two corresponding example diagrams are displayed in Fig.\ \ref{fig:diags:4l-B}.
Although the photon-induced contribution is numerically small (with at most singly-resonant
diagrams contributing), it appears at the same order and must be included in
a consistent calculation at NLO EW. The one-loop virtual corrections naturally comprise the same
partonic channels as the LO. In Fig.\ \ref{fig:diags:4l-V} we give some illustrative
example diagrams. This includes hexagon graphs connecting
all initial- and final-state particles with both massive and massless propagators.
For the real-emission corrections new partonic channels open up, here in particular
\begin{equation}
  q\,\bar{q}\to e^+e^-\mu^+\mu^-\,\gamma\,,\quad
  \gamma\,\qbarbracket\to e^+e^-\mu^+\mu^-\,\qbarbracket\quad\text{and}\quad
  \gamma\,\gamma\to e^+e^-\mu^+\mu^-\,\gamma\quad \text{at}\quad \order{\alpha^5}\;.\nnb
\end{equation}
It needs to be noted that the $\gamma q$-initiated channels
contain collinear divergences that cancel corresponding poles in
both the $q\bar{q}$- and $\gamma\gamma$-channel in the virtual corrections
and, thus, link both LO production modes. Illustrative examples of
such contributions are depicted in Fig.\ \ref{fig:diags:4l-R}.

\begin{figure}[t!]
  \centering
   \includegraphics[height=\diagheight]{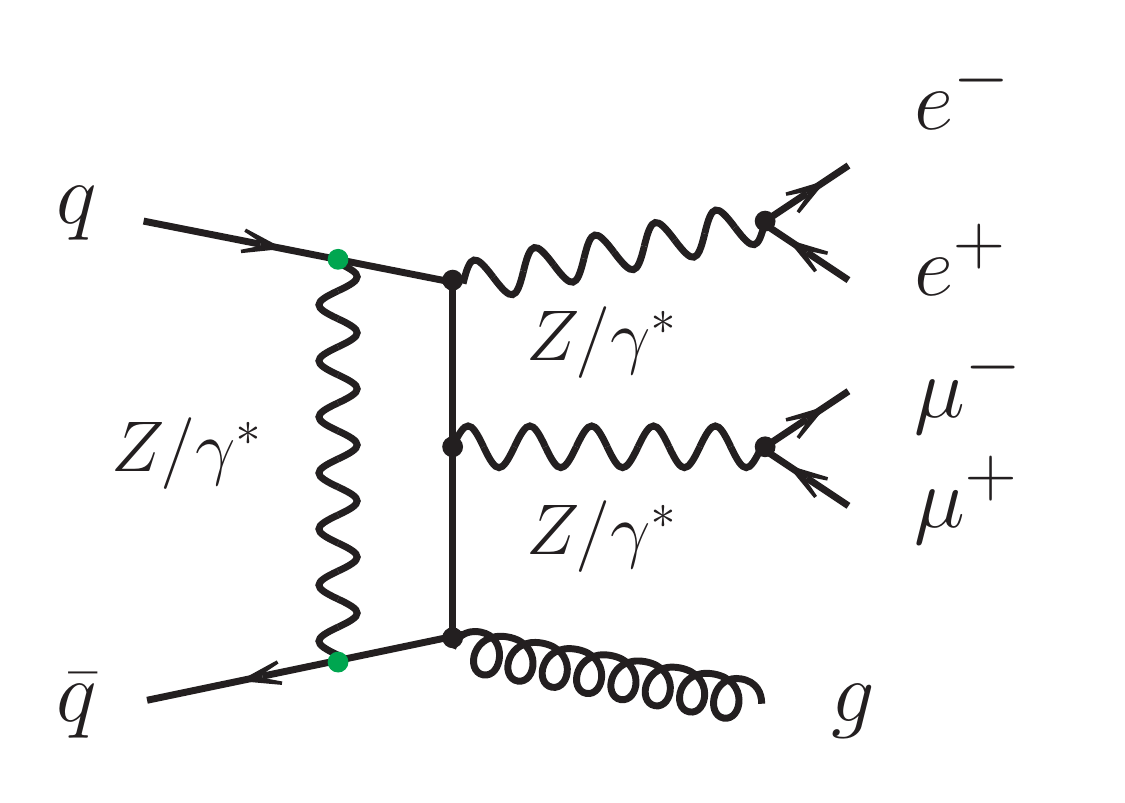}
  \hspace*{\diagsepB}
  \includegraphics[height=\diagheight]{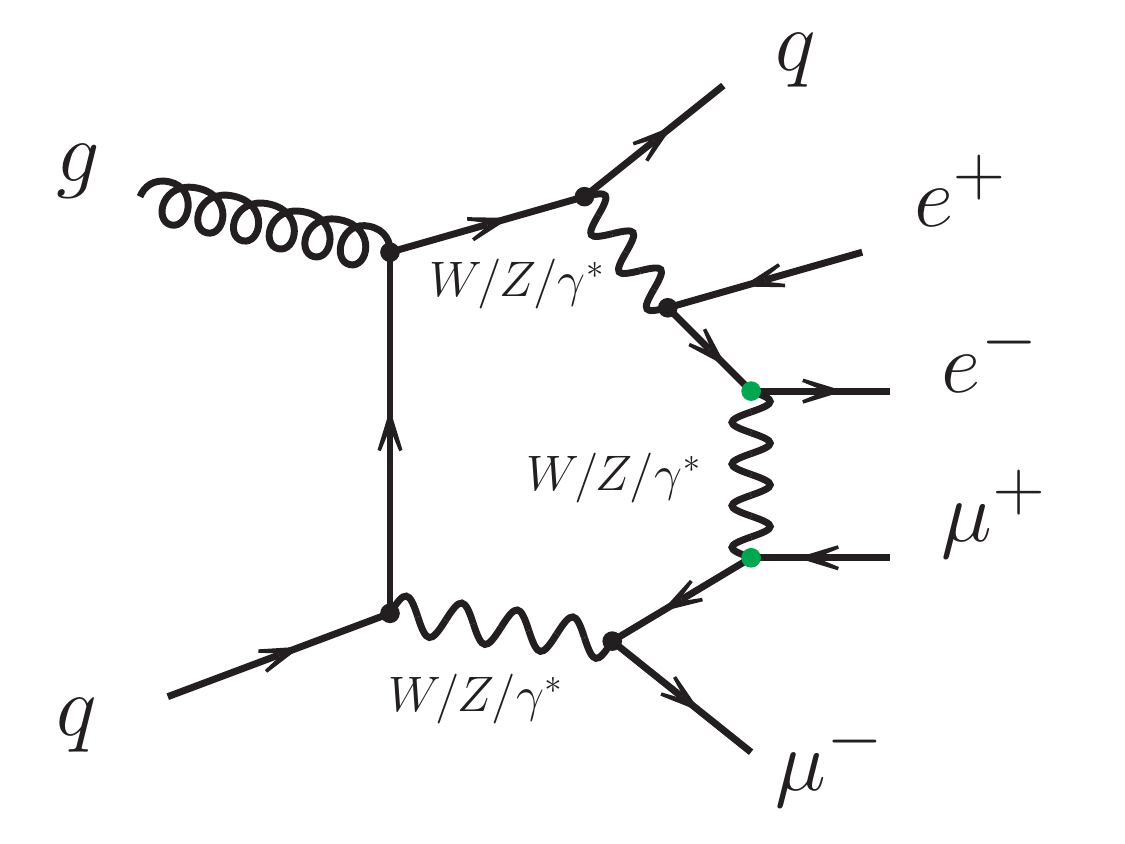}
  \hspace*{\diagsepB}
  \includegraphics[height=\diagheight]{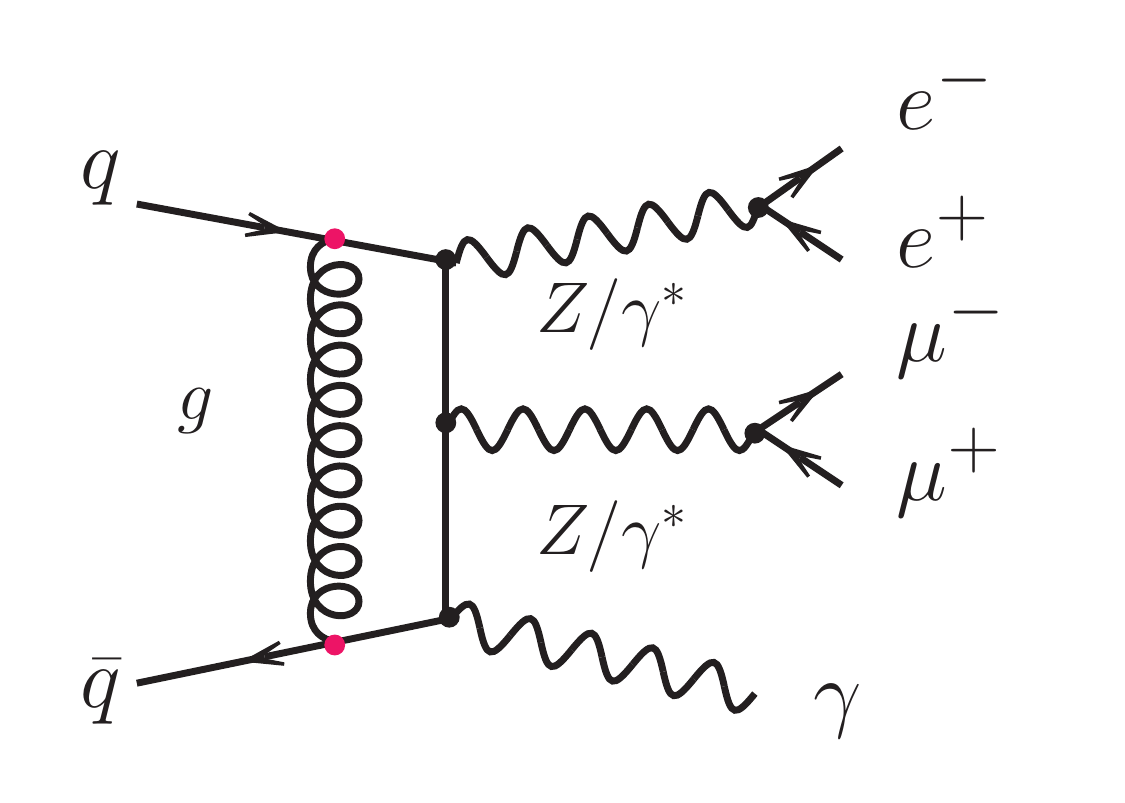}
   \hspace*{\diagsepB}
  \includegraphics[height=\diagheight]{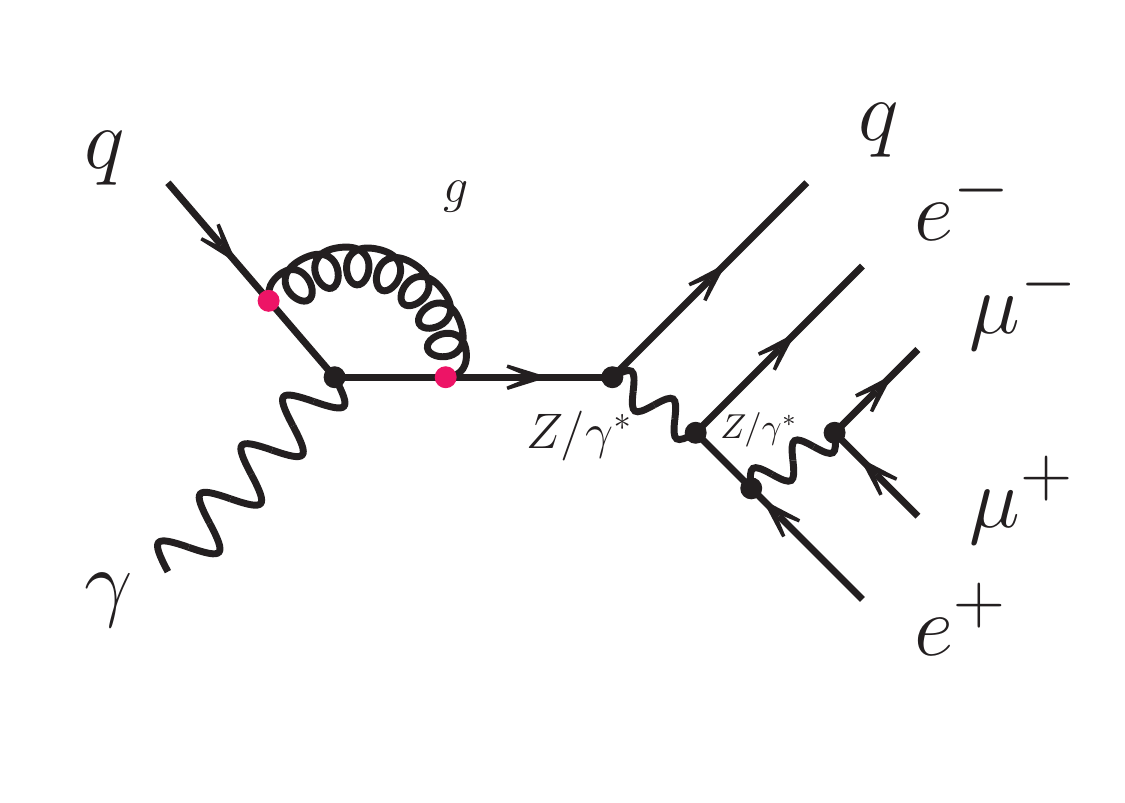}
  \caption{%
    Example one-loop diagrams contributing to $pp\to e^+e^-\mu^+\mu^-j$
    at $\order{\alpha_s\alpha^5}$.
  }
  \label{fig:diags:4lj-V}
\end{figure}

\begin{figure}[t!]
  \centering
  \includegraphics[width=0.35\textwidth]{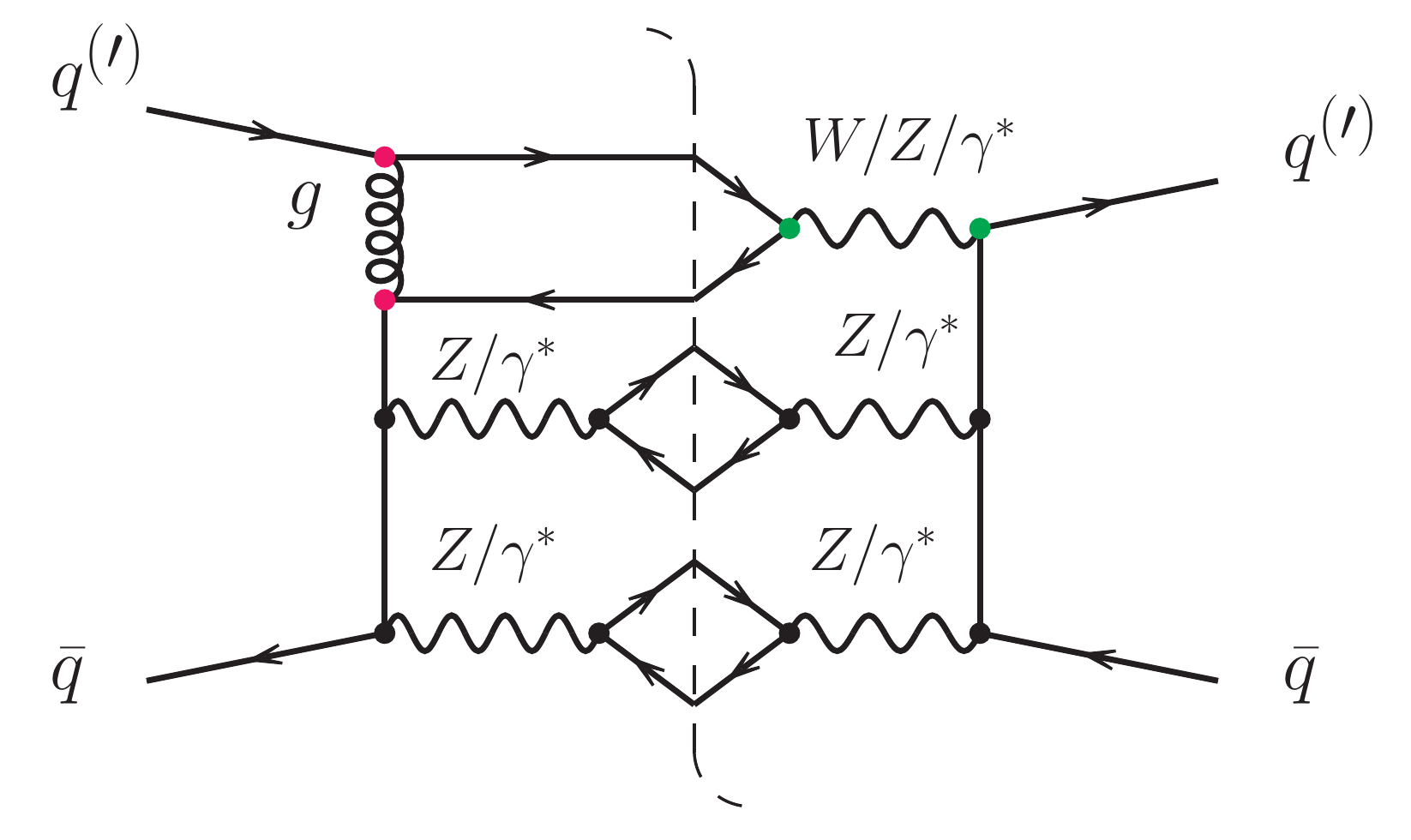}
  \includegraphics[width=0.35\textwidth]{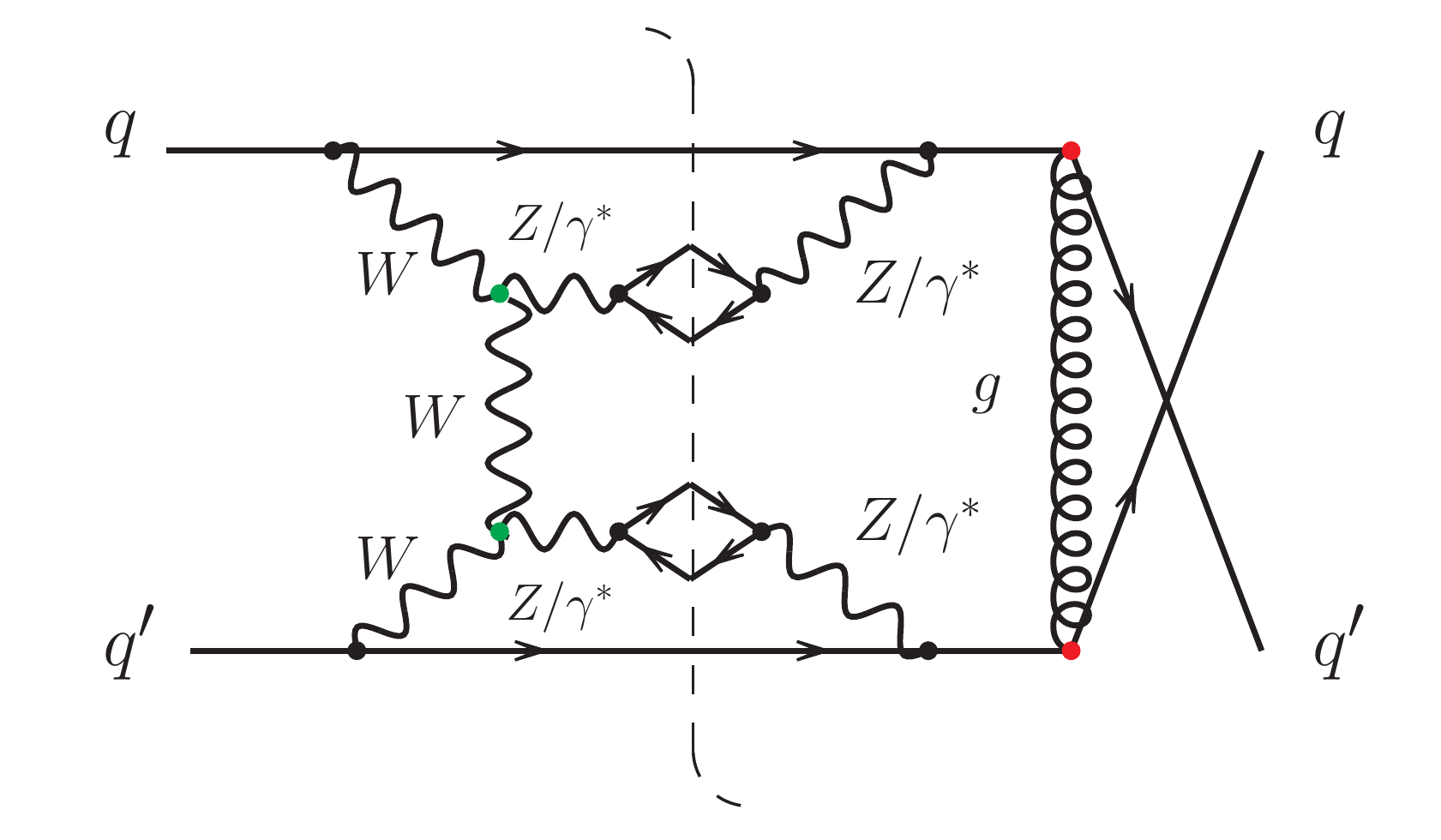}
  \caption{%
    Example real-emission QCD--EW interference contributions in
    $\qbarbracket\qbarbracket\to e^+e^-\mu^+\mu^-\qbarbracket\qbarbracket$.}
  \label{fig:diags:4ljj-int}
\end{figure}

\subsubsection*{Jet-associated production}

The leading-order channels contributing to the associated production of
$e^+e^-\mu^+\mu^-$ with at least one additional parton which can
constitute a jet are given by
\begin{equation}
  \begin{split}
    &
    q\,\bar{q}\to e^+e^-\mu^+\mu^-\,g\quad\text{and}\quad
    g\,\qbarbracket\to e^+e^-\mu^+\mu^-\,\qbarbracket\quad \text{at}\quad \order{\alpha_s\alpha^4}\;.\nnb
  \end{split}
\end{equation}
Corresponding Born-level diagrams can be easily visualised by
attaching an additional external gluon to $q\bar{q}$-initiated graphs as the one depicted in Fig.\ \ref{fig:diags:4l-B}.
The set of partonic channels contributing to the one-jet process at the one-loop level contains
\begin{equation}
  \begin{split}
    &
    q\,\bar{q}\to e^+e^-\mu^+\mu^-\,g\,,\quad
    g\,\qbarbracket\to e^+e^-\mu^+\mu^-\,\qbarbracket\quad \text{and}\quad\\
    &
    q\,\bar{q}\to e^+e^-\mu^+\mu^-\,\gamma\,,\quad
    \gamma\,\qbarbracket\to e^+e^-\mu^+\mu^-\,\qbarbracket\quad \text{at}\quad \order{\alpha_s\alpha^5}\;.\nnb
  \end{split}
\end{equation}
In contrast to the inclusive case, new channels emerge, where the external
gluon gets replaced by a photon. So besides the canonical EW loops inserted in the LO
channels, also QCD-loop corrections to channels with external photons, contributing
at tree-level only in subleading orders, appear.
See Fig.\ \ref{fig:diags:4lj-V} for examples of one-loop amplitudes corresponding to the
two types of processes.
New channels also open for the real corrections:
\begin{equation}
  \begin{split}
    &
    q\,\bar{q}\to e^+e^-\mu^+\mu^-\,g\,\gamma\,,\quad
    g\,\qbarbracket\to e^+e^-\mu^+\mu^-\,\gamma\,\qbarbracket\,,\quad\\
    &
    \gamma\,\qbarbracket\to e^+e^-\mu^+\mu^-\,g\,\qbarbracket\,,\quad
    \gamma\,g\to e^+e^-\mu^+\mu^-\,q\,\bar{q}\quad\text{and}\quad
    \qbarbracket\,\qbarbracket\to e^+e^-\mu^+\mu^-\,\qbarbracket\,\qbarbracket
    \quad \text{at}\quad \order{\alpha_s\alpha^5}\;.\nnb
  \end{split}
\end{equation}
The last process, involving four external quarks and no external
photon, deserves an explicit discussion.
This contribution is completely infrared finite and, thus,
separable from the other processes.
It consists of an interference of diagrams of $\order{g_s^2 e^4}$
and $\order{e^6}$, commonly referred to as QCD production
and EW production, respectively.
Due to the colour algebra involved they typically entail
$s$-/$t$-channel or $t$-/$u$-channel interferences.
Two examples are given in Fig.\ \ref{fig:diags:4ljj-int}.
It is noteworthy that, due to their nature as interference terms,
they are of indeterminate sign, and are typically small for inclusive
observables.
However, as they are the only contributions at this order which
can contain two valence quarks as initial states, \textit{e.g.}
$uu$, $ud$, or $dd$, they can still be quite sizeable in the TeV range.

\subsubsection*{Loop-induced production}

\begin{figure}[t!]
  \captionsetup[subfigure]{justification=centering}
  \begin{subfigure}{0.5\textwidth}
    \centering
    \includegraphics[height=\diagheight]{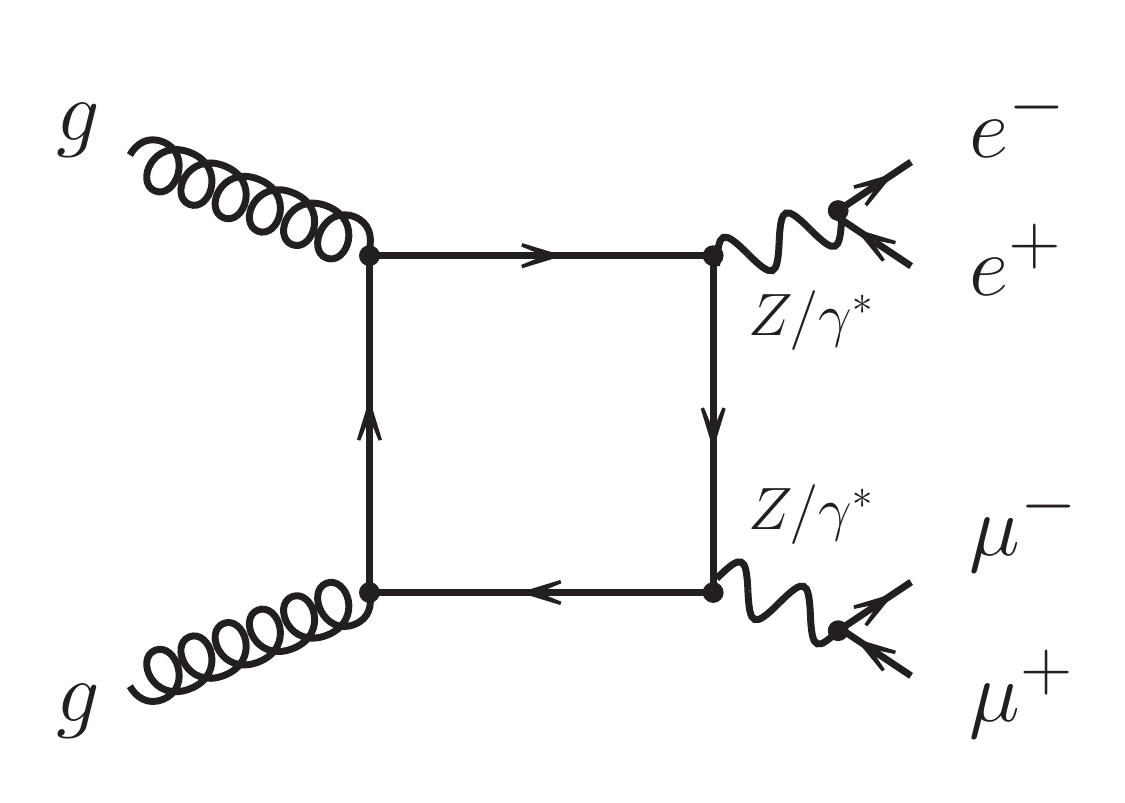}\\[2mm]
    \includegraphics[height=\diagheight]{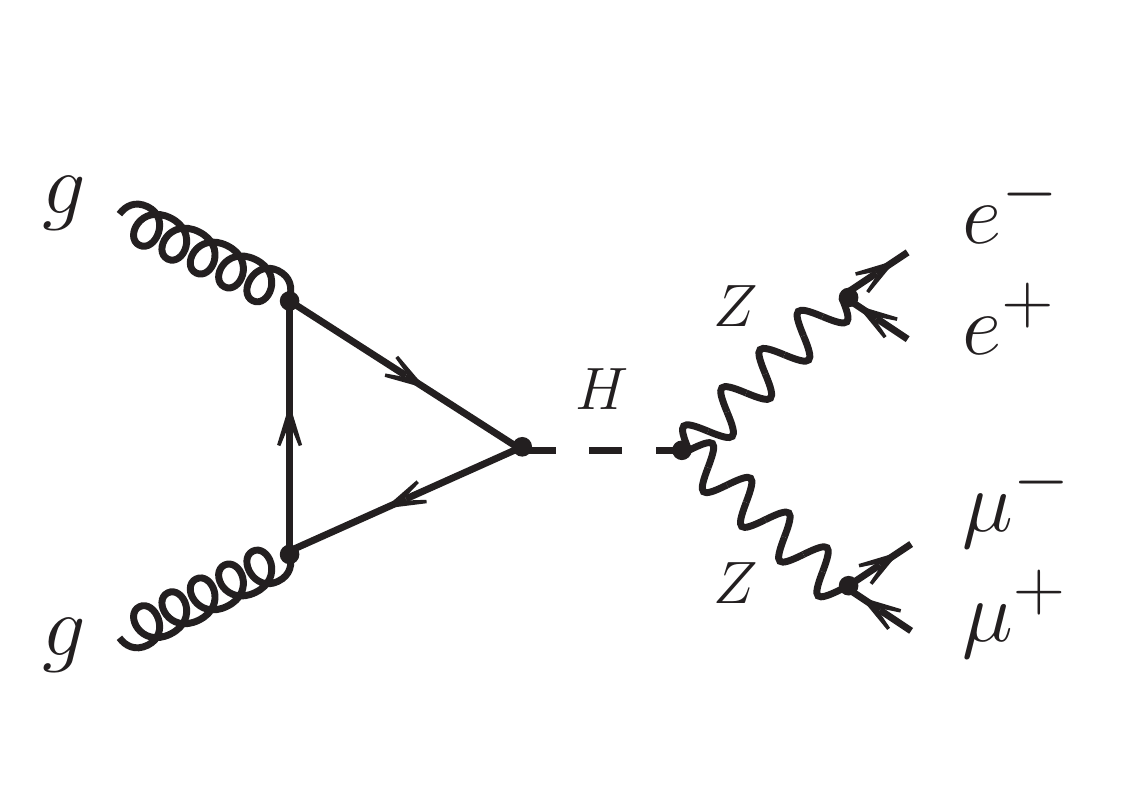}
    \hspace*{\diagsepB}
    \includegraphics[height=\diagheight]{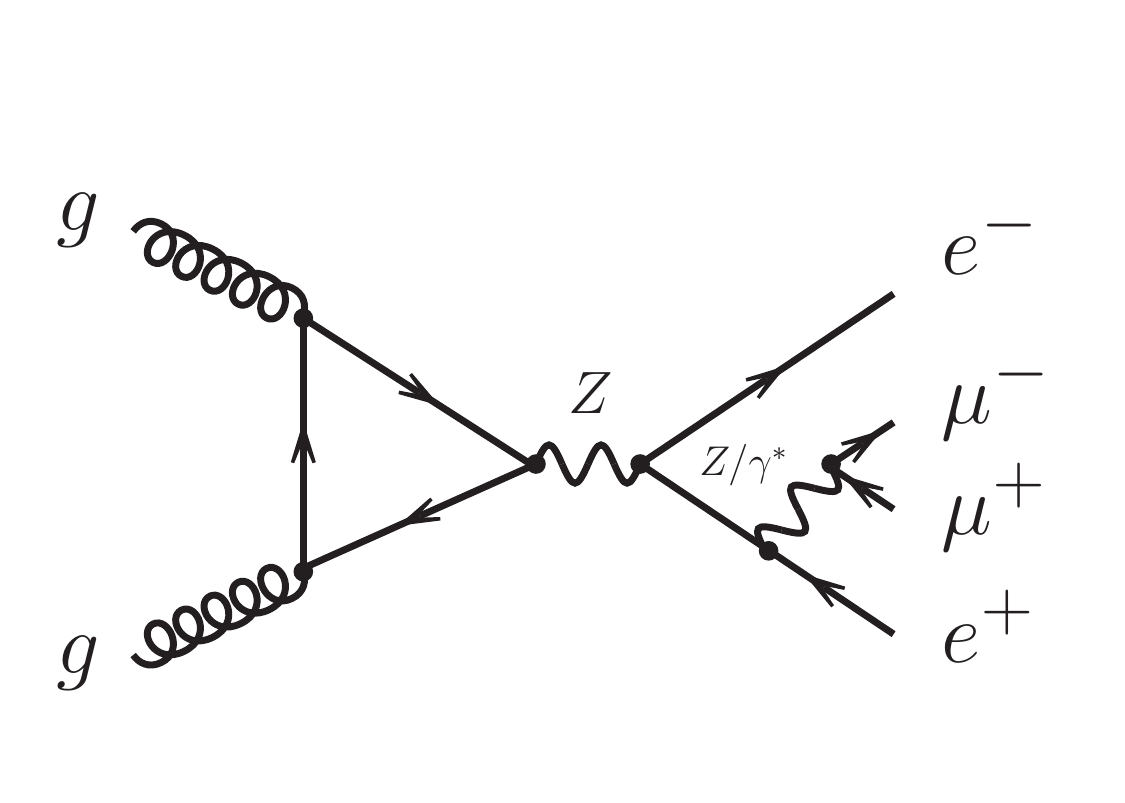}
    \caption{}
    \label{fig:diags:4l-LI}
  \end{subfigure}
  \hfill
  \begin{subfigure}{0.5\textwidth}
    \centering
    \includegraphics[height=\diagheight]{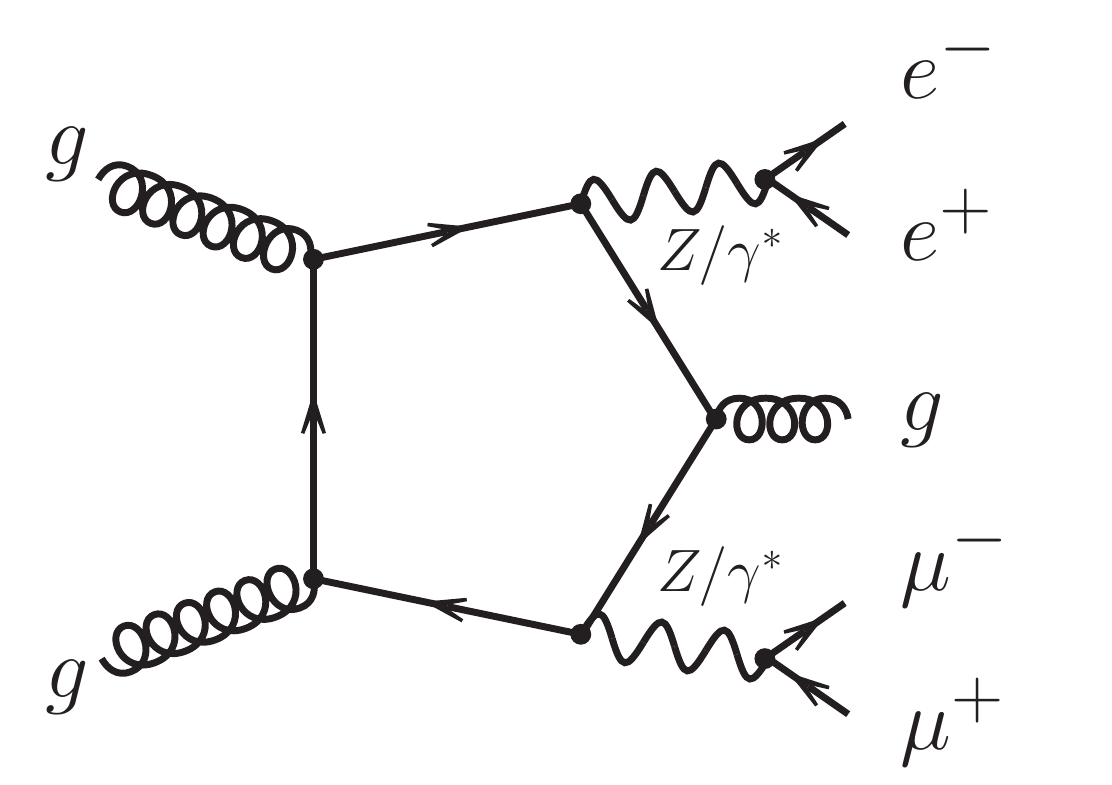}
    \hspace*{\diagsepB}
    \includegraphics[height=\diagheight]{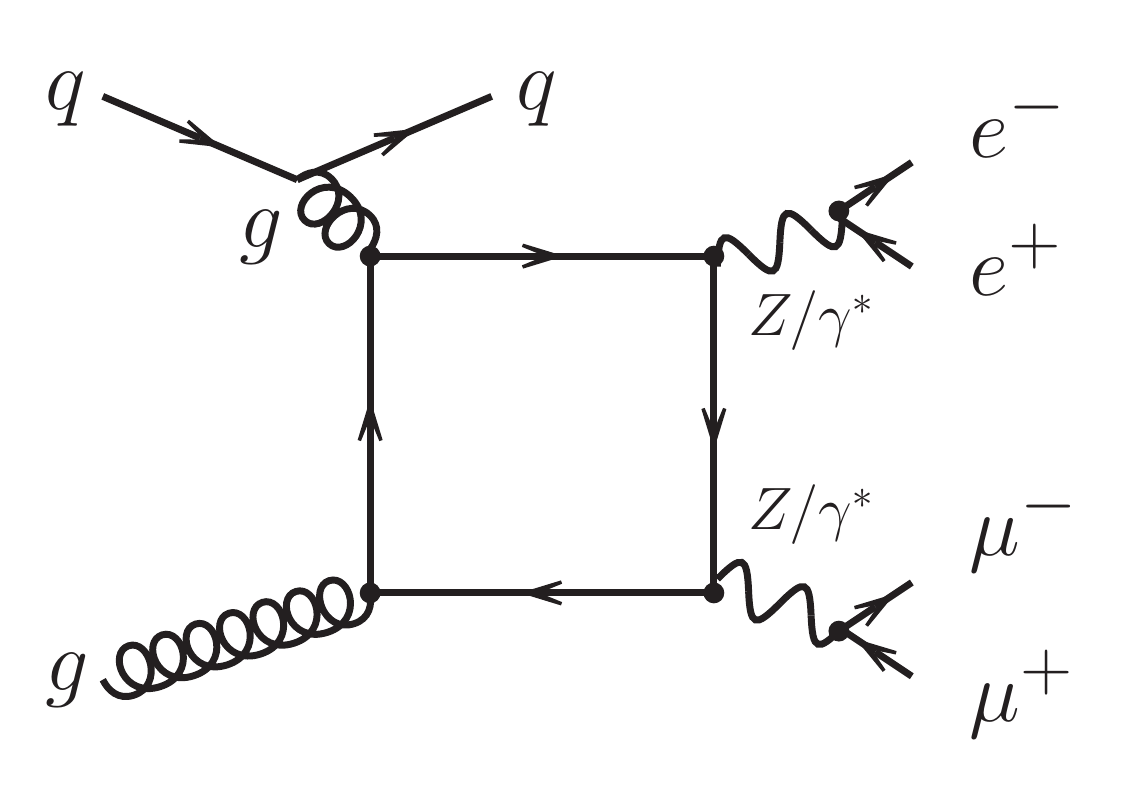}\\[2mm]
    \includegraphics[height=\diagheight]{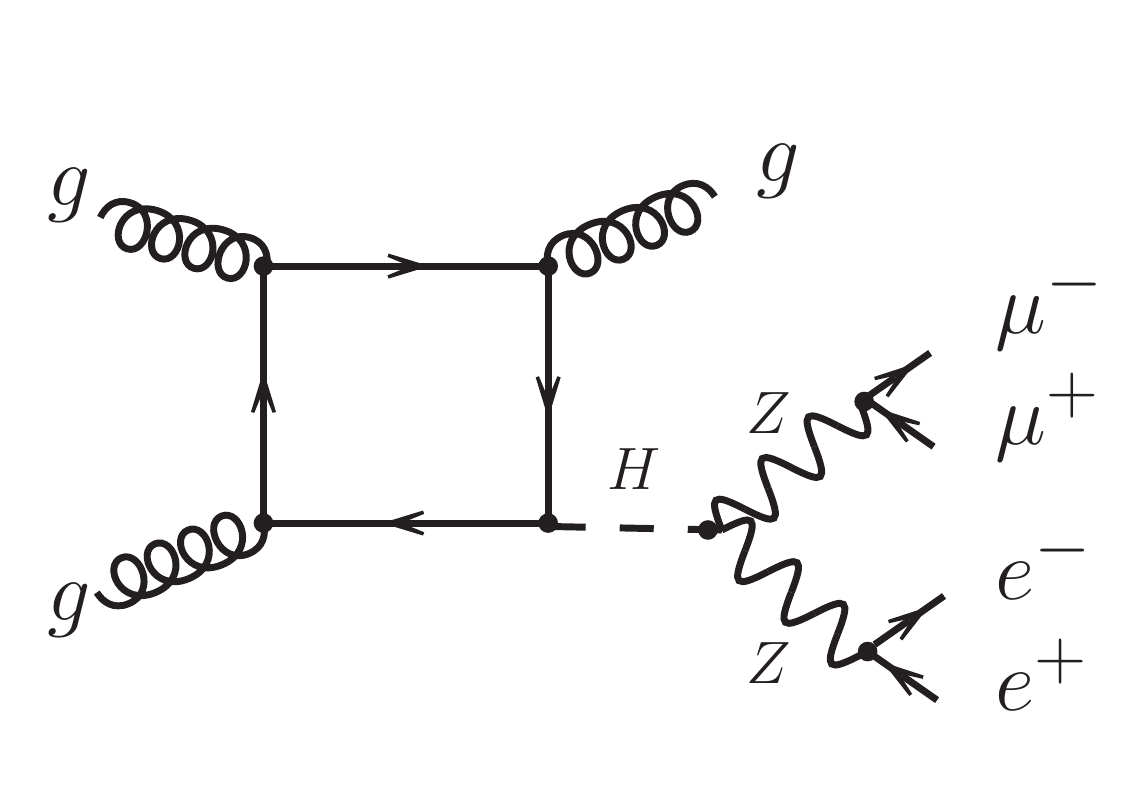}
    \hspace*{\diagsepB}
    \includegraphics[height=\diagheight]{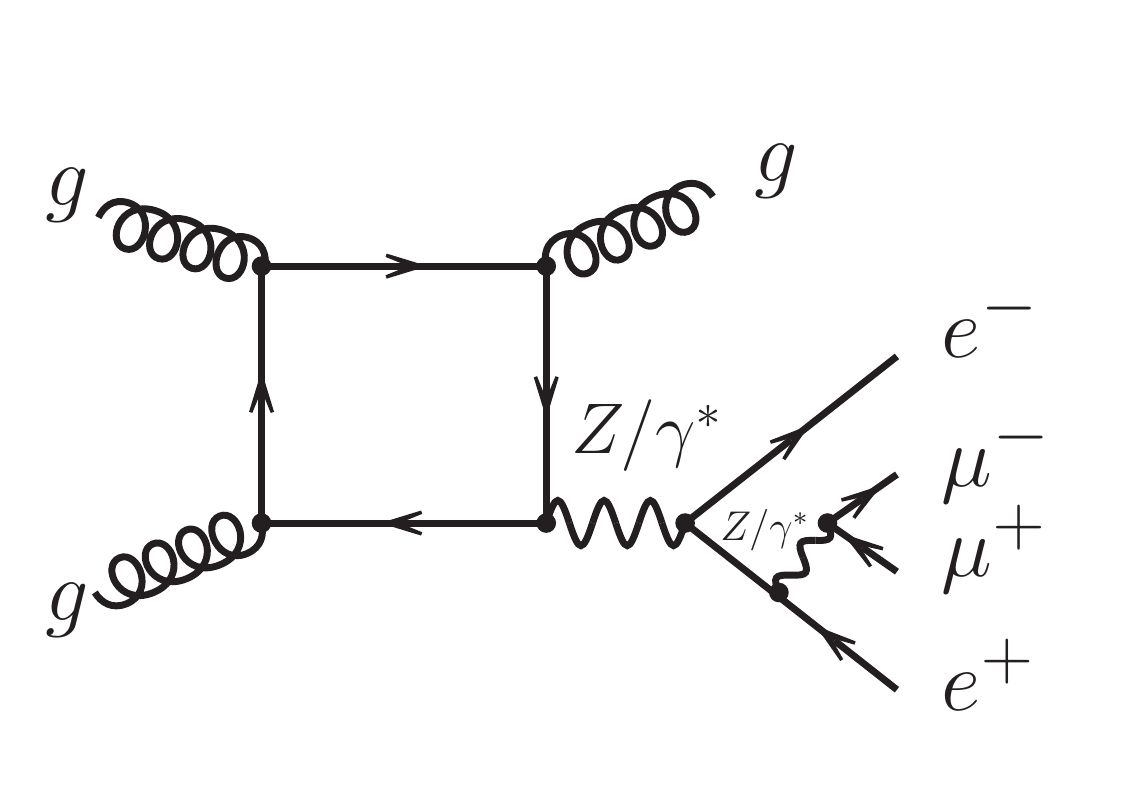}
    \caption{}
    \label{fig:diags:4lj-LI}
  \end{subfigure}\\[3mm]
  \caption{%
    Example loop-squared contributions to $pp\to e^+e^-\mu^+\mu^-$ (a)
    and $pp\to e^+e^-\mu^+\mu^-j$ (b).
  }
  \label{fig:diags:4l-4lj-LI}
\end{figure}

Although technically only a NNLO correction, gluon-induced loop-squared
contributions have a numerically important phenomenological impact,
both for inclusive and one-jet observables.
Therefore, in addition to the standard quark-induced four-lepton production
processes, we consider the gluon-induced loop-squared contributions as
one component of our multijet-merging approach.
Here, immediately the question of systematically separating both production
modes arises, as they coexist at higher orders that are at least
approximated in any calculation matched to a parton shower.
To this end, keeping in mind the parton shower's generating functional
of Eq.~\eqref{eq:def:Fcal} and following \cite{Cascioli:2013gfa},
we include all diagrams where all four leptons couple through one (or more)
electroweak gauge boson to the closed quark loop, but not any other
quarks in the process.
This excludes in particular diagrams where one lepton pair is radiated
off an external quark.
This means, that while
\begin{equation}
  g\,g\to e^+e^-\mu^+\mu^- \quad\text{at}\quad\order{\alpha_s^2\alpha^4}
  \nnb
\end{equation}
solely constitutes inclusive loop-induced four-lepton production,
\begin{equation}
  g\,g\to e^+e^-\mu^+\mu^-\,g\,,\quad
  g\,\qbarbracket\to e^+e^-\mu^+\mu^-\,\qbarbracket\quad\text{and}\quad
  q\,\bar{q}\to e^+e^-\mu^+\mu^-\,g \quad\text{at}\quad\order{\alpha_s^3\alpha^4}
  \nnb
\end{equation}
contribute for the jet-associated final state. Fig.\ \ref{fig:diags:4l-4lj-LI}
contains example Feynman diagrams contributing to the loop-induced modes
of hadronic $e^+e^-\mu^+\mu^-$ (a) and $e^+e^-\mu^+\mu^-j$ (b) production.
This comprises triangle-, box- and pentagon-type contributions.
We include light quarks and top quarks
in the closed fermion loops and allow for both double- and
single-resonant as well as Higgs-boson mediated topologies.
It is important to note, that in the single-resonant diagrams
the $Z$-boson couples through its axial component in triangle-like
and its vector component in box-like topologies.

Electroweak corrections to these loop-induced processes
are of two-loop complexity, they have not been calculated yet,
and consequently we do not consider them in this paper.

\FloatBarrier

\subsection{Numerical inputs and event-selection cuts}\label{sec:numinputs}
All calculations shown in this work are performed in the \Sherpa{}+\OpenLoops/\Recola
\cite{Bothmann:2019yzt,Gleisberg:2008ta,Buccioni:2019sur,Cascioli:2011va,
  Actis:2016mpe,Biedermann:2017yoi}
framework, allowing for a fully automated calculation of cross sections and
observables at next-to-leading order in the strong and electroweak sector
of the Standard Model.
In this framework, renormalised QCD and EW virtual corrections are provided
by \OpenLoops \cite{Buccioni:2019sur,Cascioli:2011va} and 
\Recola \cite{Actis:2016mpe} for the standard
$4\ell$ and $4\ell+j$ and all loop-induced processes.
Both programs use the \Collier tensor-reduction library \cite{Denner:2016kdg}.
In addition, \OpenLoops also uses \CutTools \cite{Ossola:2007ax}
together with \OneLoop \cite{vanHameren:2010cp}.
All remaining tasks, \emph{i.e.}\ the tree-level Born and real-emission
matrix elements as well as the bookkeeping of partonic subprocesses,
phase-space integration, and the subtraction of all QCD and QED infrared
singularities, are performed by \Sherpa
using the \Amegic and \Comix matrix-element generators \cite{Krauss:2001iv,
  Gleisberg:2008fv,Schonherr:2017qcj,Gleisberg:2007md}.
\Sherpa in combination with \OpenLoops and/or \Recola has been employed
successfully in a range of different calculations at NLO EW
\cite{Kallweit:2014xda,Kallweit:2015dum,
  Kallweit:2017khh,Biedermann:2017yoi,Lindert:2017olm,Chiesa:2017gqx,
  Greiner:2017mft,Gutschow:2018tuk,Schonherr:2018jva,Reyer:2019obz,
  Brauer:2020kfv,Gutschow:2020cug}
and has been validated against other tools in \cite{Bendavid:2018nar}.
For off-shell inclusive $ZZ$ production in particular, we have recomputed the
results of \cite{Biedermann:2016yvs,Biedermann:2016lvg,Gutschow:2020cug}
and excellent agreement was found.
The analyses for the following results have been implemented in \Rivet~\cite{Buckley:2010ar,Bierlich:2019rhm}.

We present predictions for proton--proton collisions at $\sqrt{s} = 13 \TeV$.
For the masses and widths we use the following values
\cite{ParticleDataGroup:2020ssz,Heinemeyer:2013tqa}
\begin{center}
  \begin{tabular}{rclrcl}
    $\MWOS$ & \shortequal & 80.385\,\text{GeV}\hspace*{20mm}&
    $\GWOS$ & \shortequal & 2.085\,\text{GeV} \\
    $\MZOS$ & \shortequal & 91.1876\,\text{GeV}&
    $\GZOS$ & \shortequal & 2.4952\,\text{GeV} \\
    $M_h$ & \shortequal & 125.0\,\text{GeV}&
    $\Gamma_h$ & \shortequal & 0.00407\,\text{GeV} \\
    $m_t$ & \shortequal & 173.2\,\text{GeV}&
    $\Gamma_t$ & \shortequal & 0\;. \\
  \end{tabular}
\end{center}
All other particles are treated as massless, in particular
we are working in the five-flavour scheme. Note, we throughout assume the CKM
quark-mixing matrix to be diagonal.
The pole
masses and widths used in the computation are obtained from the given on-shell (OS)
values for the $W$ and $Z$~bosons according to~\cite{Bardin:1988xt}
\begin{equation}
  M_{\text{V}} = \frac{\MVOS}{\sqrt{1+(\GVOS/\MVOS)^2}}\;,\qquad
  \Gamma_{\text{V}} = \frac{\GVOS}{\sqrt{1+(\GVOS/\MVOS)^2}}\;,
\end{equation}
with ${\text{V}}=W, Z$.
We work in the complex-mass scheme~\cite{Denner:1999gp,Denner:2006ic}, where the complex
masses and the weak mixing angle are given by
\begin{equation}
  \mu_i^2=M_i^2-\mr{i}M_i\Gamma_i
  \qquad\text{and}\qquad
  \sin^2\theta_w=1-\frac{\mu_W^2}{\mu_Z^2}\;.
\end{equation}
Per default we use the $G_\mu$ input parameter and
renormalisation scheme \cite{Denner:2000bj,Dittmaier:2001ay} with
\begin{equation}
  \GF    = 1.16637\times 10^{-5}\GeV^{-2}
  \qquad \text{and}  \qquad
  \alpha_{\GF} =
  \frac{\sqrt{2}}{\pi}\, G_\mu \MW^2 \left( 1 - \frac{\MW^2}{\MZ^2} \right)
  \;.
\end{equation}
In order to gauge the impact of the choice for the EW renormalisation scheme, we also
consider results obtained in the $\alpha(\MZ^2)$ scheme, with
\begin{equation}
  \alpha(\MZ^2) = 1/128.802
\end{equation}
as an input parameter, and renormalise the amplitudes accordingly.
All other parameters remain unchanged. The difference between these
two choices can be seen as a (partial) missing-higher-order uncertainty.
See Refs.~\cite{Denner:2019vbn,Chiesa:2019nqb,Brivio:2021yjb} for recent discussions
of EW input parameter and renormalisation schemes.

As parton distribution function (PDF) we use the
\texttt{NNPDF31\_nlo\_as\_0118\_luxqed} set \cite{Bertone:2017bme},
interfaced through \textsc{Lhapdf}~\cite{Buckley:2014ana}.
The extraction of the photon content is based on
\cite{Manohar:2016nzj}.
The value of the strong coupling is chosen accordingly, \emph{i.e.}\
\begin{equation}
  \alphas(M^2_Z)= 0.118\;.
\end{equation}

For the event selection, we first
combine the charged leptons with collinear photons
within a standard cone of radius $R = 0.1$.
The so dressed leptons $\ell$ are then required
to pass the following selection cuts
for their transverse momentum, rapidity and distance to other dressed leptons $\ell'$:
\begin{align}
  p_{\mr{T},\Pl} > 20 \GeV, \qquad \qquad |y_{\Pl}|<2.5, \qquad \qquad \Delta R_{\Pl \Pl'} > 0.1\,.
\end{align}
These criteria define the fiducial phase space of our analysis
for inclusive setups, such as the $e^+e^-\mu^+\mu^-$ fixed-order calculation
or the multijet-merged one.

For the fixed-order evaluation of $e^+e^-\mu^+\mu^-j$, as well as in the analysis
of jet observables for the multijet-merging samples, the additional jet(s) $j$ are defined with the anti-$k_t$
algorithm~\cite{Cacciari:2008gp} with $R=0.4$ and
\begin{align}
  p_{\mr{T},j} > 30 \GeV, \qquad \qquad |y_{j}|<4.5, \qquad \qquad \Delta R_{\Pl j} > 0.4\,.
\end{align}
All particles that are not part of the previously defined dressed leptons
are considered as input to the jet algorithm.

\subsection{Fixed-order results}
\label{sec:fo}

In this section we validate the approximations introduced in Sec.~\ref{sec:setup:ew}
applied to $pp\to e^+ e^- \mu^+ \mu^-$ and $pp\to e^+ e^- \mu^+ \mu^-j$
at LO against the respective full NLO EW results.
The \EWvirt and \EWsud computations include YFS soft-photon resummation
as discussed in Sec.~\ref{sec:setup:ew},
in order to account for real-emission kinematics and effects.
We further provide predictions based on matching the full NLO EW result
with NLL EW Sudakov resummation, as an improved description including
the all-orders resummed EW corrections in the region where they are large.

For these calculations,
the renormalisation and factorisation scales are set
in accordance with Refs.~\cite{Kallweit:2019zez,Brauer:2020kfv}, \emph{i.e.}\
\begin{equation*}
 \muR = \muF = \tfrac{1}{2}\left(E_{\mr{T},ee}+E_{\mr{T},\mu\mu}\right)\,,
\end{equation*}
with the transverse energies of the two vector bosons given by
\begin{equation*}
  E_{\text T, \ell\ell} = \sqrt{m_{\ell\ell}^2 + p^2_{\mr{T},\ell\ell}}\,.
\end{equation*}
Here, $m_{\ell\ell}$ and $p_{\mr{T},\ell\ell}$ denote the invariant mass
and the transverse momentum of the off-shell charged dressed lepton pair.

Further, all inclusive cross sections are given both in the $G_\mu$ scheme
and in the $\alpha (\MZ^2)$ scheme, with the former being our default choice
for the final predictions presented in Sec.~\ref{sec:merged}.
In addition to the inclusive phase space, we also consider dedicated additional selections
focusing on the EW Sudakov region where our matched result
formally improves on the pure NLO EW one, and both approximations are
in their respective regime of validity.
Here, we give the $G_\mu$ scheme results only, because we are interested on the effects of
the resummation in this region.
Finally, we provide differential distributions, including the full NLO EW
results in the \Gmu\ and \amz\ scheme in order to discuss their input-parameter
and renormalisation scheme dependence.

\subsubsection*{Inclusive $e^+e^-\mu^+\mu^-$ production}

We start by discussing the process $pp\to e^+ e^- \mu^+ \mu^-+X$, for which
QCD and EW corrections are known to NNLO and NLO, respectively~\cite{Cascioli:2014yka,Grazzini:2015hta,Heinrich:2017bvg,Kallweit:2018nyv,Biedermann:2016yvs,Biedermann:2016lvg,Kallweit:2017khh,Chiesa:2018lcs,Grazzini:2019jkl}. We use this setup
as an initial benchmark for the approximations described in Sec.~\ref{sec:setup:ew}.
We report results for the inclusive fiducial cross section in the various
calculational schemes in Tab.~\ref{table:foxs0j}. All approximate treatments for
EW corrections as well as the inclusion of the full set of NLO EW corrections
reduce the cross section.

Examining the cross sections and higher-order EW corrections summarised in
Tab.~\ref{table:foxs0j},
the different impact of the change of input-parameter and renormalisation scheme
on the LO and NLO EW results is the most striking observation.
At the leading order, \emph{i.e.}\ ${\cal{O}}(\alpha^4)$, the
change of EW scheme amounts to a simple rescaling of the parameter $\alpha$ used in
the calculation, as this has different values in the schemes. The values of all
masses, widths, and $\sin\theta_w$ remain unchanged. In particular, here
this change of scheme amounts to an \SI{11.3}{\percent} difference in the
inclusive cross section at LO.
At NLO, the scheme dependence is not quite so straight-forward.
While the one-loop, mass factorisation and real-emission contributions
are again simply rescaled, the scheme-dependent renormalisation
contributions differ significantly in magnitude and structure.
The result is the expected much smaller, and now oppositely signed,
scheme dependence of \SI{-3.8}{\percent}.
The relative NLO corrections of \SI{-6.8}{\%} and \SI{-19.4}{\%}
in the \GF\ and \amz\ schemes, respectively, also
demonstrate the superior adequacy of the \GF\ scheme, caused
by its effective partial accounting for the leading universal
renormalisation effects originating from the $\rho$-parameter.
While this behaviour is well reproduced by the \EWvirt\ approximation
due to its inclusion of the renormalisation terms,
the \EWsud\ approximation follows the LO behaviour in its scheme dependence.
Apart from the parameter renormalisation (PR) logarithms
\cite{Denner:2000jv,Denner:2001gw} which are however generally not
the dominant terms, \EWsud\ features no additional scheme-dependence
compensation.
Finally, the \NLOEWNLLEWsudexp\ prediction largely coincides
with the scheme dependence of the NLO EW calculation as the
influence of the resummed EW Sudakov exponent is minimal in
the inclusive phase space.

Considering only the \Gmu\ scheme, the relative NLO EW corrections
are well reproduced, within about \SI{1}{\%}, by both the \EWvirt\
and \EWsud\ approximations when complemented with the soft-photon
resummation.
This is because the total correction in this process is driven
in roughly equal parts by negative one-loop corrections and
the negative impact of energy-loss due to real-photon radiation.
The latter process is only described at $\order{\alpha}$ accuracy
while the resummation includes the impact of higher-order emissions
further reducing the cross section.
We have checked that truncating the resummation to $\order{\alpha}$
-- \emph{i.e.}\ allow at most a single
photon to be emitted and expand the form-factor accordingly --
results in a much closer reproduction of the exact result.
However, given that the \EWvirt\ and \EWsud\ approximations are tailored
to the high-energy regime only, their close reproduction
of inclusive observables is to some degree accidental.
Finally, the relative correction of the \NLOEWNLLEWsudexp matched
result follows the NLO EW one closely as Sudakov logarithms are
small and their resummation does not lead to noticeable effects.

\begin{table}[h!]
  \centering
  \sisetup{table-format = +2.1, table-align-exponent = false, table-align-text-post = false}
  \resizebox{\textwidth}{!}{%
  \begin{tabular}{@{}llCSSSSS@{}} \toprule
                    \multicolumn{2}{c}{%
                        $pp\to e^+e^-\mu^+\mu^-$
                      }
                    & \multicolumn{1}{c}{%
                        fiducial cross section
                      }
                    & \multicolumn{5}{c}{%
                        corrections to LO
                      } \\ \cmidrule(r){1-2}\cmidrule(lr){3-3}\cmidrule(l){4-8}
    Scheme          & Region
                    & {LO}
                    & {NLO EW}
                    & {$\text{LO}+\EWvirt{}+\text{YFS}$}
                    & {$\text{LO}+\EWsud{}+\text{YFS}$}
                    & {$\text{LO}+\EWsudexp{}+\text{YFS}$}
                    & {\NLOEWNLLEWsudexp}
                    \\ \midrule
    $\Gmu$          & inclusive
                    & \textbf{\hc 9.819\,fb}
                    & -6.8 {\,\%}
                    & -7.9 {\,\%}
                    & -7.3 {\,\%}
                    & -7.2 {\,\%}
                    & -6.7 {\,\%}
                    \\ \addlinespace
    $\amz$ &
                    & {10.928\,fb}
                    & -19.4 {\,\%}
                    & -20.2 {\,\%}
                    & -7.7  {\,\%}
                    & -7.6  {\,\%}
                    & -19.3 {\,\%}
                    \\ \addlinespace
    $\delta^{\amz}_{\Gmu}$ &
                    & \hc  11.3\,\%
                    &  -3.8 {\,\%}
                    &  -3.6 {\,\%}
                    & 10.8  {\,\%}
                    & 10.8  {\,\%}
                    &  -3.7 {\,\%}
                    \\ \midrule
    $\Gmu$          & high energy
                    & \textbf{$4.27\cdot 10^{-3}$\,fb}
                    & -42 {\,\%}
                    & -45 {\,\%}
                    & -39 {\,\%}
                    & -33 {\,\%}
                    & -36 {\,\%}
                    \\ \bottomrule
  \end{tabular}
  }
  \caption{
    Inclusive fiducial cross sections for
    $pp\to e^+ e^- \mu^+ \mu^-$ at $\sqrt{s}=13\,\TeV$
    at LO for the \Gmu\ and the \amz\ scheme,
    along with the relative corrections
    for NLO EW, \NLOEWNLLEWsudexp{}
    and the \EWsud\ and \EWvirt\ approximations.
    The table also gives
    the relative differences $\delta^{\amz}_{\Gmu}$
    of the \amz\ scheme with respect to the default \Gmu\ scheme,
    and results for the ``high-energy'' region, which requires
    $p_{\text T, 2e}>\SI{600}{\GeV}$ in addition to the fiducial cuts.
  }
  \label{table:foxs0j}
\end{table}

Tab.\ \ref{table:foxs0j} accompanies the inclusive cross section
with a  ``high energy'' region requiring additionally
$p_{\text T,2e} > \SI{600}{\GeV}$,
thus entering the region where the Sudakov logarithms become sizeable
and dominate the total NLO EW corrections.
As expected, the resummation of the Sudakov logarithms is important here,
giving a \SI{6}{\percent} smaller correction with respect to LO in the
\NLOEWNLLEWsudexp calculation as compared to the NLO EW result
(or a \SI{10}{\percent} increase of the cross section relative to it).

\begin{figure}[t!]
      \centering
      \includegraphics[width=0.47\textwidth]{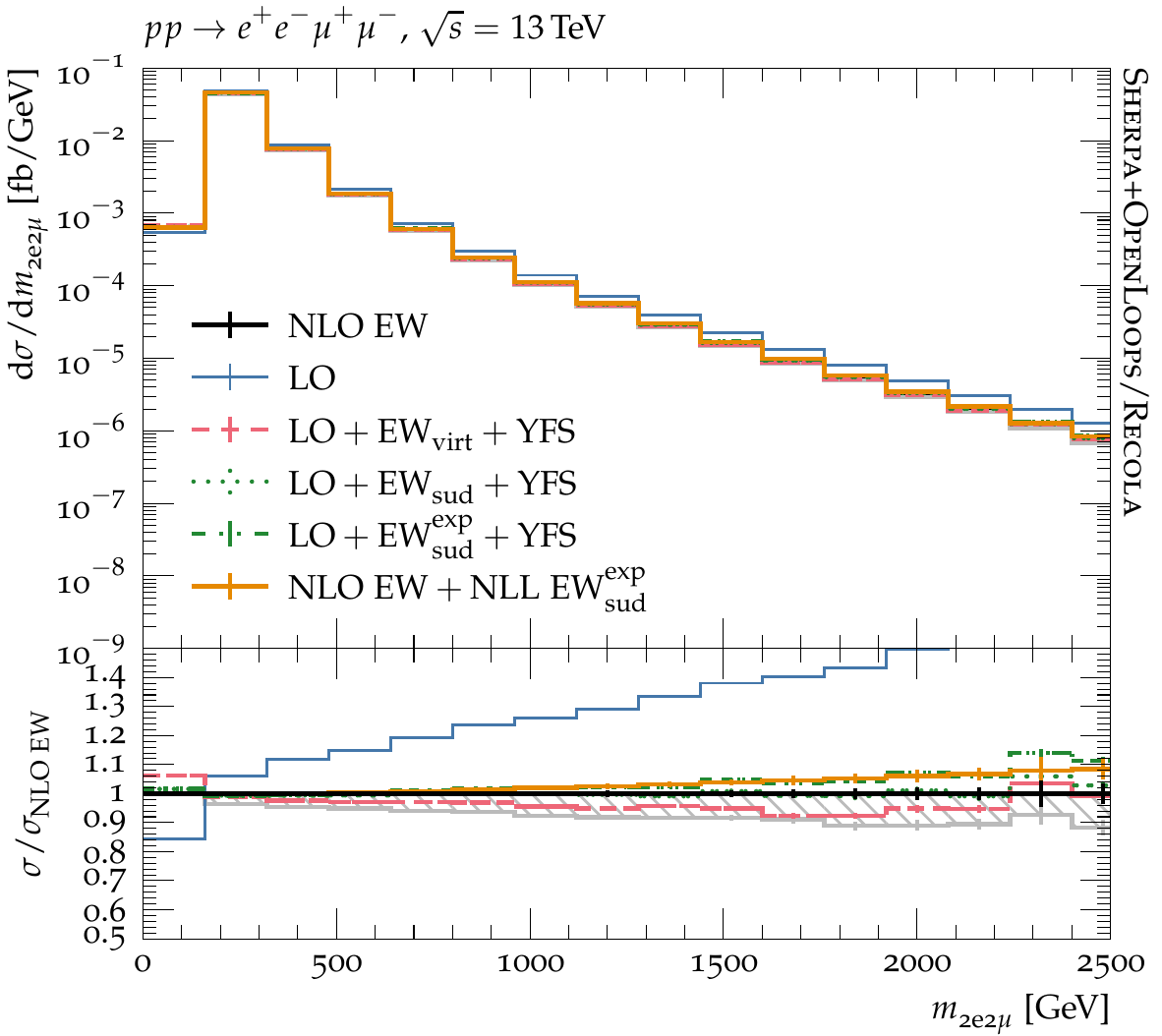} \hfill
      \includegraphics[width=0.47\textwidth]{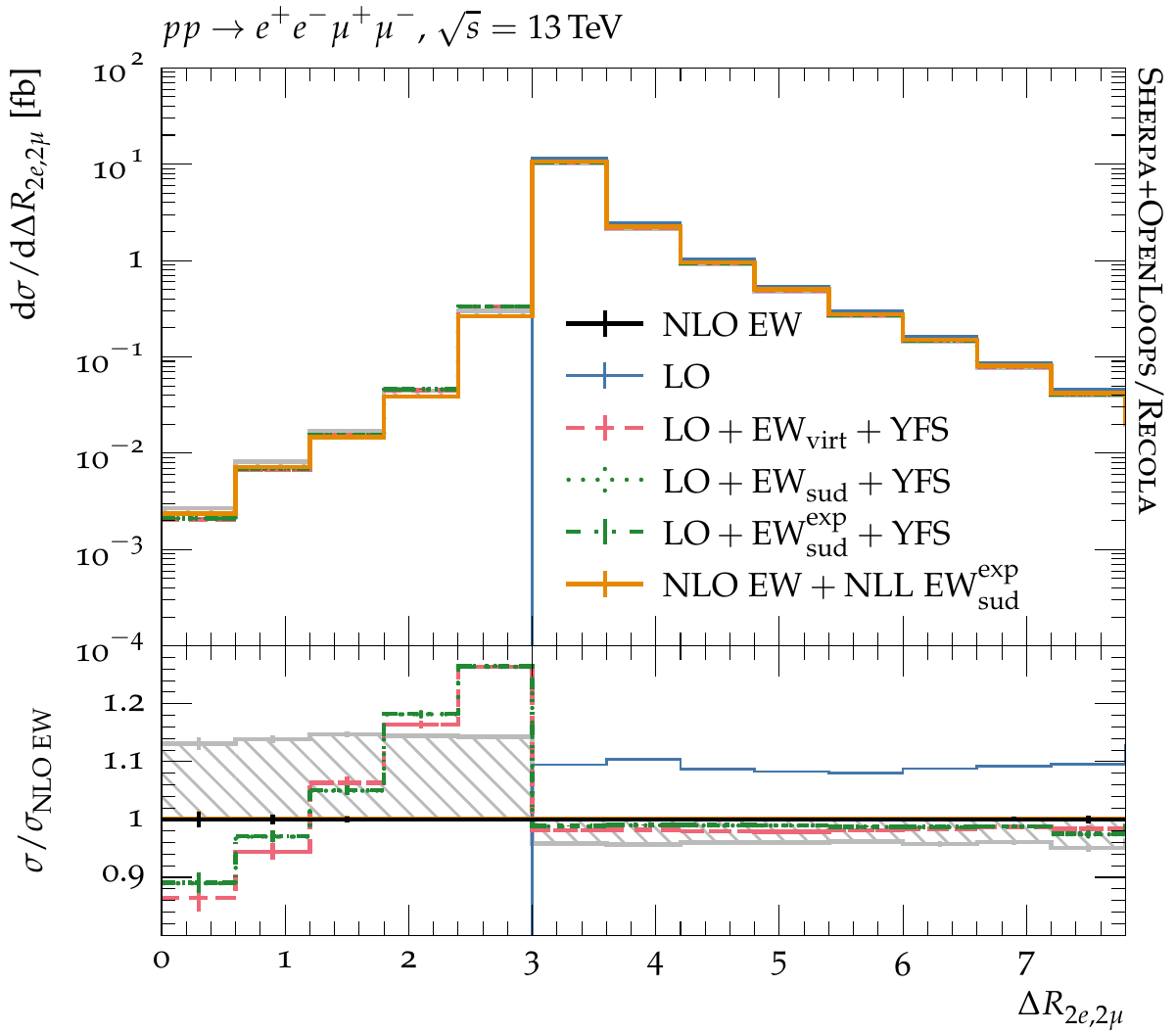} \\
      \includegraphics[width=0.47\textwidth]{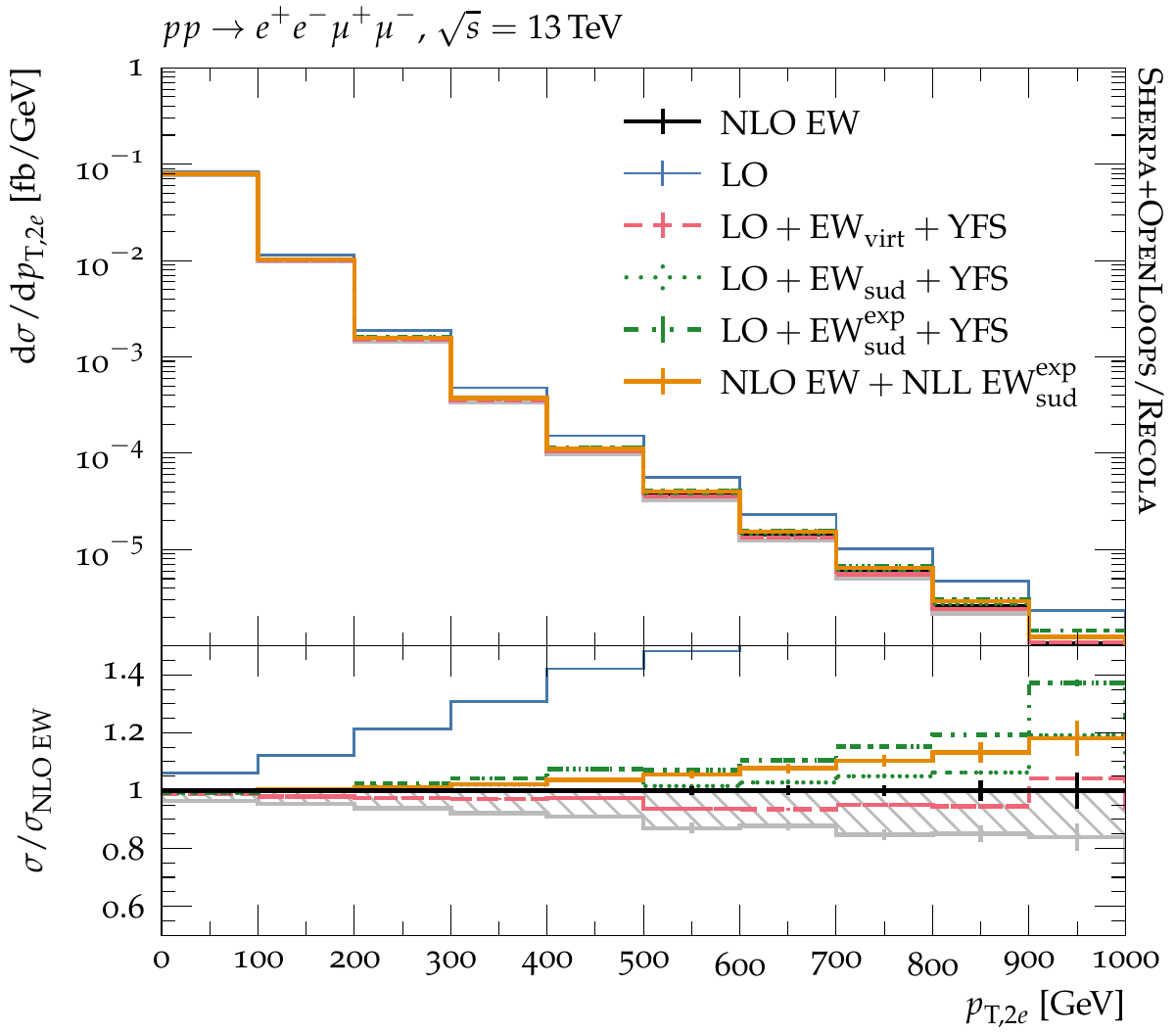} \hfill
      \includegraphics[width=0.47\textwidth]{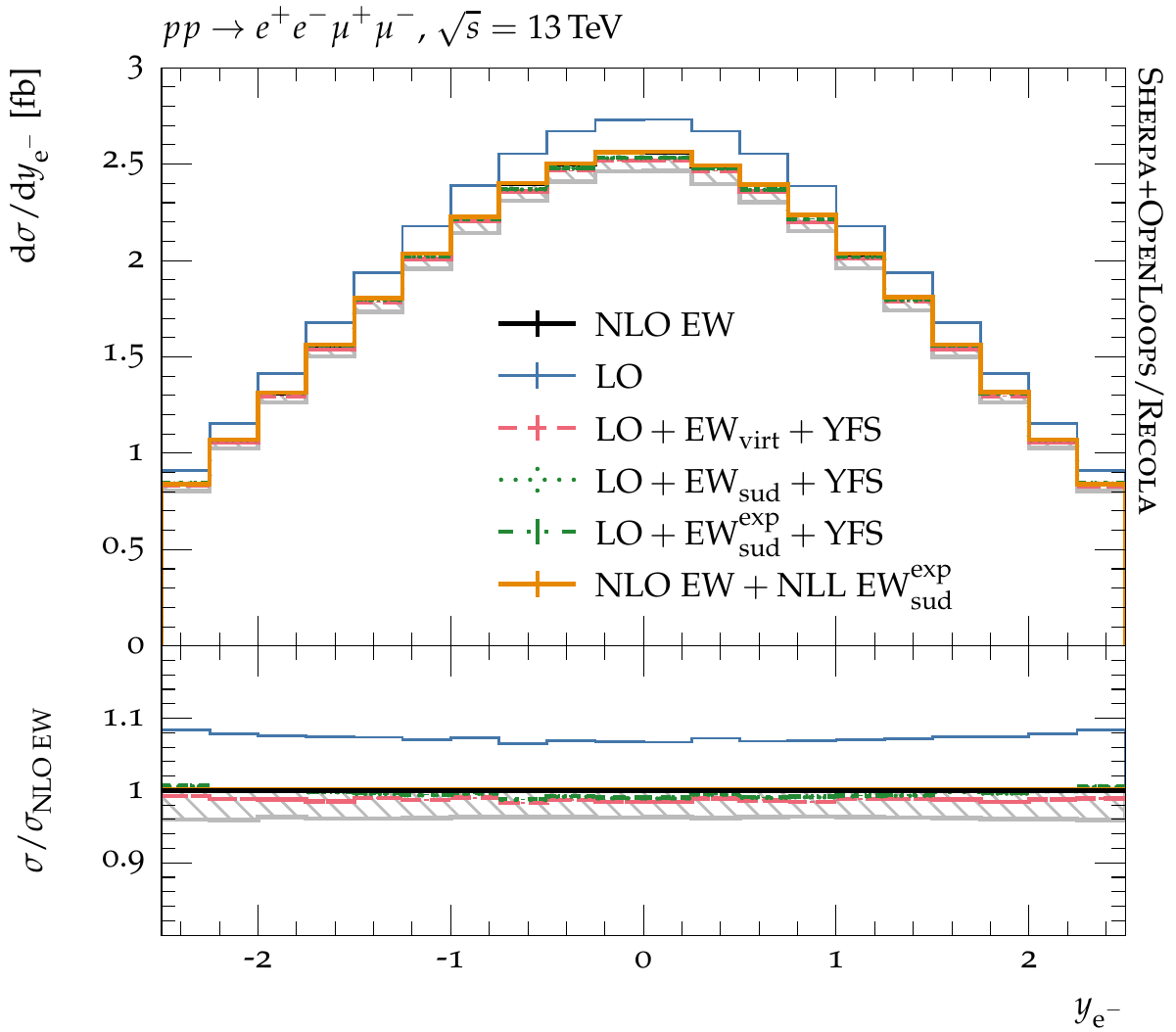}
      \caption{%
        Observable distributions for the $pp \to e^+e^- \mu^+\mu^-$ process.
        From top left to bottom right we show: the four-lepton invariant mass $m_{2e2\mu}$,
        the $Z$-boson distance $\Delta R_{2e,2\mu}$, the transverse momentum of the
        di-electron pair $p_{\text T,2e}$, and the rapidity of the electron $y_{\text{e}^-}$.
        Results are given at LO and NLO EW and compared to approximative EW calculations.
        The NLO EW is given for the $G_\mu$ (black line) and $\alpha(\MZ^2)$ (grey line)
        renormalisation schemes, and the span between the two is marked by a hatched band.
        All predictions are calculated using \Sherpa{}{}+\OpenLoops/\Recola.
      }
      \label{fig:fo_validation_0j}
\end{figure}
We show differential distributions obtained in the various calculational schemes
in Fig.~\ref{fig:fo_validation_0j}, where in addition to the nominal predictions
we indicate the NLO EW scheme dependence with a grey hatched band. The observables
considered are the invariant mass of the four-lepton system $m_{2e2\mu}$, the
$Z$-boson distance $\Delta R_{2e,2\mu}$, the transverse momentum of the di-electron
pair $p_{\text T,2e}$, and the electron rapidity $y_{\text{e}^-}$.

We start by noticing that the overall good agreement between the
$\EWvirt$ approximation and the full NLO EW observed for the total cross section is
also found for all the distributions. The only significant difference comes from
phase-space regions dominated by real-photon radiation, such as $\Delta
R_{2e,2\mu} < \pi$.
There one can see the impact of resumming soft photons through
YFS versus treating them at fixed order, which exhibits the main advantage of
including YFS resummation. We have indeed checked that if we expand the YFS
resummation to $\order{\alpha}$, as discussed above, we
reproduce the NLO EW result throughout, as a result of the inclusion on exact
NLO QED corrections in the YFS resummation. A similar overall good agreement can be seen
in the Sudakov approximation.

To further discuss the impact and the effects of the EW approximations we need
to distinguish between energy-dependent observables, such as the invariant mass of the
four leptons and the $p_{\mr{T}}$ of the electron pair,
and energy-independent observables, such as the separation of the two lepton
pairs and the rapidity of the electron.
The former class shows, as expected, a suppression in the high-energy tail of
distributions that feature a similar shape for both observables. The size of the
suppression is, however, different, but we find that overall the various
approximations are all within $5$--\SI{10}{\percent} of the exact NLO EW result.
The matched \NLOEWNLLEWsudexp result, shown here for the first time, has the expected behaviour,
interpolating from the NLO EW result at low energies to the exponentiated
Sudakov logarithms at high energies. In particular, we notice that at high energies the
resummation leads to a reduction of the suppression of about $5$--\SI{10}{\percent} compared
to the fixed-order Sudakov approximation.
The latter class of observables, on the other hand, has no energy dependence,
and encapsulates the $k$-factors of the total cross-section table uniformly
distributed across the available phase space. The only deviation from this, as discussed
above, is the region sensitive to additional real radiation, \emph{e.g.}\
$\Delta R_{2e,2\mu} < \pi$ or $m_{2e2\mu}<2\MZ$.

\FloatBarrier

\subsubsection*{Jet-associated production}

We now turn our attention to the production of two lepton pairs associated
by an $R=0.4$ anti-$k_t$ jet with $p_{\mr{T},j} > 30 \GeV$, \emph{see}\ Sec.~\ref{sec:numinputs}.
To the best of our knowledge, this is the first time that NLO EW corrections are
shown for this specific process, though the technology is readily
available. Nevertheless, the comparison between the full NLO EW and the various
approximations requires more care, as the full NLO EW result contains QCD
corrections to lower-order Born terms as well as QCD--EW interference
contributions, see Figs.~\ref{fig:diags:4lj-V} and \ref{fig:diags:4ljj-int}
respectively, which are not captured by either approximation.
However, as stressed before, such interference contributions are typically small
for inclusive observables. Nonetheless, with the QCD--EW interference terms
in particular being the only contributions at this order
which can contain two valence quarks as initial state, \textit{e.g.}\
$uu$, $du$, or $dd$, they can be quite sizeable in the TeV range and potentially spoil
the quality of the \EWsud and \EWvirt approximations.

\begin{table}[h!]
  \centering
  \sisetup{table-format = +2.2(1), table-align-exponent = false, table-align-text-post = false}
  \resizebox{\textwidth}{!}{%
  \begin{tabular}{@{}llCSSSSS@{}} \toprule
                    \multicolumn{2}{c}{%
                        $pp\to e^+e^-\mu^+\mu^-j$
                      }
                    & \multicolumn{1}{c}{%
                        fiducial cross section
                      }
                    & \multicolumn{5}{c}{%
                        corrections to LO
                      } \\ \cmidrule(r){1-2}\cmidrule(lr){3-3}\cmidrule(l){4-8}
    Scheme          & Region
                    & {LO}
                    & {NLO EW}
                    & {$\text{LO}+\EWvirt{}+\text{YFS}$}
                    & {$\text{LO}+\EWsud{}+\text{YFS}$}
                    & {$\text{LO}+\EWsudexp{}+\text{YFS}$}
                    & {\NLOEWNLLEWsudexp}
                    \\ \midrule
    $\Gmu$          & inclusive
                    & \textbf{5.170\,fb}
                    & -6.6 {\,\%}
                    & -8.5 {\,\%}
                    & -6.9 {\,\%}
                    & -6.7 {\,\%}
                    & -6.4 {\,\%}
                    \\ \addlinespace
    $\amz$ &
                    & {5.754\,fb}
                    & -19.2 {\,\%}
                    & -20.6 {\,\%}
                    & -6.9 {\,\%}
                    & -6.7 {\,\%}
                    & -19.0 {\,\%}
                    \\ \addlinespace
    $\delta^{\amz}_{\Gmu}$ &
                    & \hc  11.29\,\%
                    &  -3.7 {\,\%}
                    &  -3.4 {\,\%}
                    & 11.3 {\,\%}
                    & 11.3 {\,\%}
                    & -3.7 {\,\%}
                    \\ \midrule
    $\Gmu$          & high energy
                    & \textbf{$6.64\cdot 10^{-3}$\,fb}
                    & -33 {\,\%}
                    & -37 {\,\%}
                    & -30 {\,\%}
                    & -25 {\,\%}
                    & -29 {\,\%}
                    \\ \bottomrule
  \end{tabular}
  }
  \caption{
    Inclusive fiducial cross sections for
    $pp\to e^+ e^- \mu^+ \mu^- j$ at $\sqrt{s}=13\,\TeV$
    at LO for the \Gmu\ and the \amz\ scheme,
    along with the relative corrections
    for NLO EW, \NLOEWNLLEWsudexp{}
    and the \EWsud\ and \EWvirt\ approximations.
    The table also gives
    the relative differences $\delta^{\amz}_{\Gmu}$
    of the \amz\ scheme with respect to the default \Gmu\ scheme,
    and results for the ``high-energy'' region, which requires
    $p_{\text T, 2e}>\SI{600}{\GeV}$ in addition to the fiducial cuts.
  }
  \label{table:foxs1j}
\end{table}

The discussion for this process follows closely that for the four-lepton
final state. In Tab.~\ref{table:foxs1j} we report results
for the inclusive fiducial cross sections
at LO, NLO EW, LO+\EWvirt{}+YFS, LO+\EWsud{}+YFS,
LO+\EWsudexp{}+YFS, and \NLOEWNLLEWsudexp, once again,
both for the \Gmu\ and the \amz\ scheme, and similar conclusions can
be drawn here.
Notably, the size of the NLO EW corrections with respect to the LO are only
slightly reduced, which confirms the known behaviour that EW corrections with
extra QCD jets are roughly of the same size. In particular we find that they are
about \SI{-6.6}{\percent} and \SI{-19.2}{\percent} with respect to the LO result
for the \Gmu\ and the \amz\ schemes, respectively.
In addition, we find a slightly reduced agreement between the NLO EW and the \EWvirt
approximation, amounting to roughly \SI{-2}{\percent}
with respect to the NLO EW result
for both the \Gmu\ and the \amz\ scheme. While numerically small,
this may be attributed to the fact that, while the QCD corrections to lower-order
Born contributions are included, the additional four-quark channels discussed
are not present in the \EWvirt approximation.
Here, the \EWsud approximation shows a better level of agreement,
despite more contributions get discarded in this approximation.
Hence, the quality of reproduction of the exact
result is to some degree accidental for both approximations,
as they are
tailored to account for the high-energy regime in particular.
Nonetheless, it is reassuring that, at least in the \Gmu\ scheme,
both reproduce the exact result qualitatively quite well.
In similar fashion, the resummed Sudakov approximation changes
the fixed-order result very little, both in the pure Sudakov
approximation and the matched \NLOEWNLLEWsudexp calculation.

\begin{figure}[t!]
      \centering
      \includegraphics[width=0.47\textwidth]{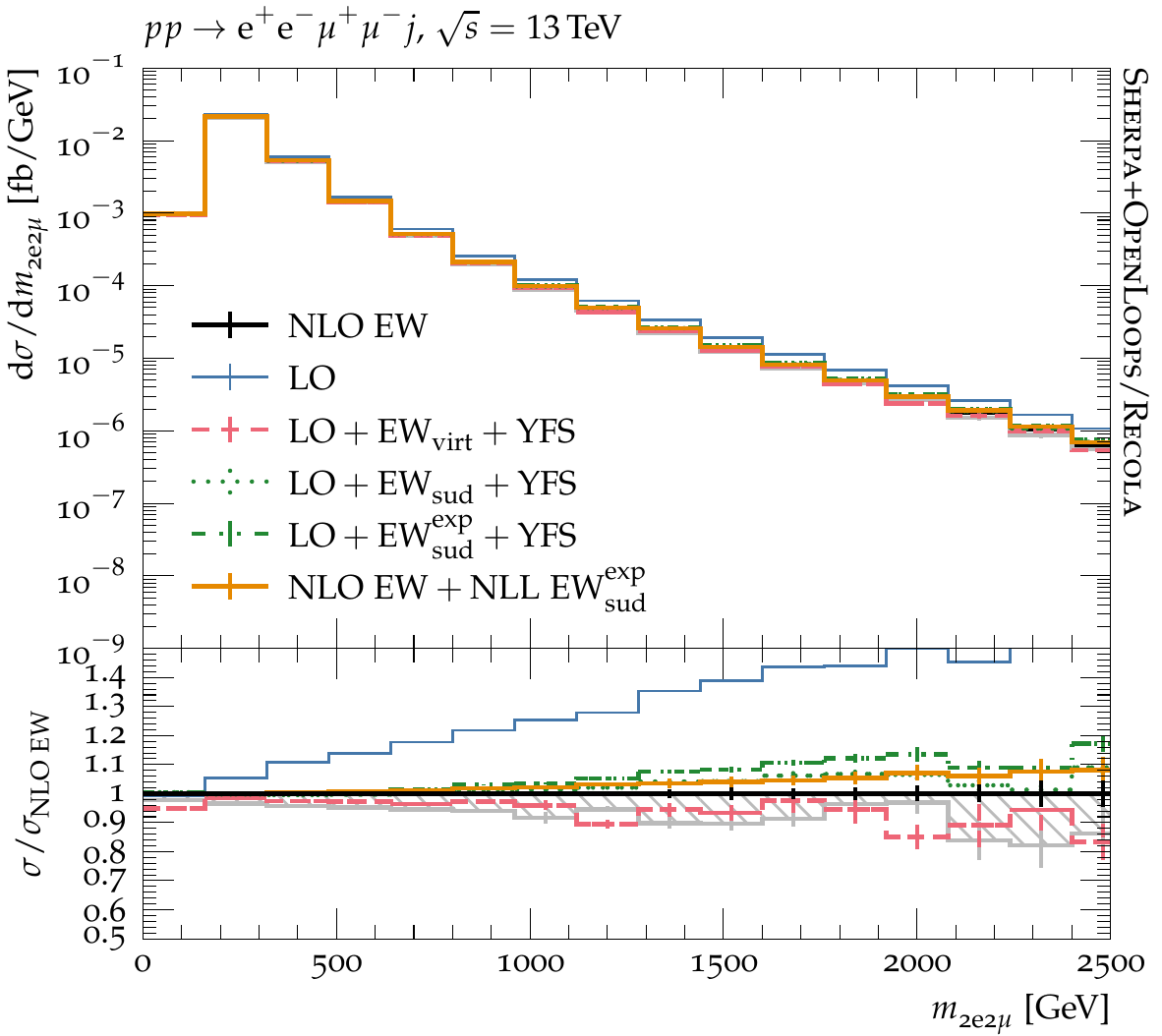} \hfill
      \includegraphics[width=0.47\textwidth]{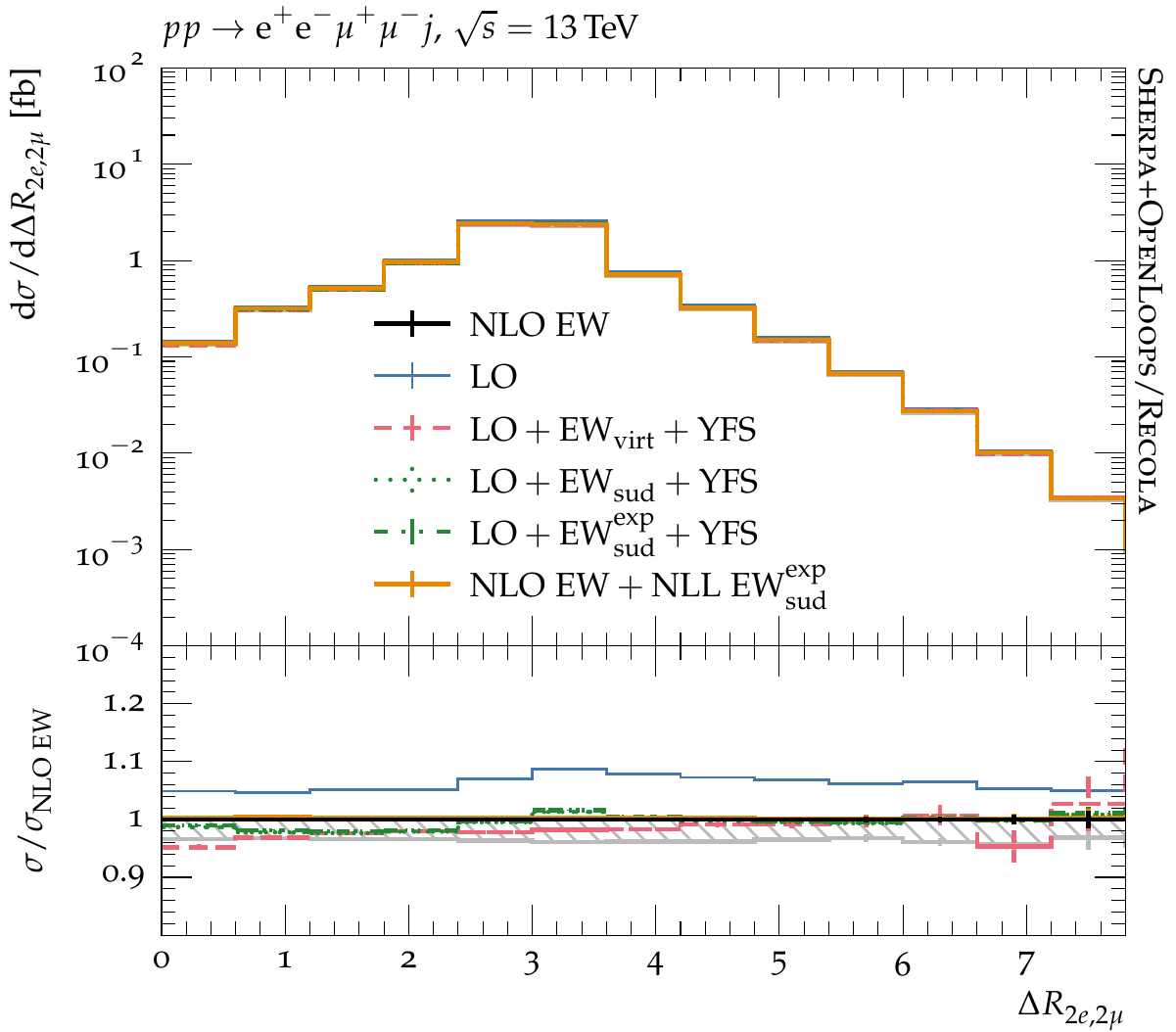} \\
      \includegraphics[width=0.47\textwidth]{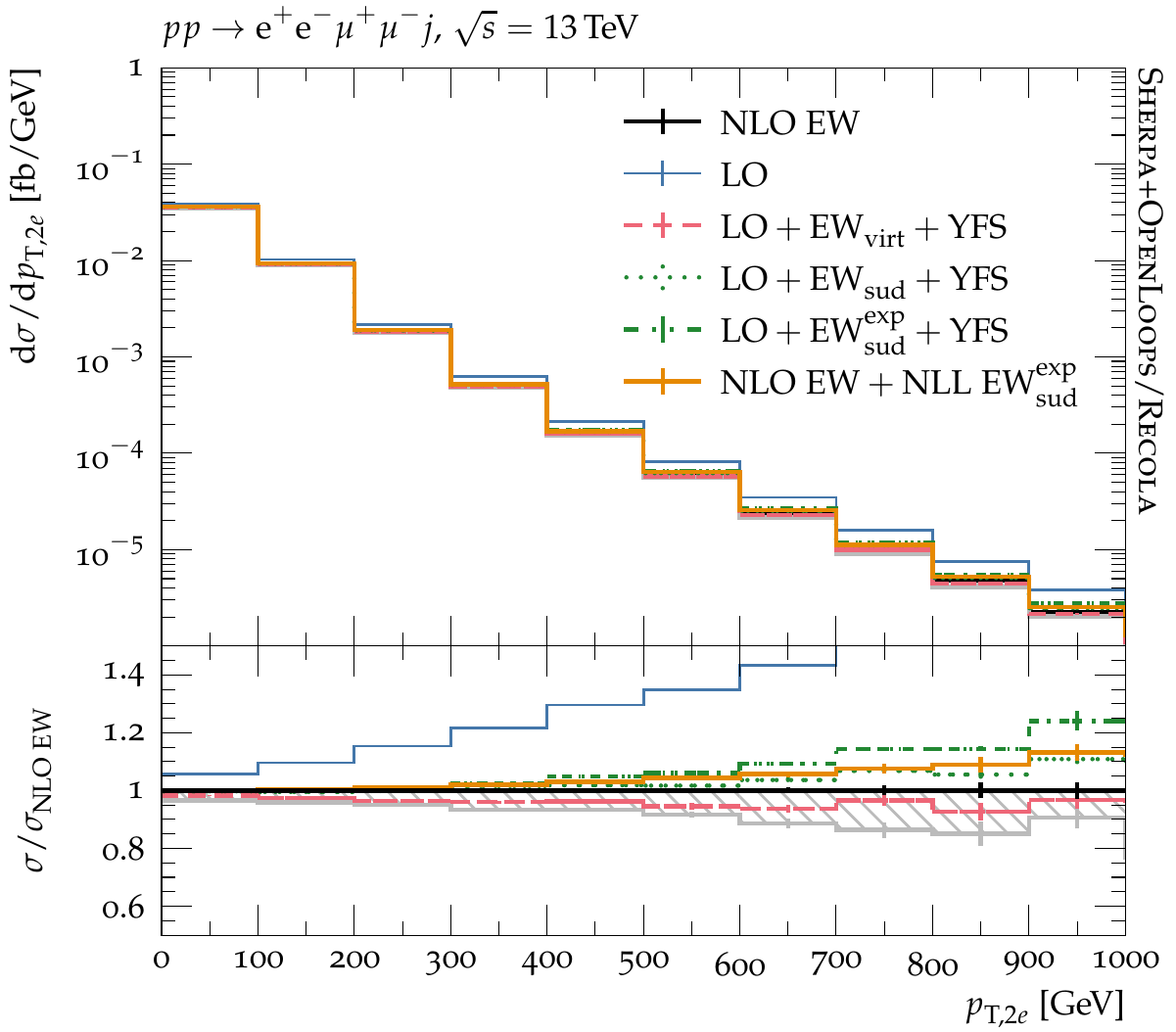} \hfill
      \includegraphics[width=0.47\textwidth]{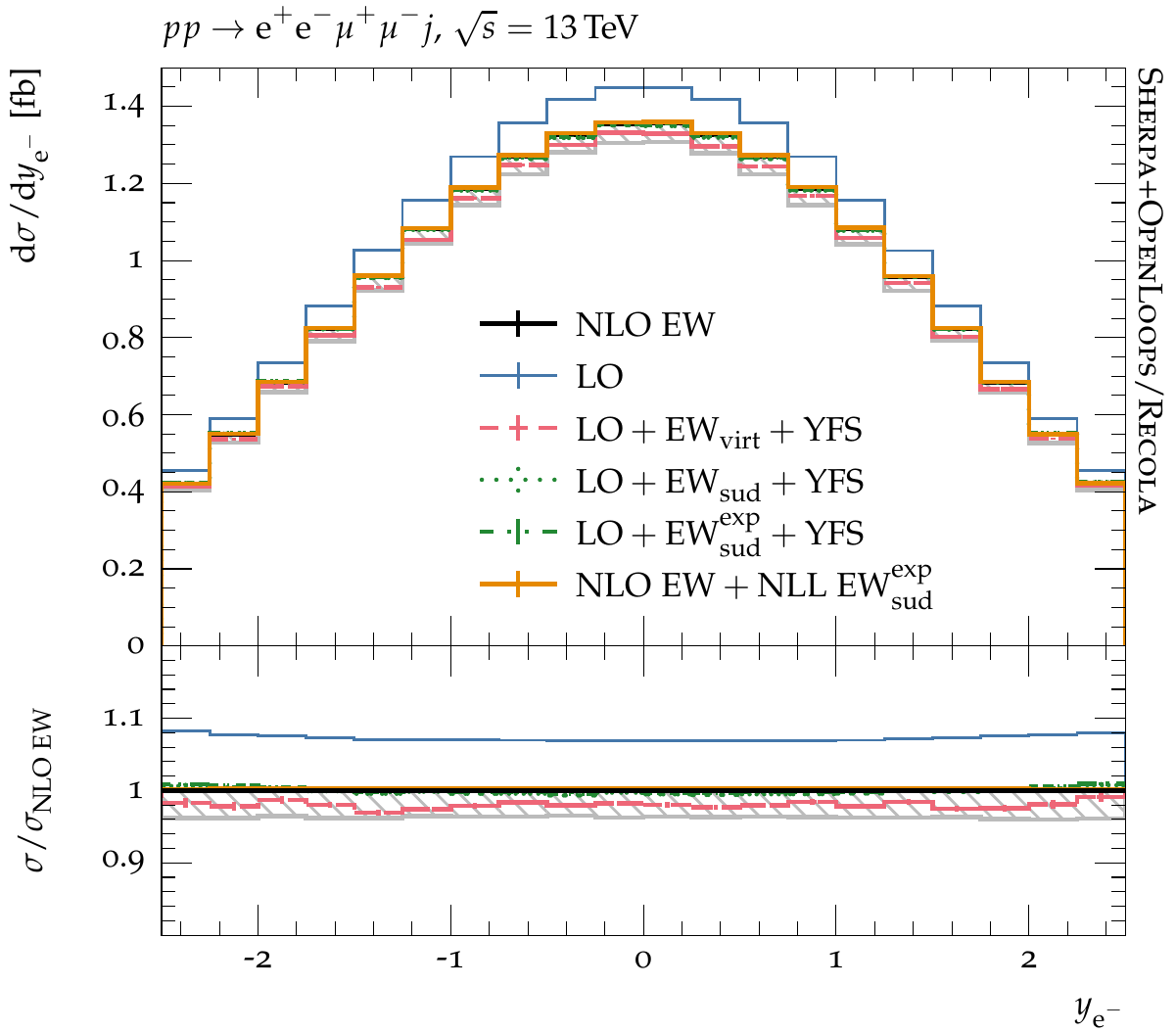}
      \caption{%
        Leptonic observable distributions as shown in Fig.~\ref{fig:fo_validation_0j} but for
        the $pp \to e^+e^- \mu^+\mu^{-} j$ process. }
      \label{fig:fo_validation_1j_0jobs}
\end{figure}

In Fig.~\ref{fig:fo_validation_1j_0jobs}
we now show the same four observables as for the 0-jet case.
In comparison,
we find a small reduction of the EW scheme dependence,
slightly reduced NLO EW corrections,
and a very similar behaviour of the approximation
with respect to the NLO EW results.
This suggests a factorisation of the logarithmic corrections
of the \EWvirt\ and \EWsud approximation with respect to
additional QCD emissions.
This can be related to the fact that the sum of EW
charges of the external lines remains unaffected
by QCD corrections~\cite{Brauer:2020kfv}.

Due to the presence of the jet,
the $\Delta R_{2e,2\mu}<\pi$ region is now already populated
at LO. This results in a good agreement between the NLO EW calculation
and the approximations also in this region,
removing the discontinuity seen in the inclusive $ZZ$ results.
This is also true for the scheme dependence of the NLO EW calculation,
since now the entire observable range receives NLO contributions.
In fact, we observe a nearly constant NLO correction of about
\num{5}--\SI{10}{\percent}.

\begin{figure}[t!]
      \centering
      \includegraphics[width=0.47\textwidth]{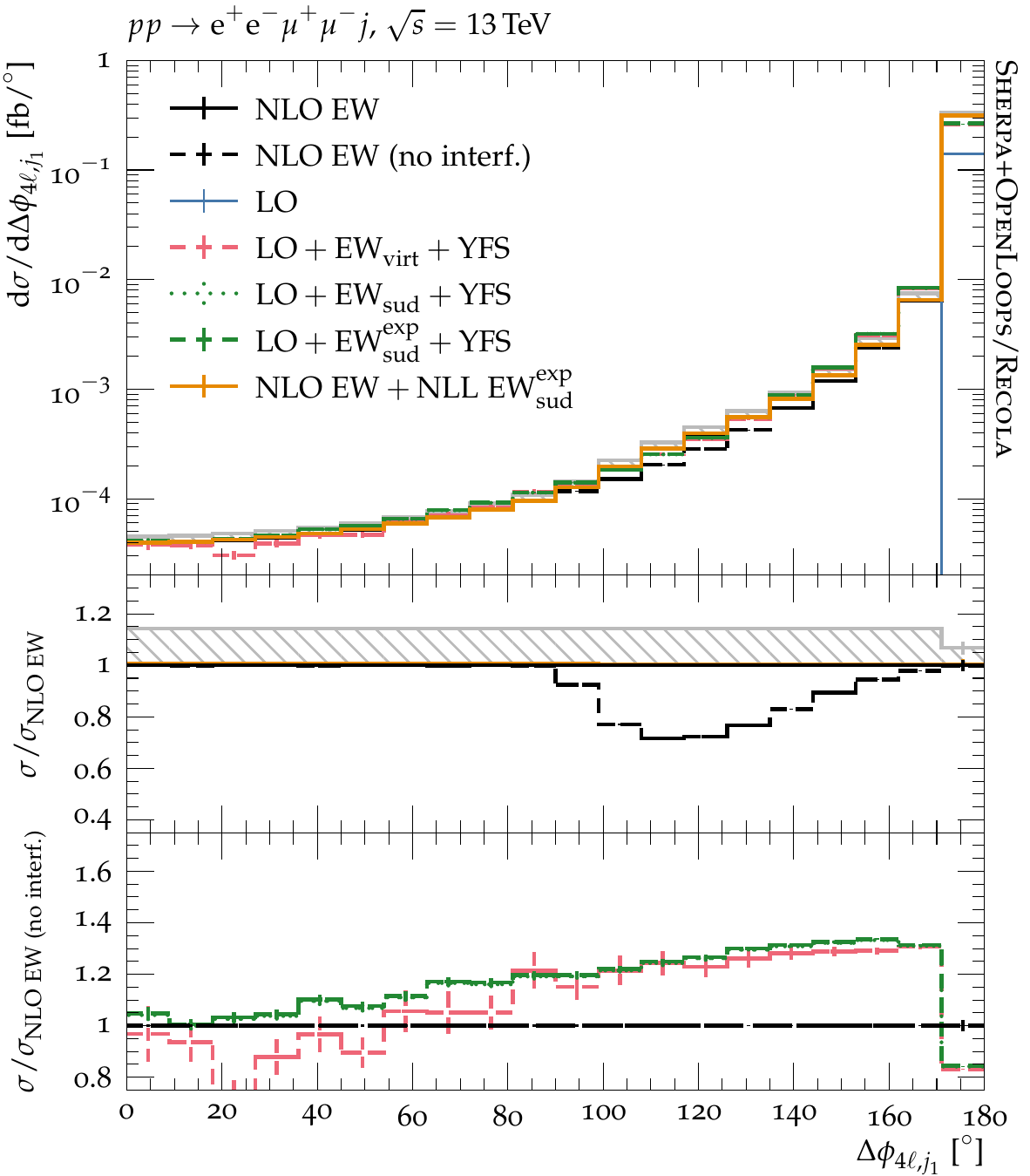} \hfill
      \includegraphics[width=0.47\textwidth]{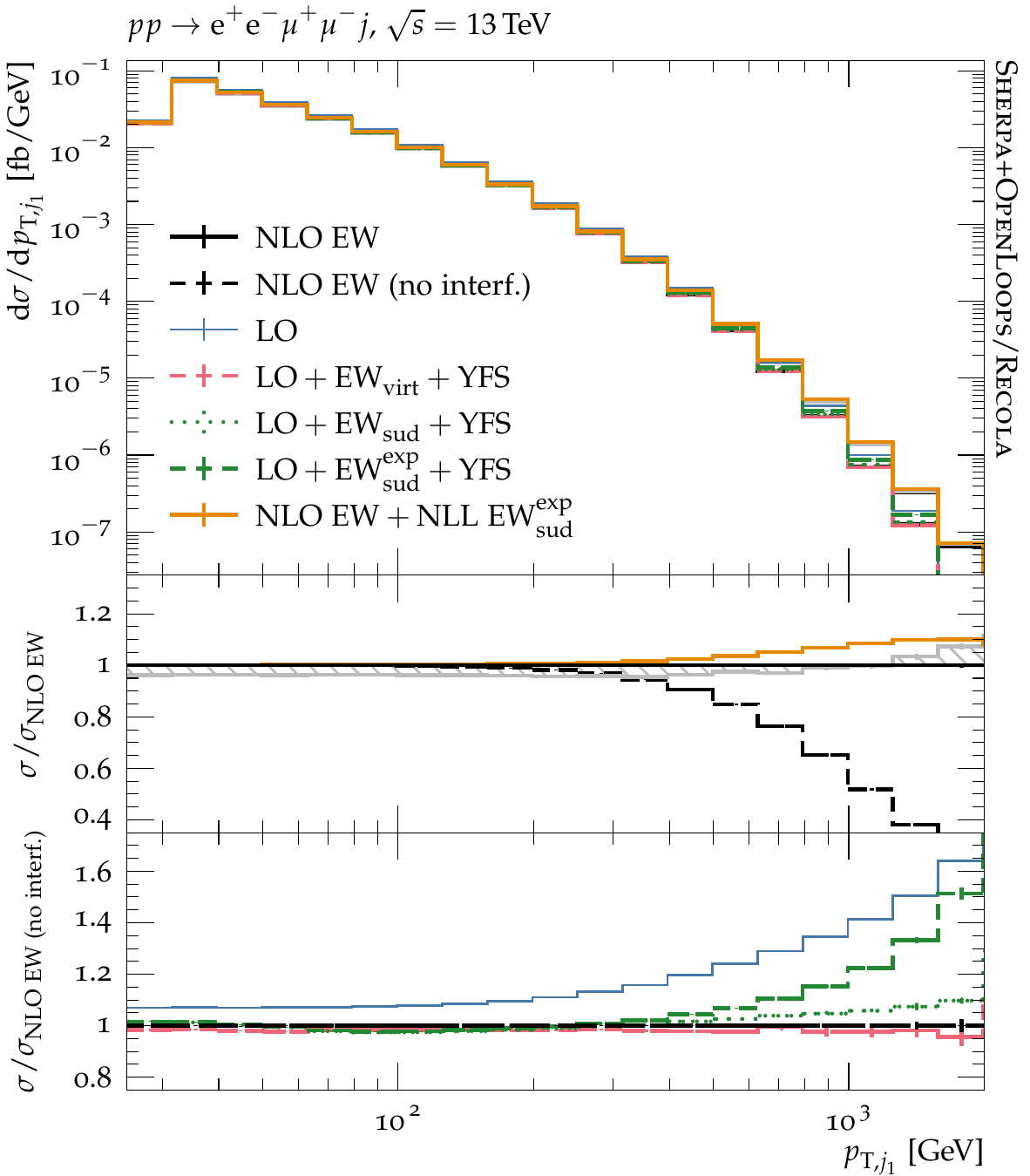}
      \caption{%
        Distributions for
        the azimuthal distance between the four-lepton system and the leading
        jet $\Delta\phi_{4\ell,j_1}$ (left),
        and the leading-jet transverse momentum $p_{\mr{T},j_1}$ (right)
        for the $pp \to e^+e^- \mu^+\mu^{-} j$ process.
        They are given at LO and at NLO EW
        and compared to approximative EW calculations.
        The NLO EW is given for the $G_\mu$ (black line) and $\alpha(\MZ^2)$ (grey line)
        renormalisation schemes, and the span between the two is marked by a hatched band.
        In addition, the NLO EW is plotted without interference terms (``no interf.'').
        All predictions are calculated using \Sherpa{}{}+\OpenLoops/\Recola.}
      \label{fig:fo_validation_1j_1jobs}
\end{figure}

In Fig.~\ref{fig:fo_validation_1j_1jobs}
we additionally present two observables for the leading (\emph{i.e.}\ highest-$p_T$) jet:
the angular separation between it and the four-lepton system $\Delta\phi_{4\ell,j_1}$,
and its transverse-momentum $p_{\text T,j_1}$.
For these observables we display an additional result,
labelled as ``NLO EW (no interf.)''.
This line represents the NLO EW result for the $G_\mu$ scheme
excluding the finite real correction coming from
the interference of diagrams of orders $\mathcal{O}(g_s^2 g^4)$
and $\mathcal{O}(g^6)$ for $ pp \to e^+ e^- \mu^+\mu^- jj$.
Note that this contribution is by construction \emph{not} included
in either the EW approximations or the YFS approach,
since it is a finite real-emission correction.

For the angular separation $\Delta\phi_{4\ell,j_1}$,
Born kinematics produce back-to-back configurations only,
\emph{i.e.} $\Delta\phi_{4\ell,j_1} = \SI{180}{\degree}$.
Hence, for the two approximations
the $\Delta\phi_{4\ell,j_1} < \SI{180}{\degree}$ region
is only populated via YFS photon emissions.
We find that the shape of the NLO EW distribution is reproduced well by YFS,
although there is a shape difference for intermediate angles
$\SI{90}{\degree} < \Delta\phi_{4\ell,j_1} < \SI{180}{\degree}$.
Comparing the full NLO EW with the ``NLO EW (no interf.)'' result we conclude
that this difference is entirely due to these interference terms, which are
missing from the LO+YFS simulation.
The resulting difference in rate is \SI{30}{\%} at large $\Delta\phi_{4\ell,j_1}$
and decreases towards zero with decreasing $\Delta\phi_{4\ell,j_1}$.

For the same reason we find a strong discrepancy between the NLO EW calculation
and the approximations for $p_{\text T, j_1}>\SI{300}{\GeV}$. Again this is entirely
due to the presence of the interference terms, and once they are removed we find
very good agreement.
In particular, we then observe an excellent agreement with the \EWvirt\ approximation,
and a good agreement with the \EWsud\ one.

With these observations we infer that adding a jet veto in order to limit
the activity of the finite contribution would allow the approximations introduced
here to be even closer to the full NLO EW results, see~\cite{Brauer:2020kfv}, where
this is studied for \EWvirt\ in the context of $WW$ and $WWj$ production.
Note that adding a jet
veto introduces further logarithms related to the jet-veto scale which would in general
need to be resummed.
Please also note, that even in the absence of jet vetoes, the picture
changes once QCD corrections, which are of the order of
\SI{100}{\%} in this regime \cite{Binoth:2009wk}, are included, rendering the
impact of the QCD--EW interference contributions less marked.

\FloatBarrier

\subsection{Multijet-merged results}
\label{sec:merged}

In this section we use the approximations validated for both 
$ZZ$ and $ZZj$ production in Sec.\ \ref{sec:fo} to incorporate 
higher-order electoroweak effects in a particle-level 
calculation, as introduced in Sec.\ \ref{sec:setup}. 
Before we present our final results, however, we are 
first proceeding with a structural analysis and validation 
of our calculation in order to understand the interplay of each 
contribution of the multijet-merged computation and their impact 
on the final result.

\subsubsection*{Structural analysis and validation}

The \MEPSatNLO predictions are obtained by merging NLO \QCD\ matrix elements
for $pp \to e^+e^-\mu^+\mu^-$ and $pp \to e^+e^-\mu^+\mu^- j$
and the QCD tree-level matrix elements for
$pp \to e^+e^-\mu^+\mu^- jj$ and $pp \to e^+e^-\mu^+\mu^-jjj$.
As a way to study the interplay between QCD and EW corrections, we also provide
results at \MEPSatLO accuracy, which we obtain by merging the same parton multiplicities
$n=0,1,2,3$, but in this case all are evaluated at LO only, \emph{i.e.}\ \order{\alpha_s^n\alpha^4}.
The merging is performed using the algorithms outlined in
Sec.\ \ref{sec:setup:merging}, with the merging scale set to
\begin{equation}
  \Qcut = 30\;\GeV\,.
\end{equation}
The renormalisation, factorisation, and resummation scales are set according
to the CKKW scale-setting prescription.
The renormalisation scale is thereby defined through
$\muR=\muCKKW$~\cite{Catani:2001cc}, with
\begin{equation}\label{eq:muRCKKW}
  \begin{split}
    \alpha_s^n(\muCKKW^2)
    =&\;\alpha_s(t_1)\cdots\alpha_s(t_n)\;,
  \end{split}
\end{equation}
where the $t_i$ are the reconstructed shower-emission scales of the $n$-jet
hard-process configuration. The scale of the inner core process, $\mucore$, is set to
\begin{equation}
  \mucore=\tfrac{1}{2}\,\left(E_{\mr{T},ee}+E_{\mr{T},\mu\mu}\right)\;,
\end{equation}
and is used to define the factorisation and resummation scales via $\muF=\muQ=\mucore$.
Lastly, note that the simulation of soft and collinear
photon emissions off the final-state leptons in the YFS formalism is enabled
for all results in this section.

We use the QCD \MEPSatNLO prediction as a reference to determine the impact of approximate
EW corrections.
We incorporate EW correction in the \EWvirt\ approximation into the underlying \MEPSatNLO calculation,
both in the additive and the multiplicative scheme, as defined in Eq.~\eqref{eq:deltaEWvirtplus}
and Eq.~\eqref{eq:deltaEWvirt}, respectively. Furthermore, we consider the \EWsud\ approximation as
an alternative.

Finally, we give the QCD \MEPSatLO prediction,
both as-is, serving as an additional reference for the inclusive fiducial cross section,
and including \EWsud corrections.
To better capture non-trivial kinematical effects, we multiply this LO accurate
prediction by the global QCD $k$-factor
given by the ratio of the inclusive fiducial cross sections
of the \MEPSatNLO and \MEPSatLO calculations.
Note that \EWvirt is not readily available for
this \MEPSatLO calculation,
given that it requires NLO EW one-loop matrix elements for all multiplicities
(\emph{i.e.}\ up to $n=3$),
and a local $k$-factor mechanism as described
for \MEPSatNLO in Eq.~\eqref{eq:kfacewvirt}
to make up for missing one-loop matrix elements
is not yet implemented for \MEPSatLO in \Sherpa.

\begin{table}[h!]
  \centering
  \sisetup{table-format = +1, table-align-exponent = false, table-align-text-post = false}
  \resizebox{\textwidth}{!}{%
  \begin{tabular}{@{}llccSSS@{}} \toprule
                      \multicolumn{2}{c}{%
                      $pp\to e^+e^-\mu^+\mu^- + \text{jets}$
                      }
                    & \multicolumn{2}{c}{%
                      fiducial cross section
                      }
                    & \multicolumn{3}{c}{%
                      corrections to $\MEPSatNLO+\text{YFS}$
                      } \\ \cmidrule{1-2}\cmidrule(lr){3-4}\cmidrule(l){5-7}
    Scheme          & Region
                    & {$\MEPSatLO + \text{YFS}$}
                    & {$\MEPSatNLO + \text{YFS}$}
                    & {$\times\,\EWvirt{}$}
                    & {$\times\,\EWsud{}$}
                    & {$\times\,\EWsudexp{}$}
                    \\ \midrule
    $\Gmu$          & inclusive
                    & {11.10\,fb} 
                    & \textbf{13.34\,fb} 
                    & -4 {\,\%} 
                    & -4 {\,\%} 
                    & -3 {\,\%} 
                    \\ \bottomrule
  \end{tabular}
  }
  \caption{
    Inclusive fiducial cross sections for $pp\to e^+ e^- \mu^+ \mu^-$ + jets
    at $\sqrt{s}=\SI{13}{\TeV}$ for \MEPSatLO and \MEPSatNLO in the
    $G_{\mu}$ scheme including YFS photon emissions.
    For the \MEPSatNLO predictions,
    relative corrections for the combination
    with the \EWvirt and \EWsud approximations are also listed.
  }
\label{table:mepsnloxs}
\end{table}

Results for the inclusive fiducial cross sections
are listed in Tab.~\ref{table:mepsnloxs}.
We first note that the \MEPSatLO cross section
is \SI{13}{\percent} larger than the pure LO result
quoted in Tab.~\ref{table:foxs0j}.
This is not unexpected due to the presence
of higher-multiplicity matrix elements,
which explicitly take into account the opening
of the $gq$- and $gg$-induced channels.
Next, the global QCD $k$-factor between \MEPSatLO
and \MEPSatNLO is \num{1.20},
further increasing the inclusive cross section.
Applying EW approximations to the \MEPSatNLO
calculation leads to a decrease of \SI{4}{\percent}
for the additive/multiplicative \EWvirt approximations
and of \SI{3}{\percent} for the \EWsud approximation.
These numbers tend to be smaller than the ones
reported for the fixed-order calculations in Tabs.~\ref{table:foxs0j}
and~\ref{table:foxs1j}, where we found a \SI{7.3}{\percent}
reduction for the LO+\EWvirt vs.\ the LO
in the $ZZ$ case in the $G_\mu$ scheme.
However, there the LO prediction did not include YFS
QED corrections, which by itself reduces this LO cross section
by about \SI{4}{\percent}, which makes up for the discrepancy.

\begin{figure}[t!]
  \centering
  \includegraphics[width=0.47\textwidth]{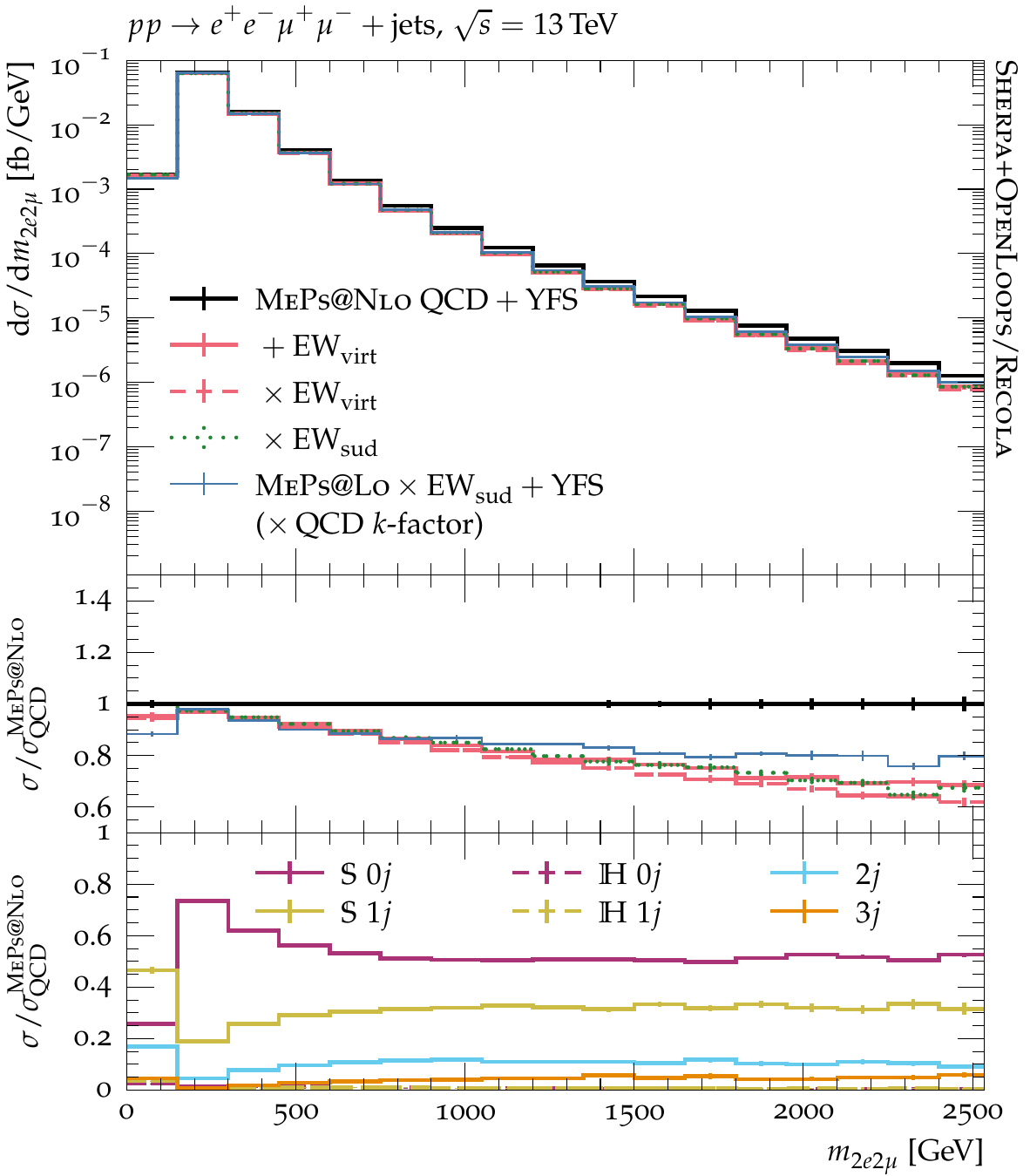} \hfill
  \includegraphics[width=0.47\textwidth]{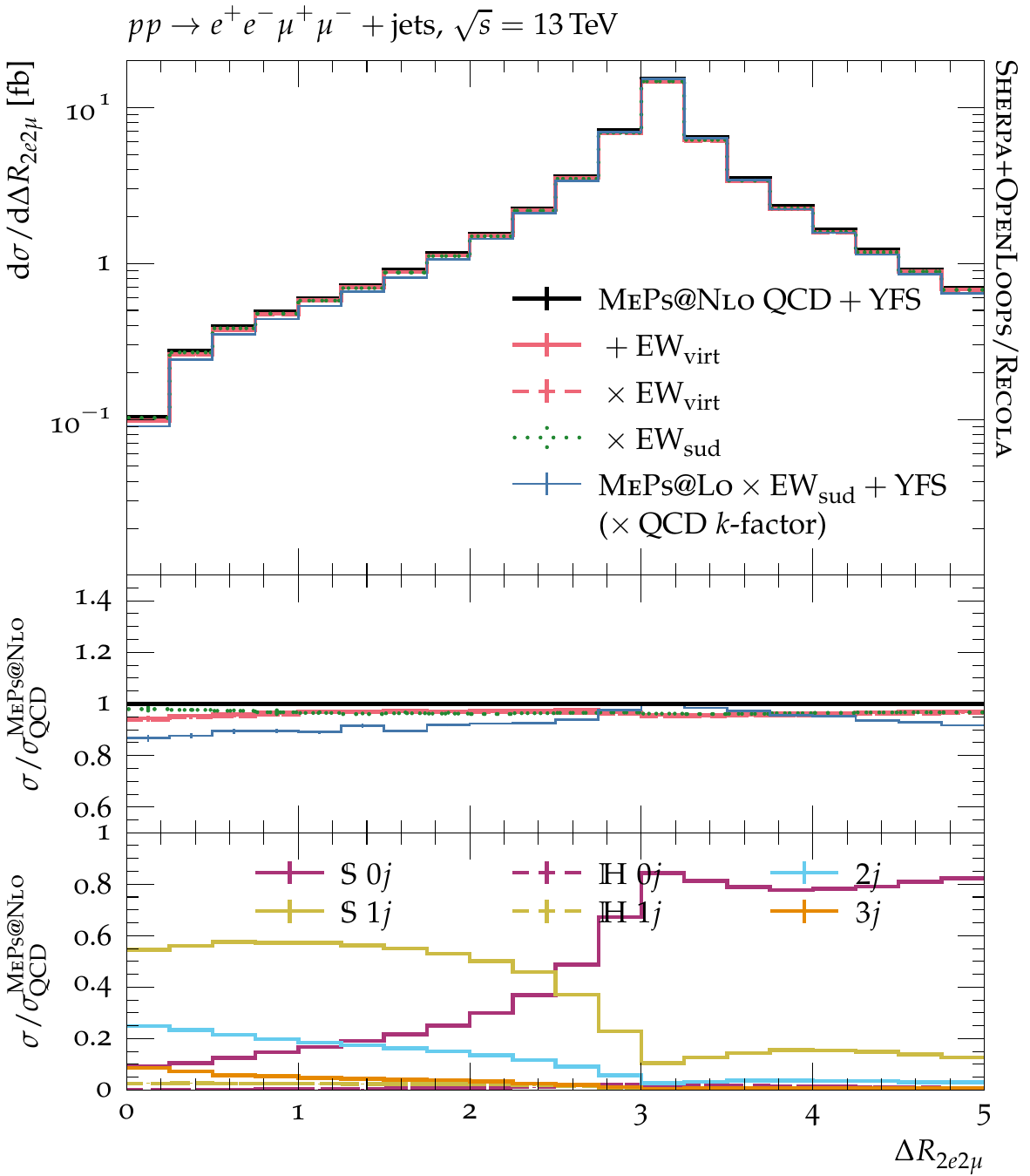} \\
  \includegraphics[width=0.47\textwidth]{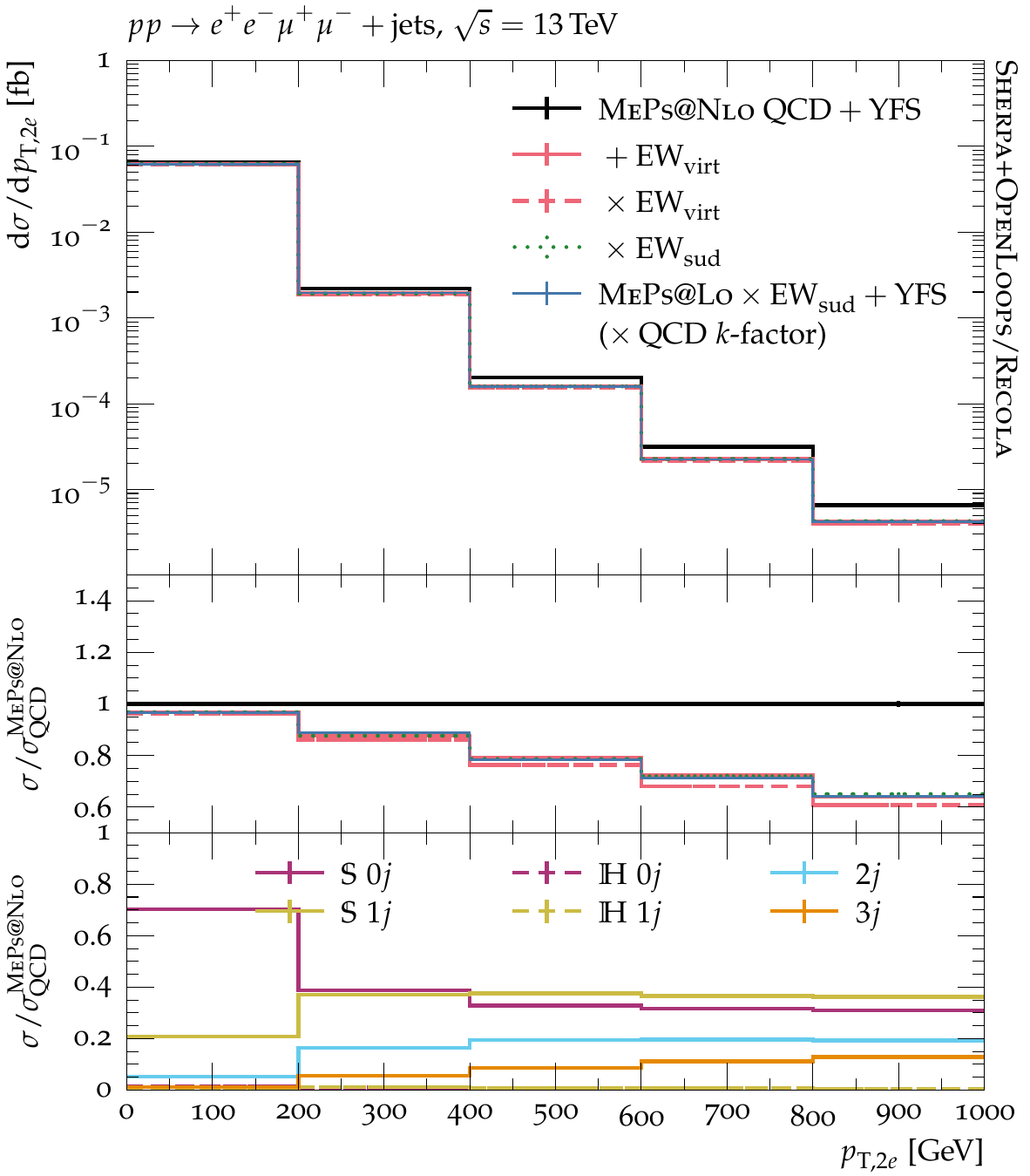} \hfill
  \includegraphics[width=0.47\textwidth]{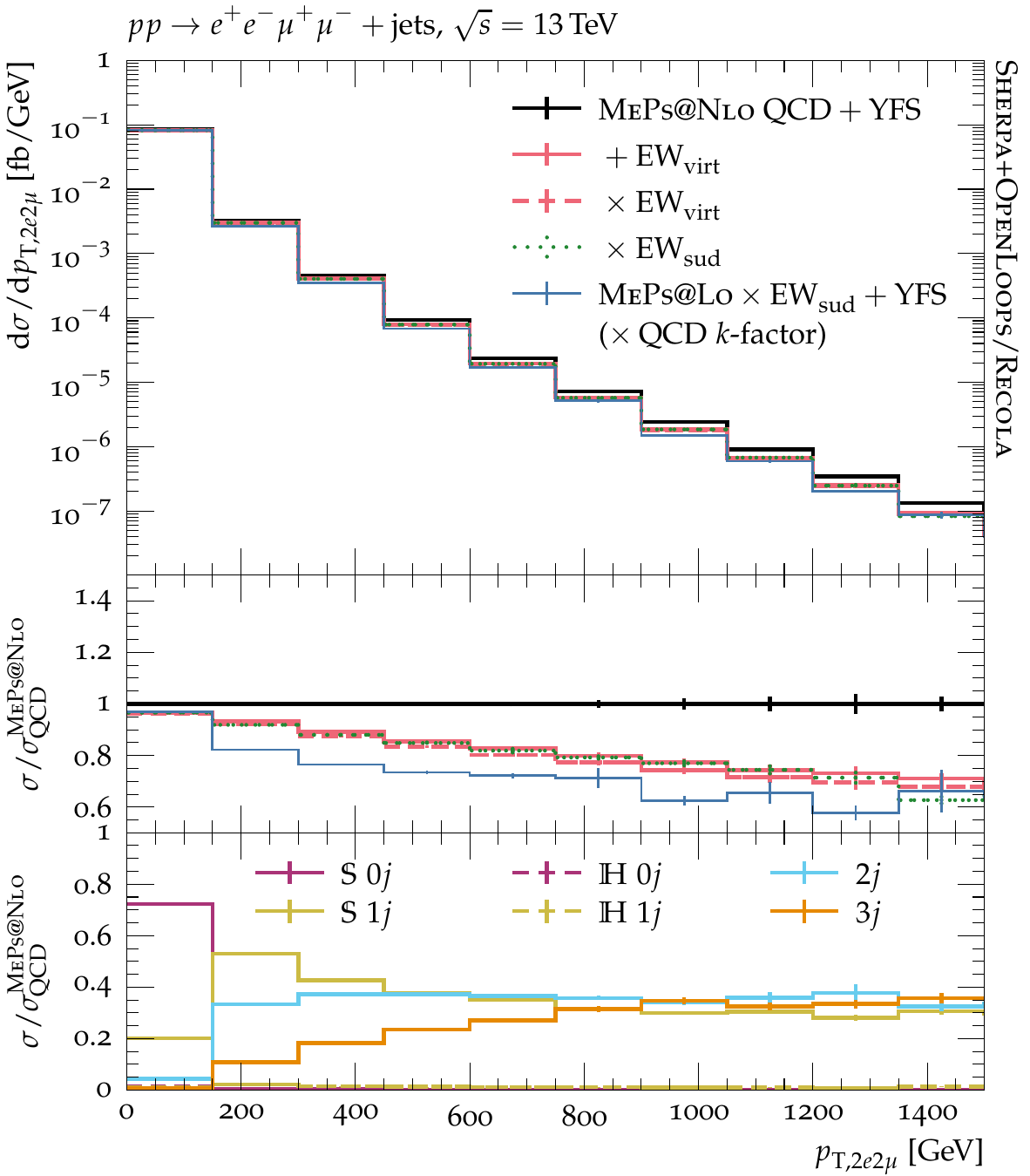}
  \caption{
    Distributions of leptonic observables
    for $pp \to e^+e^- \mu^+\mu^{-} + \text{jets}$ production.
    The baseline prediction is a \MEPSatNLO calculation in the $G_\mu$ scheme.
    On top of it, \EWvirt and \EWsud approximations are applied.
    As a reference, the \EWsud approximation is also shown
    for an underlying \MEPSatLO calculation,
    which is rescaled to the total \MEPSatNLO rate
    using the global QCD $k$-factor of 1.20.
    The four observables shown from top left to bottom right are:
    the invariant mass of the four-lepton system $m_{2e2\mu}$,
    the $Z$-boson distance $\Delta R_{2e,2\mu}$,
    the transverse momentum of the di-electron pair $p_{\mr{T},2e}$,
    and the transverse momentum of the four-lepton system $p_{\mr{T},2e2\mu}$.
    All predictions are calculated using \Sherpa{}+\OpenLoops/\Recola.
    The first ratio plot shows the relative size of the EW corrections,
    while the second one gives the relative size of the contributions
    to the \MEPSatNLO prediction.
  }
  \label{fig:mepsnlo_1}
\end{figure}

\begin{figure}[t!]
  \centering
  \includegraphics[width=0.47\textwidth]{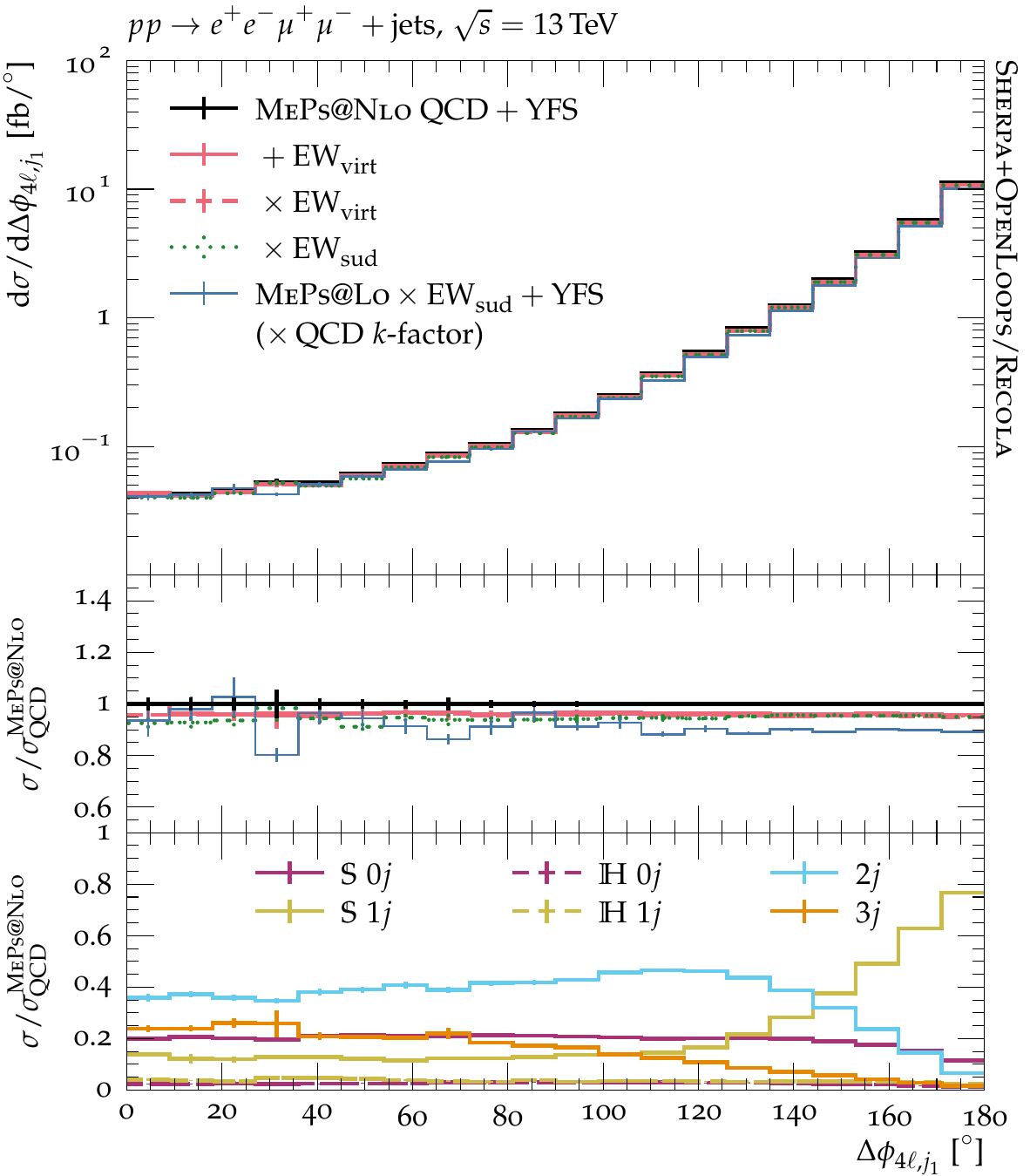} \hfill
  \includegraphics[width=0.47\textwidth]{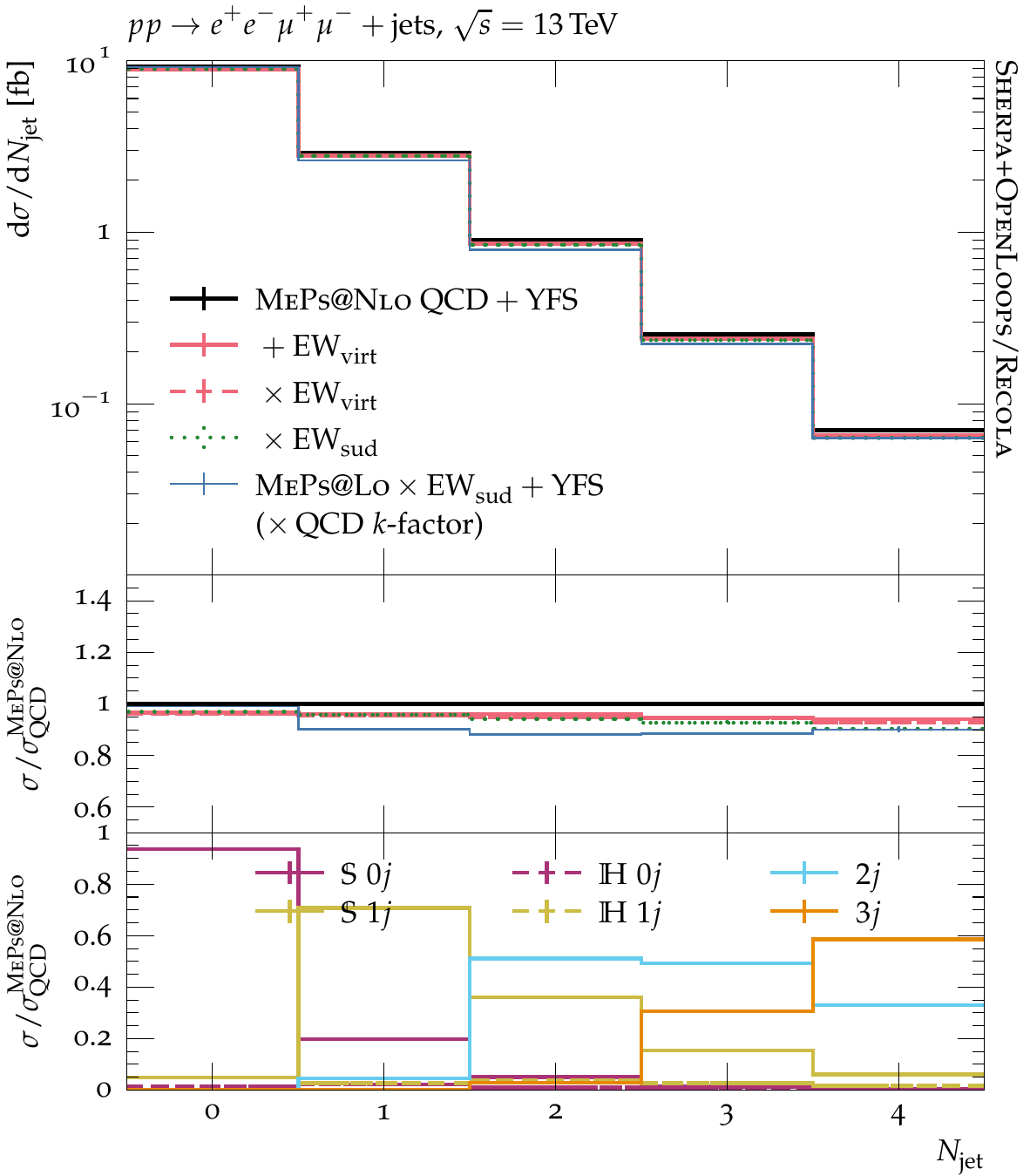} \\
  \includegraphics[width=0.47\textwidth]{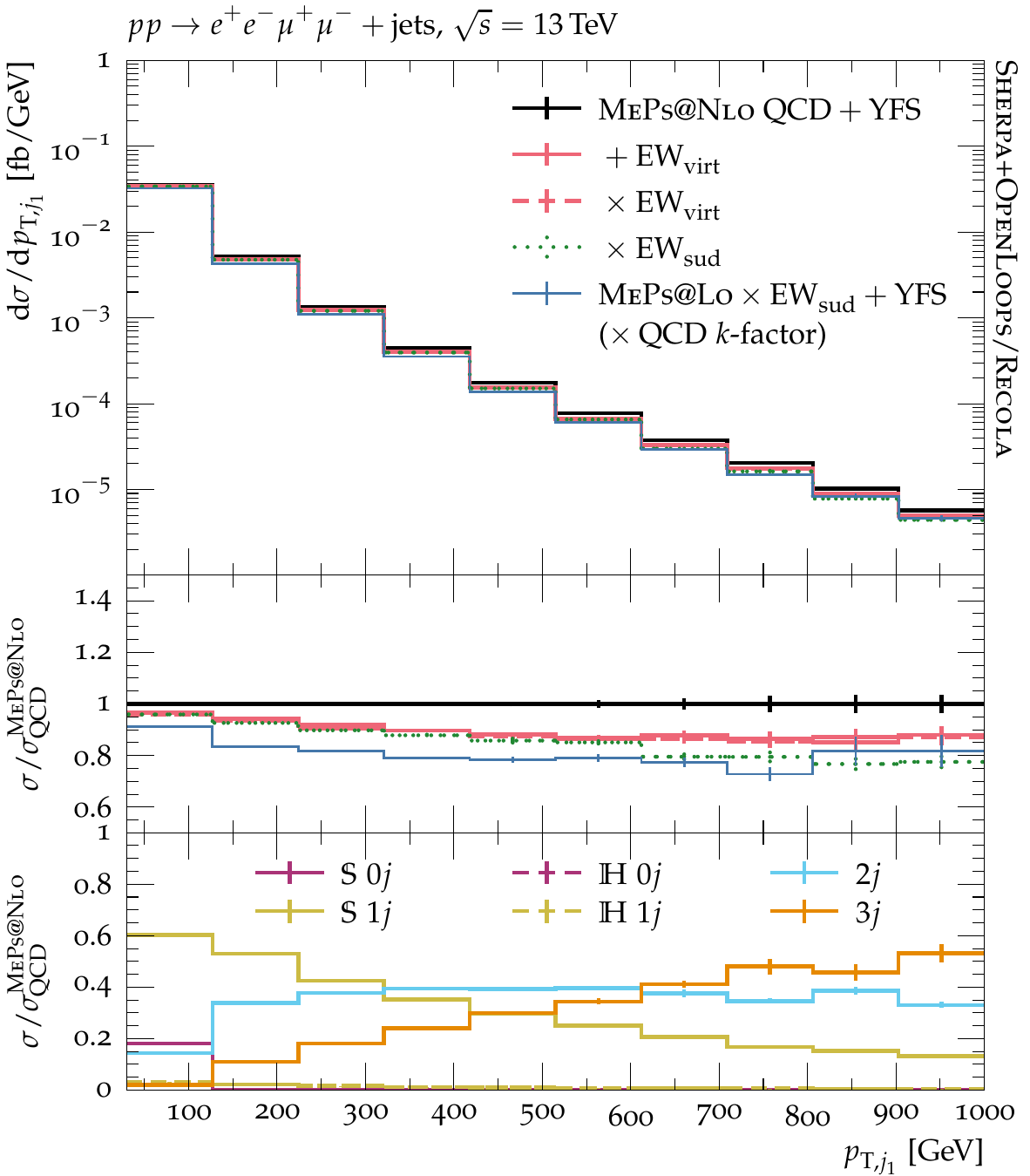} \hfill
  \includegraphics[width=0.47\textwidth]{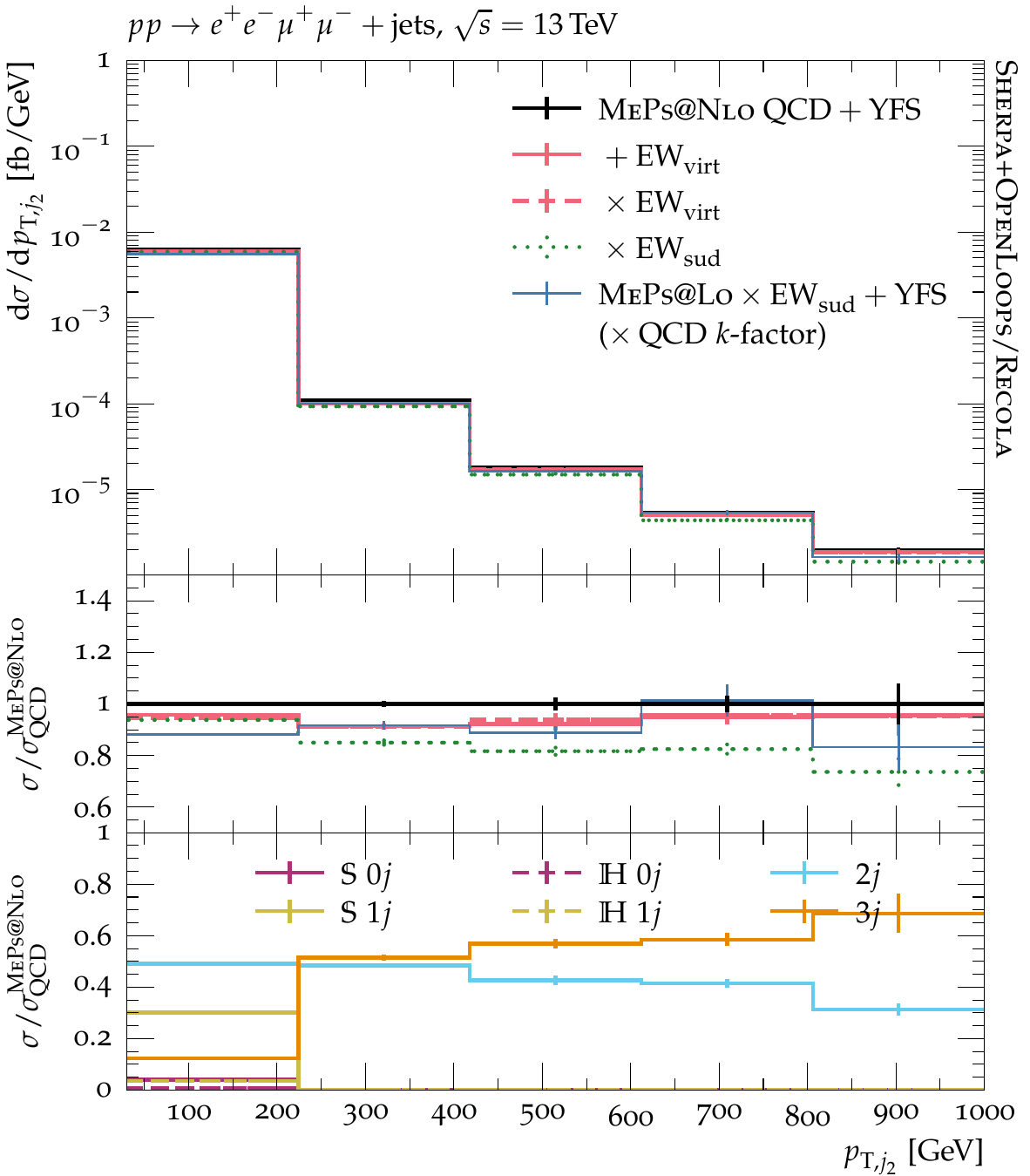}
  \caption{%
    As Fig.~\ref{fig:mepsnlo_1} but for jet observables.
    Shown are from top left to bottom right:
    the angular separation between the four-lepton system
    and the hardest jet $\Delta \phi_{4\ell,j_1}$,
    the number of jets $N_\text{jet}$,
    and the transverse momenta of the
    hardest jet $p_{\text T, j_1}$
    and second hardest jet $p_{\text T, j_2}$.
  }
  \label{fig:mepsnlo_2}
\end{figure}

Before turning to more exclusive observables,
we briefly recapitulate the key differences
of how \EWvirt and \EWsud is applied
to an underlying \MEPSatNLO calculation.
As described in Sec.~\ref{sec:setup:ew},
the two approximations
act differently on the individual
contributions that make up a multijet-merged calculation,
namely the \mm{S} and \mm{H} events of each
QCD NLO multiplicity (here $0j$ and $1j$),
and the tree-level events of the LO multiplicities (here $2j$ and $3j$).
In particular, the \EWvirt\ is not applied to \mm{H}-type events,
and approximately treats higher-multiplicity tree-level contributions
through the $k$-factor defined in Eq.~\eqref{eq:kfacewvirt},
such that the corrections for additional LO jets
in the EW Sudakov regime are incomplete.
The \EWsud approximation on the other hand is applied
to all contributions of the calculation.
Therefore,
we show and discuss in the following
not only the baseline \MEPSatNLO prediction
along with the two EW approximations,
but also the relative size of each contribution
within the \MEPSatNLO result.
This helps to better understand how the \MEPSatNLO\ results
are affected by the EW approximations,
and in particular to ensure that the \EWvirt\ approximation
is not spoiled by a large admixture of \mm{H} events.

In Fig.~\ref{fig:mepsnlo_1} we show the same lepton observables we examined
already for the fixed-order $ZZ$ and $ZZj$ predictions presented in
Sec.~\ref{sec:fo}, here for the inclusive event selection placing
no requirements on any additional jet activity,
while Fig.~\ref{fig:mepsnlo_2} shows four observables demanding
one or more additional jets beyond the inclusive lepton acceptance cuts.
The latter are the angular separation $\Delta\phi_{4\ell,j_1}$
between the four-lepton system and the leading jet,
the jet multiplicity $N_\text{jet}$,
and the transverse momenta $p_{\mr{T}, j_{1}}$
and $p_{\mr{T}, j_{2}}$
of the leading and subleading jet, respectively.

Examining Figs.~\ref{fig:mepsnlo_1} and~\ref{fig:mepsnlo_2}, we find the
same qualitative behaviour as in the fixed-order $ZZ$ and $ZZj$ results:
Distributions that have an energy scaling, such as the
invariant mass of the four-lepton system, the $p_T$ of the leptons or that of the
jets,  show enhanced EW effects at high energies, which are similarly described
by both the \EWvirt and \EWsud approximations. On the other hand, scale-less
quantities, such as the distance between the two $Z$-bosons or between the four-lepton
system and the hardest jet, exhibit roughly flat and small corrections in
both cases.
This is also the case for the $N_{\mathrm{jet}}$
distribution.
The reason can be seen in the fact that extra QCD radiation is
predominantly soft and collinear, and as such it does neither substantially change the
EW charge distribution, nor induce additional large scales.

Turning now to the various contributions of the merged calculation
(as shown in the bottom panel of each plot), we can see that
in all cases the contributions of \mm{H}-type events are below \SI{5}{\percent}.
Therefore, the \EWvirt correction is not hampered by a large admixture
of (uncorrected) \mm{H} events in the phase space explored here.
This is not surprising, because the contribution of \mm{H} events
is constrained by the $\Theta(\Qcut - Q_{n+1})$ term
in Eq.~\eqref{eq:meps-nlo-excl},
to avoid double counting with matrix elements of higher jet multiplicities.
There is however a sizeable contribution (more than \SI{20}{\percent})
of tree-level $2j$ and/or $3j$ contributions
for $\Delta R_{2e2\mu}<\pi$,
$\Delta \phi_{4\ell,j_1} < \pi$,
$N_\text{jet} > 1$
and all transverse-momentum distributions,
\emph{i.e.}\ whenever the observable favours
contributions from events with many hard and/or widely separated jets,
which are mainly generated by the $2j$ and $3j$ matrix
elements. While the \EWvirt falls back on the use of a $k$-factor
for these two contributions, we do not find evidence
that the \EWvirt and the \EWsud generally begin to deviate
from one another in the regions dominated by high-multiplicity matrix elements,
suggesting that the use of a lower multiplicity (here: $1j$)
to calculate approximate EW corrections via the $k$-factor
does not introduce a large error for the observables studied here,
as is proven in App.\ \ref{sec:app:match}.

Finally, for the \MEPSatLO{}+\EWsud calculation rescaled with the global QCD $k$-factor,
we find that it faithfully reproduces the \MEPSatNLO calculation
in phase-space regions dominated by $0j$ events,
while it otherwise deviates by up to \SI{10}{\percent} from it.
We have checked that this deviation is entirely due to the underlying QCD description.
The \EWsud corrections of both calculations are indeed nearly identical,
which is expected given the structure of the corrections as laid out in Sec.~\ref{sec:setup:ew}
and the fact that the contribution of $\mm{H}$ events is negligible;
for the \EWsud approximation it does not matter if it is applied to LO or \mm{S} events.
The finding that \EWsud gives identical corrections in both cases indeed serves
as an important cross-check of our implementation.

\FloatBarrier

\subsubsection*{Final results}

After the structural analysis of the multijet-merged calculation
with respect to the inclusion of EW corrections through the \EWvirt and \EWsud
approximation,
we are now ready to present our final results,
which are directly relevant for comparisons to data.
To this end, we adjust our predictions with respect to the previous section
as follows:
\begin{itemize}
  \item
    The \MEPSatNLO calculation remains the reference result,
    but we add a scale-variation band to estimate
    its theoretical uncertainty related to the choice of the QCD
    renormalisation and factorisation scales $\muR$ and $\muF$.
    The band is defined as the envelope of the 7-point variations~\cite{Cacciari:2003fi}
    \begin{equation*}
      \left\{(\tfrac{1}{2}\muR,\; \tfrac{1}{2}\muF),\; (\tfrac{1}{2}\muR,\muF),\;
      (\muR,\tfrac{1}{2}\muF),\; (\muR,\muF),\; (\muR,2\muF),\; (2\muR,\muF),\; (2\muR,2\muF)\right\}\,,
    \end{equation*}
    that are evaluated using on-the-fly reweighting~\cite{Bothmann:2016nao}.
    The $\alpha_\text S$ and PDF input scales in the parton shower are varied
    along with the $\muR$ and $\muF$ values used in the hard process.
    Comparing the QCD scale uncertainties with the size of the EW corrections
    allows us to assess the phenomenological relevance of the latter.
  \item
    As a second addition in the QCD sector, we provide a \MEPSatLOOP prediction,
    merging loop-induced matrix elements for $ZZ$ and $ZZj$ production at LO.
    This is added to the \MEPSatNLO calculation.
  \item
    We drop the additive \EWvirt scheme, which has been
    discussed in the previous section.
    It suffices here to consider the multiplicative scheme for further comparison.
    As before, we show the \EWsud approximation, however, now supplemented with its
    exponentiated version, $\EWsud^\text{exp}$.
\end{itemize}

\begin{figure}[t!]
  \centering
  \includegraphics[width=0.47\textwidth]{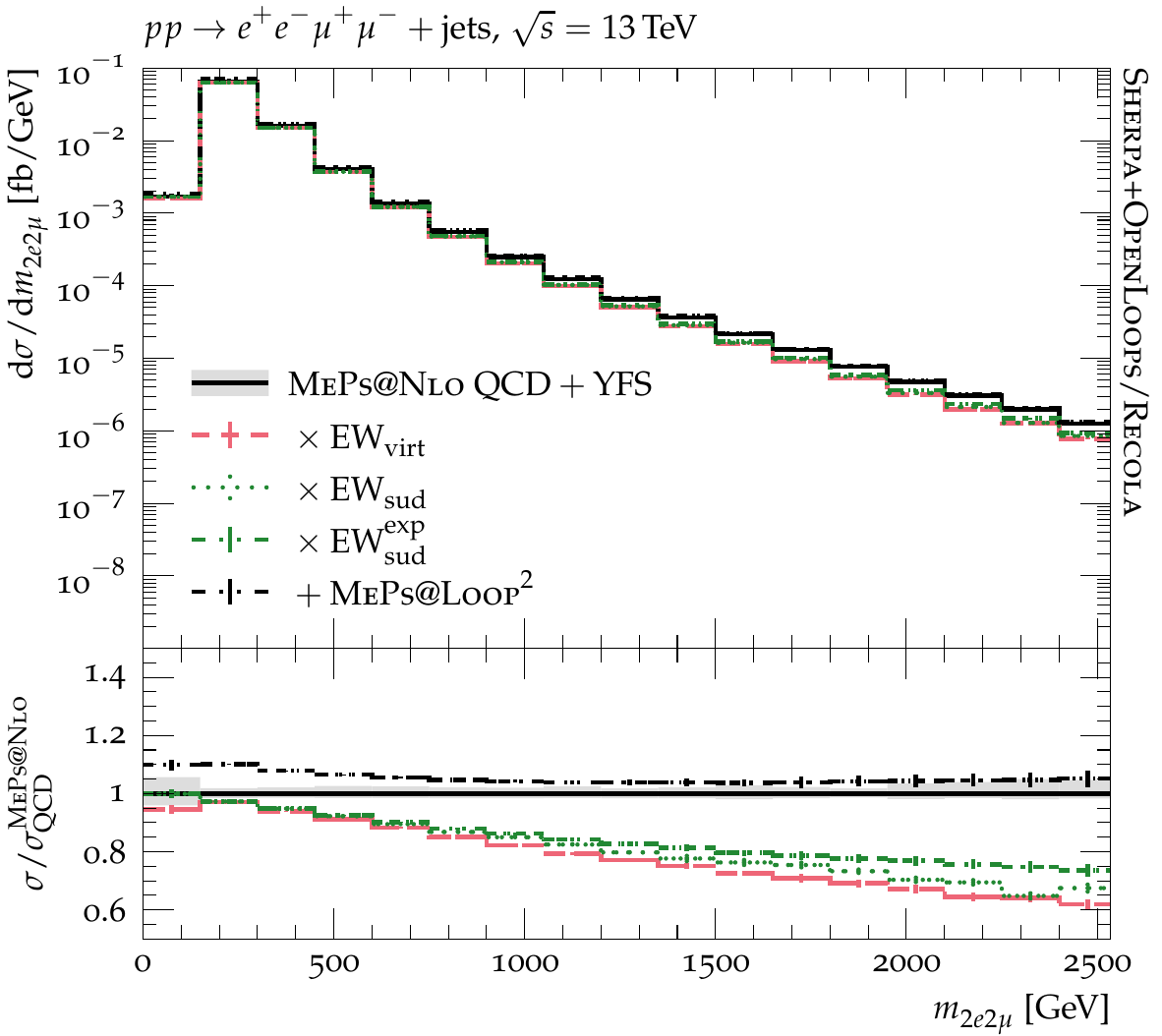} \hfill
  \includegraphics[width=0.47\textwidth]{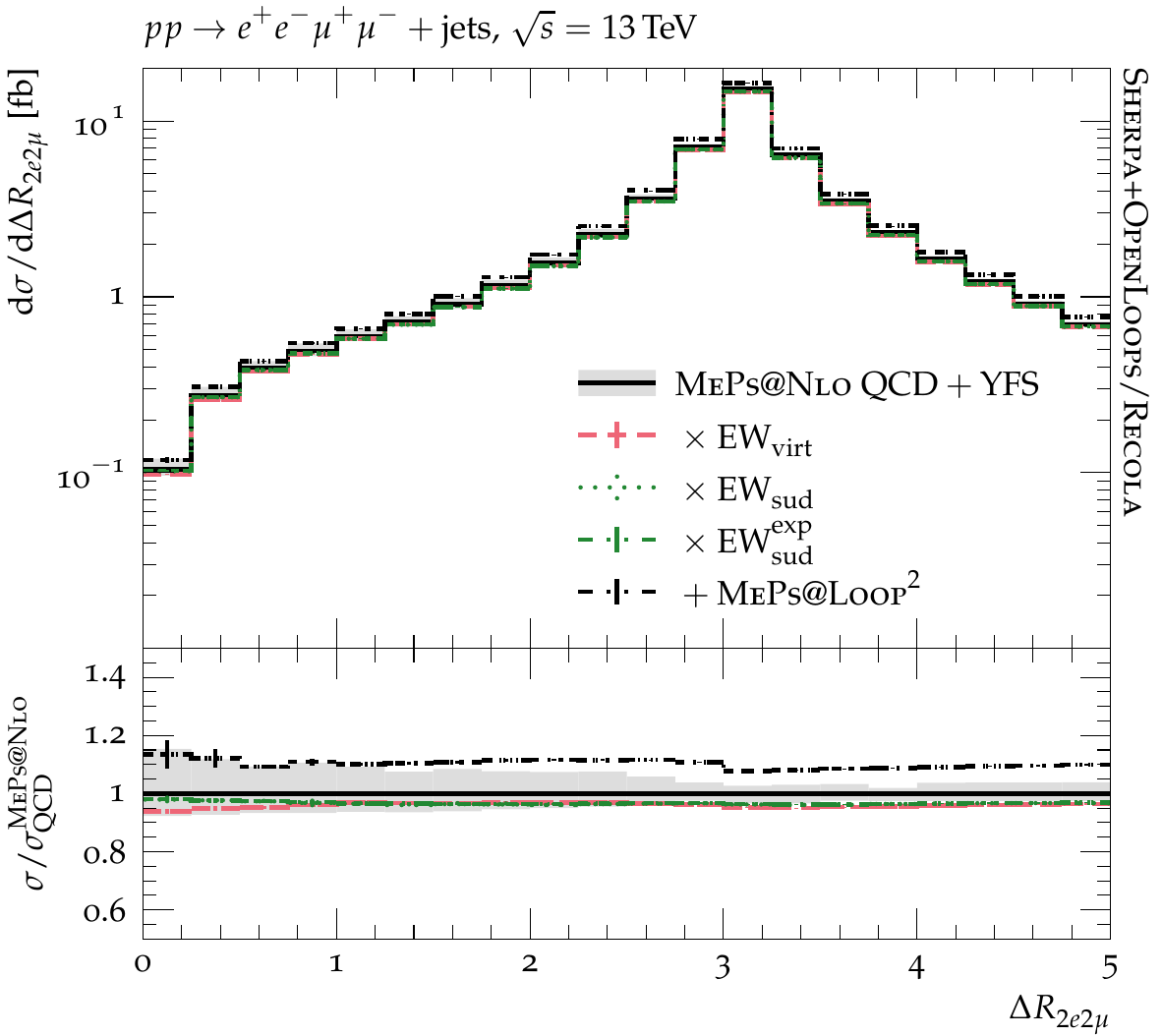} \\
  \includegraphics[width=0.47\textwidth]{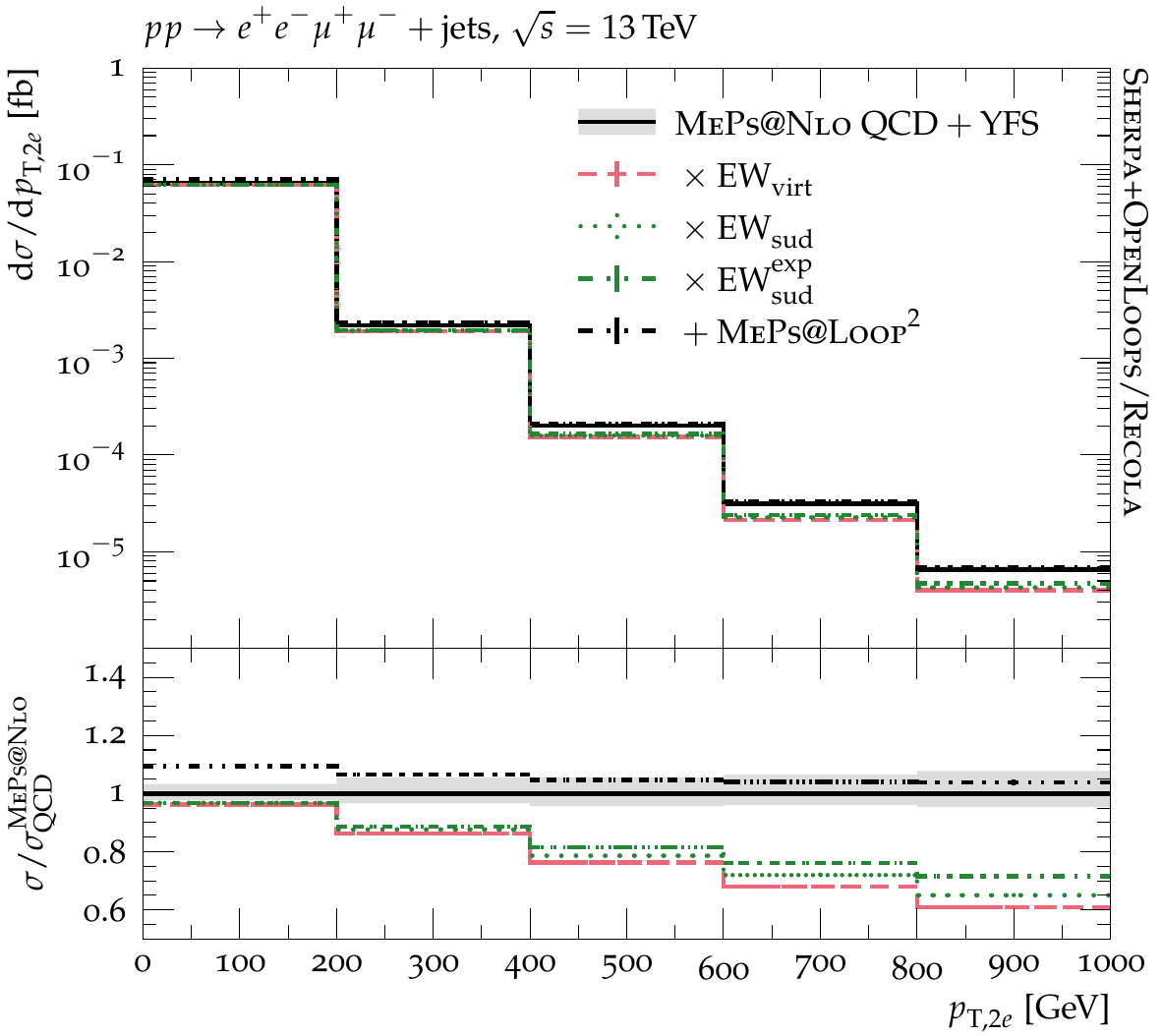} \hfill
  \includegraphics[width=0.47\textwidth]{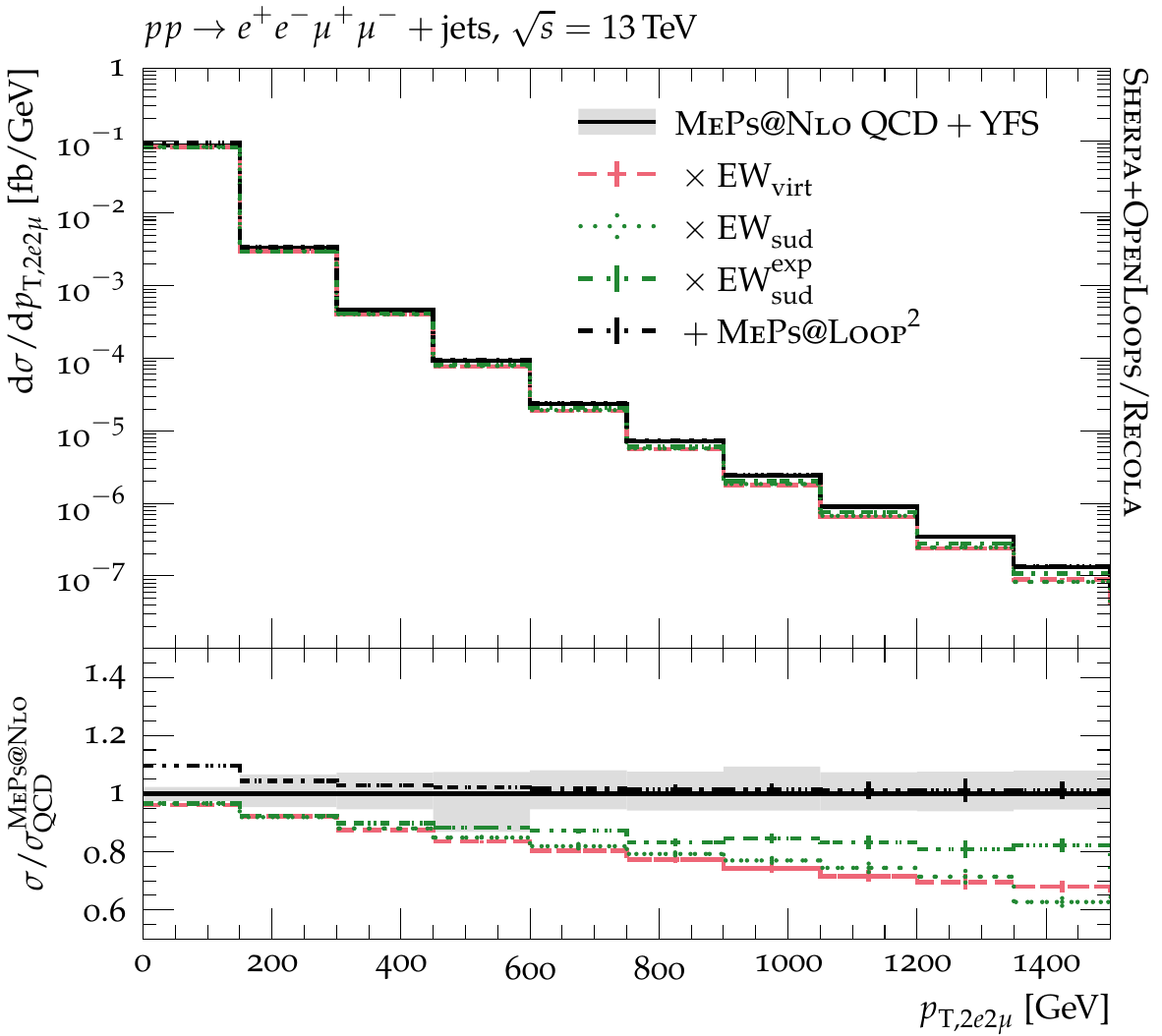}
  \caption{%
    Distributions of leptonic observables
    for $pp \to e^+e^- \mu^+\mu^{-} + \text{jets}$ production.
    The baseline prediction is given by the \MEPSatNLO result in the $G_\mu$ scheme,
    with the grey band indicating its 7-point scale-variation uncertainty.
    On top of it, loop-induced corrections and \EWvirt/\EWsud approximations are applied.
    Shown are from top left to bottom right:
    the four-lepton invariant mass $m_{2e2\mu}$,
    the $Z$-boson distance $\Delta R_{2e,2\mu}$,
    the di-electron transverse momentum $p_{\text T,2e}$,
    and four-lepton transverse momentum $p_{\text T,2e2\mu}$.
    All predictions are calculated using \Sherpa{}+\OpenLoops/\Recola.
  }
  \label{fig:mepsnlo_pheno_1}
\end{figure}

We begin by considering in Fig.~\ref{fig:mepsnlo_pheno_1} the four leptonic
observables studied before already, \emph{i.e.}\ $m_{2e2\mu}$, $p_{\text T, 2e}$,
$\Delta R_{2e2\mu}$, and $p_{\mr{T}, 2e2\mu}$. As these are all non-zero at
leading-order, in our MEPS@NLO setup they are all described
at NLO QCD accuracy, receiving additional contributions also from higher-order
terms. The decomposition of the MEPS@NLO prediction for the considered
observables can be read off from the lower panels of Fig.~\ref{fig:mepsnlo_1}.
Notably, phase-space regions that are either filled only in the presence of,
or receive large contributions from,
additional radiation from either the parton shower or
higher-multiplicity tree-level processes exhibit somewhat
larger scale uncertainties of $\order{\pm \SI{10}{\%}}$. This is in particular the case for $\Delta R_{2e2\mu}<\pi$
and $p_{\mr{T}, 2e2\mu} >\SI{100}{GeV}$, where the four-lepton system recoils against
(several) additional hard emissions. For the bulk of the considered observable
distributions, scale uncertainties are well below \SI{10}{\%}.
Accordingly, we find that they do not cover the EW corrections
for the dimensionful observables, \emph{i.e.}\ the transverse-momentum
and invariant-mass variables, beyond scales of about \SI{200}{\GeV}.
For the $\Delta R_{2e2\mu}$ observable, however,
where EW corrections are very moderate,
they are amlost entirely covered by the QCD scale-uncertainty band.

\begin{figure}[t!]
  \centering
  \includegraphics[width=0.47\textwidth]{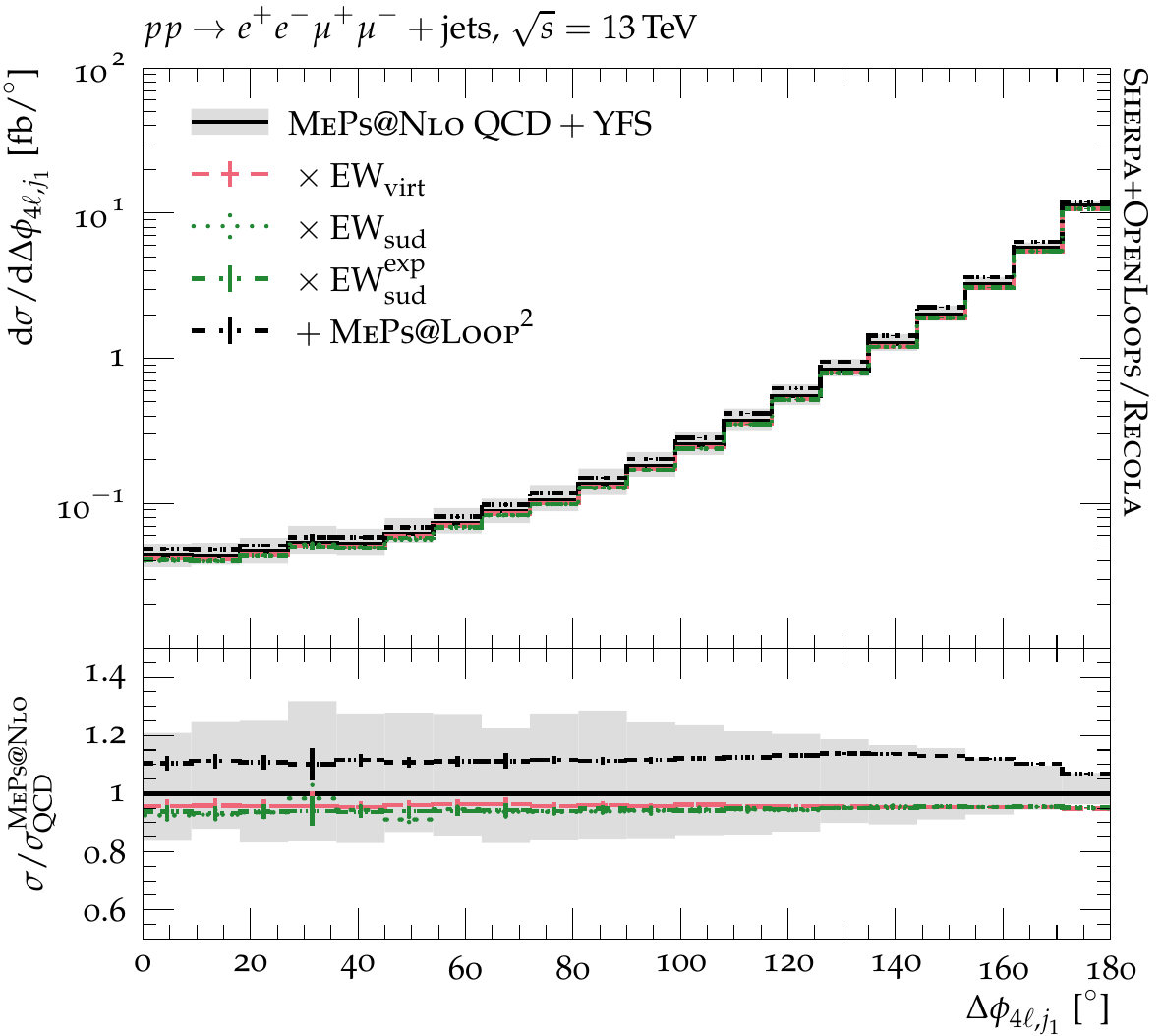} \hfill
  \includegraphics[width=0.47\textwidth]{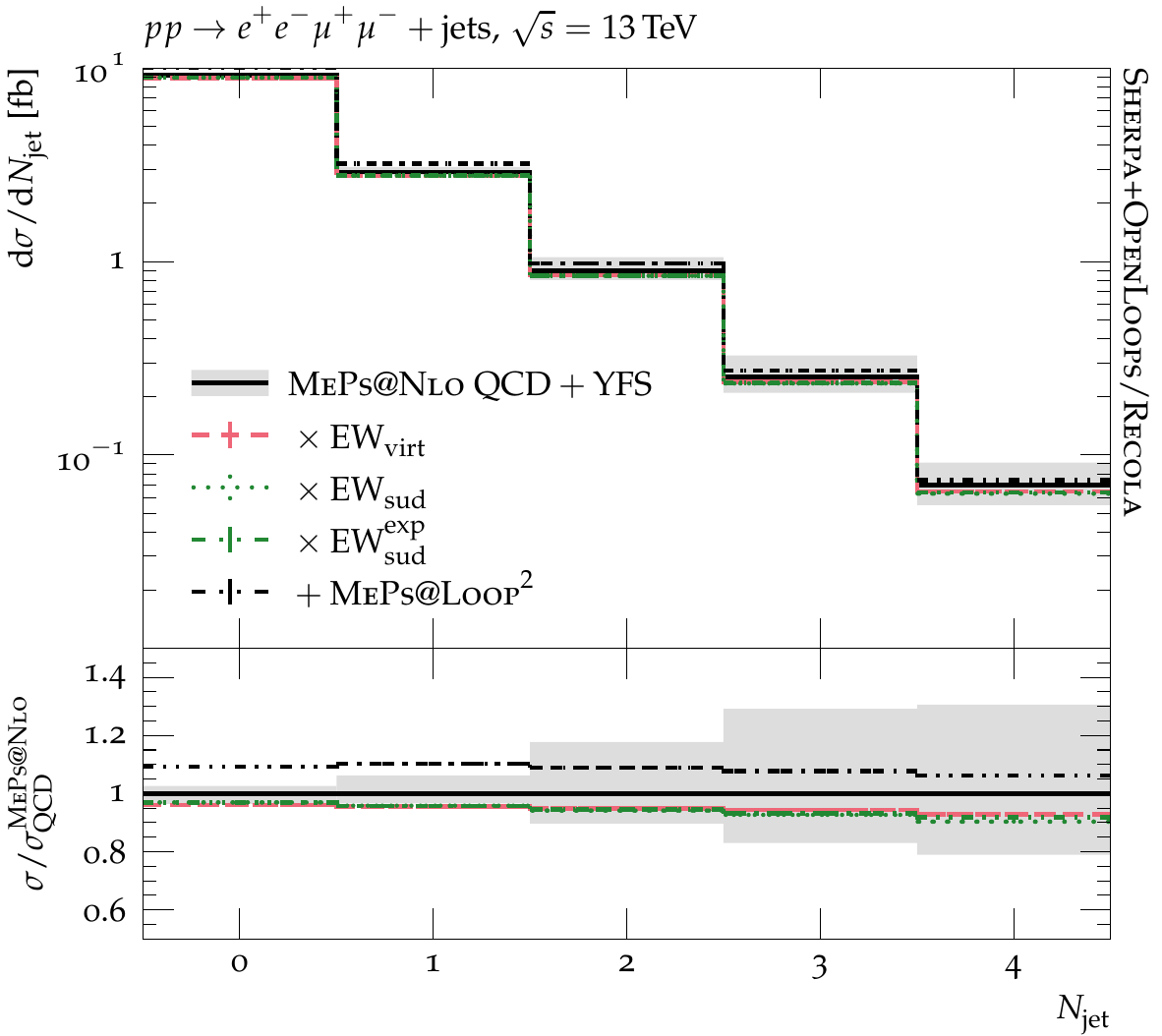} \\
  \includegraphics[width=0.47\textwidth]{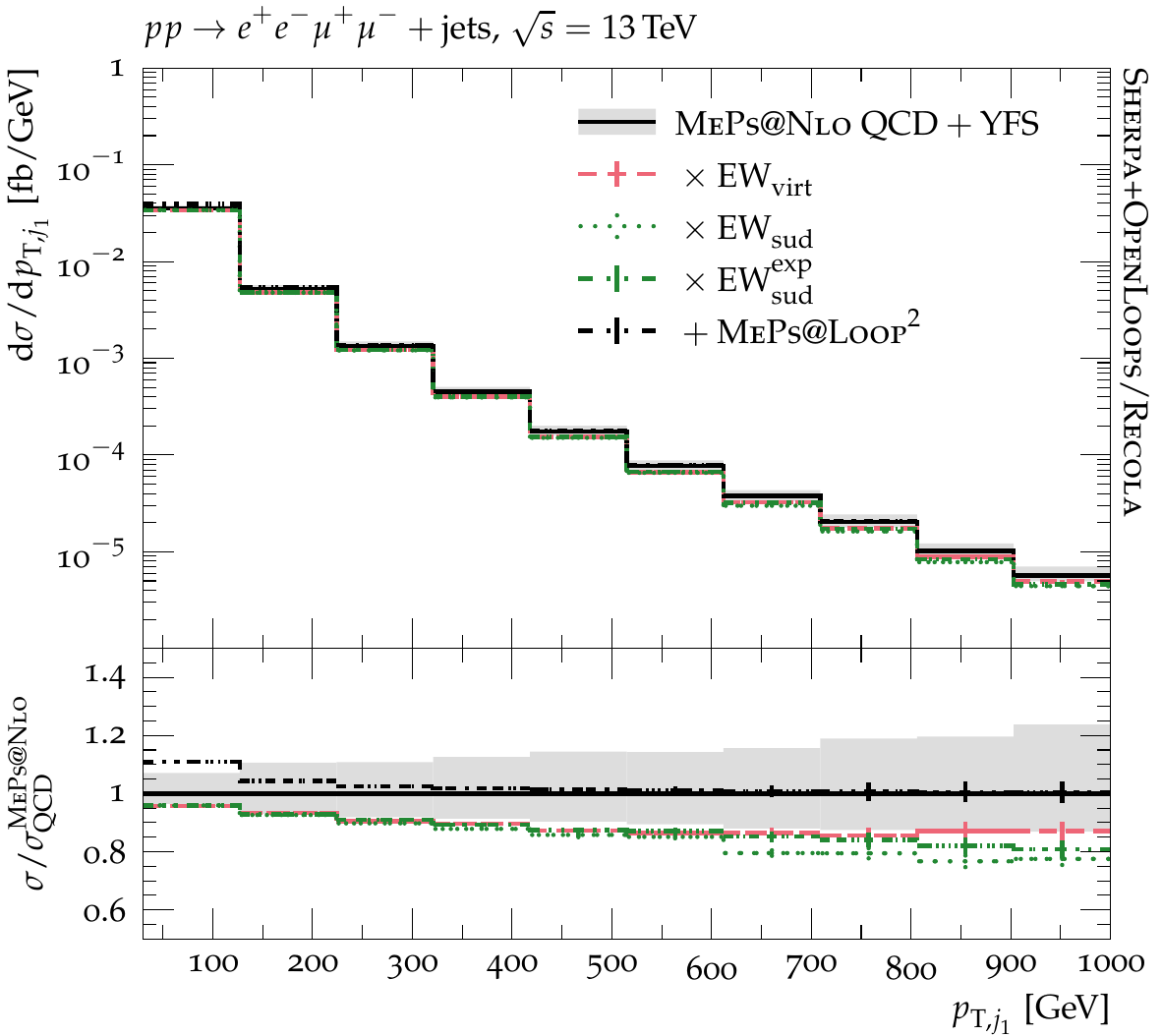} \hfill
  \includegraphics[width=0.47\textwidth]{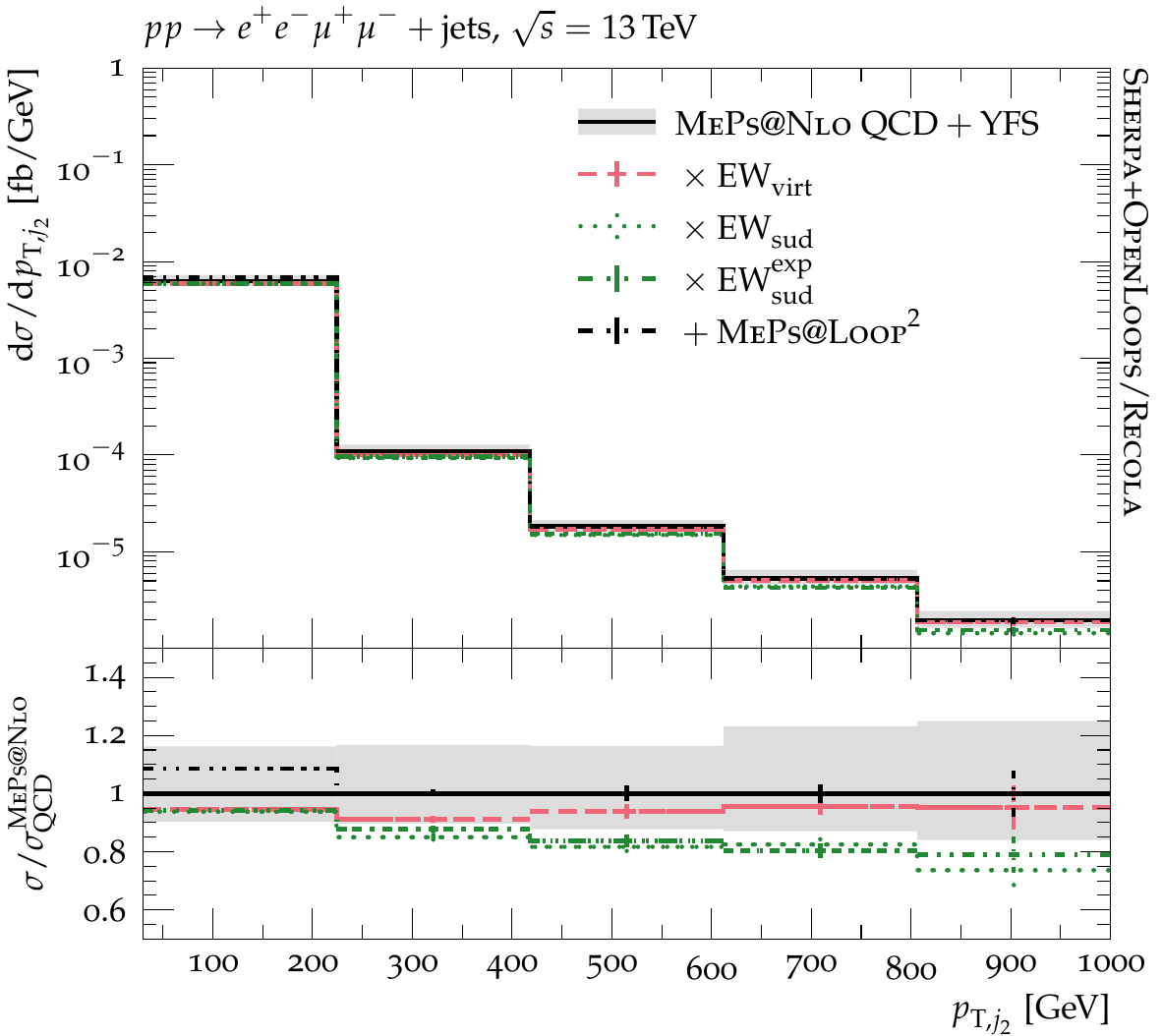}
  \caption{%
    As Fig.~\ref{fig:mepsnlo_pheno_1} but for jet observables.
    Shown are from top left to bottom right:
    the angular separation between the four-lepton system
    and the hardest jet $\Delta \phi_{4\ell,j_1}$,
    the number of jets $N_\text{jet}$,
    and the transverse momenta of the
    hardest jet $p_{\text T, j_1}$
    and second hardest jet $p_{\text T, j_2}$.
  }
  \label{fig:mepsnlo_pheno_2}
\end{figure}

In Fig.~\ref{fig:mepsnlo_pheno_2} we compile our final predictions for the
jet observables, \emph{i.e.}\ $\Delta \phi_{4\ell,j_1}$, $N_\text{jet}$,
$p_{\text T, j_1}$ and $p_{\text T, j_2}$.
Looking back at Fig.~\ref{fig:mepsnlo_2}, we find that they are all rather
dominated by the shower-evolved $2j$ and $3j$ LO matrix-element contributions,
with the exception of $\Delta \phi_{4\ell,j_1} \gtrsim \SI{150}{\degree}$, $N_\text{jet} \leq 2$,
or $p_{\text T, j_1} \lesssim \SI{100}{\GeV}$.\footnote{%
  Of course, the sample
  decomposition depends on the value used for the merging scale $\Qcut$
  and the jet threshold $p^\text{min}_{\mr{T},j}$. We here consider
  $\Qcut=p^\text{min}_{\mr{T},j}=\SI{30}{GeV}$.}
The scale-uncertainty pattern is
largely determined by the dominance of the higher-multiplicity LO contributions.
Accordingly, we observe more significant scale variations than for the more inclusive
leptonic observables. On average, the bands spread from \SI{-10}{\percent}
to $+\SI{20}{\percent}$ around the nominal \MEPSatNLO prediction.
For the phase-space region dominated by the NLO QCD contributions, this shrinks
to $\pm\SI{5}{\percent}$, while for $N_\text{jet} = 4$ it grows to $-\SI{20}{\percent}$
and $+\SI{30}{\percent}$. However, beyond $N_\text{jet} = 3$ we entirely rely on
the parton shower to generate additional jets.

Given the sizeable scale uncertainty of the \MEPSatNLO prediction, the rather mild
EW corrections observed for the $\Delta \phi_{4\ell,j_1}$ and $N_\text{jet}$ observable,
which remain below $-\SI{10}{\percent}$ throughout, are fully enclosed by the QCD
uncertainty band. However, for the transverse-momentum distributions EW Sudakov
corrections exceed the QCD uncertainties, reaching about $-\SI{20}{\%}$. This clearly
shows the necessity to include them in realistic simulations. To further reduce the
systematic uncertainties of the predictions, the two- and three-jet multiplicity
matrix elements should ideally be considered at NLO QCD as well.

For all observables considered here,
the shapes of the \MEPSatNLO and \MEPSatNLO${}+{}$\MEPSatLOOP predictions are quite similar,
with the latter being enhanced by about 5--\SI{10}{\percent}.
The exception are the hard tails for the
$p_{\text T, 2e2\mu}$, $p_\mr{T,j_1}$ and $p_\mr{T,j_2}$ distributions.
Here, the cross section is with increasing hardness
increasingly dominated
by the \MEPSatNLO contributions alone. These contain additional higher-multiplicity
LO QCD matrix elements that are also the adequate sequel for the loop-induced
sample, as long as no two-jet loop-induced contribution is included.
The addition of the \MEPSatLOOP prediction
has no sizeable effect on the overall QCD scale-uncertainty band
of the \MEPSatNLO prediction (beyond the rescaling induced by the increased rate).

As for the exponentiated \EWsud approximation
we find that it gives nearly identical results
compared to the \MEPSatNLO{}+\EWsud one,
due to the moderate absolute \EWsud correction for the studied observables.

\FloatBarrier

\section{Conclusions}\label{sec:conclusions}

To adequately address precision measurements of final states produced
in hadronic collisions, the inclusion of higher-order corrections
in corresponding theoretical predictions based on perturbation theory
is required. Besides the dominant QCD contributions, typically considered at NLO
or NNLO, also EW corrections should ideally be included, certainly when
attempting to describe phase-space regions involving high-energetic particles.
In this limit the exact NLO EW corrections factorise from the production process,
leading to logarithms of the involved kinematic invariants that can eventually
be resummed to all orders.

In this article we discussed the means to compute EW corrections at exact NLO accuracy,
as well as the high-energy virtual EW (\EWvirt) and Sudakov (\EWsud) approximations,
in the \Sherpa framework. We have presented a simple matching formula for the combination
of resummed NLL EW Sudakov corrections with the exact $\order{\alpha}$ result. This
\NLOEWNLLEWsudexp calculation enables us to account for the dominant EW corrections both in the bulk of the
inclusive cross section as well as for high-energetic kinematic configurations.
Properly modelling multiple QCD parton emissions off the initial and final state
of the process requires the inclusion of parton-shower simulations, \emph{i.e.}\ the
matching and merging of QCD matrix elements with a QCD parton cascade. Such methods
for LO and NLO matrix elements are readily available in \Sherpa, referred to as
\MEPSatLO and \MEPSatNLO, respectively. However, the consistent inclusion of (approximate)
EW corrections in such shower-evolved computations is an area of active theoretical
development.
In our work we have presented the methods available in \Sherpa for the largely automated
incorporation of EW corrections in \MEPSatLO\ and \MEPSatNLO\ calculations. This includes in
particular the \EWvirt\ approximation for $\order{\alpha}$ corrections, and the more recently implemented
\EWsud\ scheme for including (resummed) NLL EW contributions. Both approaches can
consistently be combined with QED soft-photon resummation for final-state leptons,
treated in the YFS formalism.

As an highly non-trivial application and benchmark for the various calculational schemes
we analysed $ZZ$ and $ZZj$ production under LHC conditions, thereby including off-shell
and non-resonant contributions to the $e^+e^-\mu^+\mu^-$
and $e^+e^-\mu^+\mu^-j$ channels, respectively.
For the first time we have presented results for detailed kinematical distributions for
$pp\to e^+e^-\mu^+\mu^-j$ at NLO EW. In a first analysis we studied the quality of the
\EWvirt\ and \EWsud\ schemes in approximating the exact NLO EW results, with one-loop
amplitudes provided by \OpenLoops and \Recola. For both the
$ZZ$ and the $ZZj$ channel we found a correction of the inclusive fiducial cross section
from LO to NLO of about $\SI{-7}{\%}$ when using the \Gmu\ input-parameter and renormalisation scheme.
This effect is well reproduced by the \EWvirt\ and \EWsud\ approximations when combined
with the YFS soft-photon resummation. While for the
considered inclusive cross sections the exponentiation of the NLL Sudakov logarithms has
only a very mild effect, it significantly alters the production rate of high-energetic event
configurations, \emph{e.g.}\ when applying a hard cut on the transverse momentum of one
of the $Z$ bosons. Through matching to the exact NLO EW result these higher-order EW
effects can be incorporated in the combined prediction of \NLOEWNLLEWsudexp{} accuracy.

Motivated by the good quality of the EW virtual and Sudakov approximation for the
diboson-production processes, we then considered their inclusion in \MEPSatNLO\
predictions based on the NLO QCD matrix elements for the zero- and one-jet channel,
supplemented by the tree-level two- and three-jet processes. Both for the inclusive
production rates and for the considered differential distributions, we found good agreement
of the resulting \MEPSatNLO{}+\EWsud\ and \MEPSatNLO{}+\EWvirt\ predictions. Only for
phase-space regions dominated by the higher-order tree-level contributions did we observe
systematic deviations. These can be traced back to the limitation of the \EWvirt\
approximation to model the EW corrections for these tree-level contributions
through an effective $k$-factor corresponding to the highest-multiplicity available
at NLO, here the one-jet process. In the \EWsud\ approximation, however, higher-order
tree-level processes with kinematics in the high-energy regime receive the complete
NLL EW Sudakov factor. Having analysed in detail the anatomy of the multijet-merged
samples with EW corrections included, we presented our final phenomenological
predictions, thereby also including an uncertainty estimate corresponding to scale
variations and taking into account the QCD loop-induced contributions to $ZZ$ and $ZZj$
production. The loop-induced channels contribute at the order of $\SI{10}{\%}$
of the total production rate only. Accordingly, neglecting their EW corrections seems well
justified. However, we could clearly show that EW corrections for the direct
production mode---in particular for high-energetic event configurations---amount to
up to $\SI{-40}{\%}$, and thereby largely exceed the QCD scale uncertainty of
the \MEPSatNLO computation.

Accordingly, the inclusion of EW corrections in theoretical predictions is of paramount
importance for upcoming measurements of diboson production at the LHC in Run~3 and the
subsequent High-Luminosity phase.
This applies both to fixed-order calculations and to fully differential multijet-merged
Monte Carlo simulations to be used in precision Standard Model studies.
The inclusion is similarly important for analyses
in which these processes contribute as Standard Model background to New Physics
searches.also Consequently, the methods developed here, due to their automation in the \Sherpa framework,
can and should be applied to any other process.
They will be publicly available with \Sherpa-3.

\clearpage
\section*{Acknowledgements}
This work is supported by funding from the European Union's Horizon
2020 research and innovation programme as part of the Marie
Sk\l{}odowska-Curie Innovative Training Network MCnetITN3
(grant agreement no.~722104).
The work of DN
is supported by the ERC Starting Grant 714788 REINVENT.
MS is funded by the Royal Society through a University Research Fellowship
(URF\textbackslash{}R1\textbackslash{}180549) and an Enhancement Award 
(RGF\textbackslash{}EA\textbackslash{}181033 and 
 CEC19\textbackslash{}100349).
EB, SLV, and SS acknowledge funding from BMBF (contracts 05H18MGCA1 and 05H21MGCAB) and
support by the Deutsche Forschungsgemeinschaft (DFG, German Research Foundation) - project number 456104544.

\appendix
\input{text/appendix-ewsud}
\input{text/appendix-match}

\bibliographystyle{amsunsrt_modp}
\bibliography{references}

\end{document}

%% file: macros.tex
\let\spreprint\empty
\newcommand{\preprint}[1]{\def\spreprint{\protect#1}}
\let\sinstitute\empty

\makeatletter
\renewcommand{\maketitle}{\begingroup
  \null\thispagestyle{empty}%
    \ifx\spreprint\empty
      \vskip 5ex
    \else
      \flushright\large\spreprint\vskip 2ex
    \fi
    \vskip 5ex
    \flushleft
      {\sffamily\bfseries\huge\@title}\vskip 2ex
      \@author\vskip 2ex
      \ifx\sinstitute\empty
      \else
        {\small\sinstitute}
      \fi
    \vskip 5ex
  \endgroup
}
\makeatother

\renewenvironment{abstract}{\begin{center}
  {\large\sffamily\bfseries Abstract: }
  \begin{minipage}[t]{0.75\textwidth}
}{\end{minipage}\end{center}\vskip 10ex}

\allsectionsfont{\normalfont\sffamily\bfseries}
\subsectionfont{\normalfont\sffamily\bfseries}
\subsubsectionfont{\normalfont\sffamily\bfseries}

\makeatletter
\DeclareRobustCommand*{\bfseries}{%
  \not@math@alphabet\bfseries\mathbf
  \fontseries\bfdefault\selectfont
  \boldmath
}
\makeatother

\newcommand{\Pl}{\ell}

\def\be{\begin{equation}}
\def\ee{\end{equation}}

\newcommand{\diagheight}{80pt}

\newcommand{\diagsepB}{-10pt}

\newcommand{\MWOS}{\ensuremath{M_W^\text{OS}}\xspace}
\newcommand{\MW}{\ensuremath{M_W}\xspace}
\newcommand{\MZOS}{\ensuremath{M_Z^\text{OS}}\xspace}
\newcommand{\MZ}{\ensuremath{M_Z}\xspace}

\newcommand{\GZOS}{\ensuremath{\Gamma_Z^\text{OS}}\xspace}

\newcommand{\GWOS}{\ensuremath{\Gamma_W^\text{OS}}\xspace}

\newcommand{\GeV}{\ensuremath{\,\text{GeV}}\xspace}
\newcommand{\TeV}{\ensuremath{\,\text{TeV}}\xspace}

\newcommand{\alphas}{\ensuremath{\alpha_\text{s}}\xspace}
\newcommand{\order}[1]{\ensuremath{\mathcal{O}{\left(#1\right)}}\xspace}

\newcommand{\Deltathr}{\ensuremath{\Delta_\text{thr}}}

\newcommand{\GF}{\ensuremath{G_\mu}}
\newcommand{\Gmu}{\GF}
\newcommand{\amz}{\ensuremath{\alpha(\MZ^2)}}

\newcommand{\MVOS}{\ensuremath{M_{\text{V}}^\text{OS}}\xspace}%
\newcommand{\GVOS}{\ensuremath{\Gamma_{\text{V}}^\text{OS}}\xspace}%

\newcommand{\newc}{\newcommand}
\newc{\bi}{\begin{itemize}}
\newc{\ei}{\end{itemize}}
\newc{\benu}{\begin{enumerate}}
\newc{\eenu}{\end{enumerate}}
\newc{\bc}{\begin{center}}
\newc{\ec}{\end{center}}
\newc{\bfig}{\begin{figure}}
\newc{\efig}{\end{figure}}
\newc{\qbar}{\bar{q}}
\newc{\go}{\tilde{g}}
\newc{\PB}{\textsc{Powheg-Box}}
\newc{\Powheg}{P\protect\scalebox{0.8}{OWHEG}\xspace}

\newcommand{\Alpgen}{{\textsc{Alpgen}}\xspace}
\newcommand{\Recola}{{\textsc{Recola}}\xspace}
\newcommand{\Sherpa}{{\textsc{Sherpa}}\xspace}

\newcommand{\Rivet}{{\textsc{Rivet}}\xspace}
\newcommand{\Amegic}{A\protect\scalebox{0.8}{MEGIC}\xspace}
\newcommand{\Comix}{C\protect\scalebox{0.8}{OMIX}\xspace}
\newcommand{\OpenLoops}{O\protect\scalebox{0.8}{PEN}L\protect\scalebox{0.8}{OOPS}\xspace}

\newcommand{\Gosam}{G\protect\scalebox{0.8}{O}S\protect\scalebox{0.8}{AM}\xspace}

\newcommand{\Collier}{\textsc{Collier}\xspace}
\newcommand{\CutTools}{\textsc{CutTools}\xspace}
\newcommand{\OneLoop}{O\protect\scalebox{0.8}{NE}L\protect\scalebox{0.8}{OOP}\xspace}

\newcommand{\MadLoop}{M\protect\scalebox{0.8}{AD}L\protect\scalebox{0.8}{OOP}\xspace}
\newcommand{\mgfive}{\textsc{\small MG5\_aMC@NLO}\xspace}
\newcommand{\MCFM}{M\protect\scalebox{0.8}{CFM}}
\newcommand{\NLOX}{N\protect\scalebox{0.8}{LO}X\xspace}

\newcolumntype{.}{D{.}{.}{-1}}
\newcolumntype{d}[1]{D{.}{.}{#1}}
\newcolumntype{C}{>{\centering\arraybackslash}p{0.25\textwidth}}

\newlength{\unitcharwidth}
\newcommand{\hc}{\hspace*{\unitcharwidth}}

\newcommand{\QCD}{\ensuremath{\text{QCD}}}

\newcommand{\MEPSatLO}{\text{\textsc{MePs@Lo}}\xspace}
\newcommand{\MEPSatNLO}{\text{\textsc{MePs@Nlo}}\xspace}

\newcommand{\MEPSatLOOP}{\text{\textsc{MePs@Loop\ensuremath{^2}}}\xspace}
\newcommand{\SMCatNLO}{\text{\textsc{S-Mc@Nlo}}\xspace}

\newcommand{\LOEWSudYFS}{LO${}+{}$EW$_{\text{sud}}+{}$YFS}
\newcommand{\LOPlusEWVirtYFS}{LO${}+{}$EW$_{\text{virt}}+{}$YFS}

\newcommand{\mr}[1]{\ensuremath{\mathrm{#1}}}
\newcommand{\mc}[1]{\ensuremath{\mathcal{#1}}}
\newcommand{\mb}[1]{\ensuremath{\mathbf{#1}}}
\newcommand{\mm}[1]{\ensuremath{\mathbb{#1}}}
\newcommand{\muR}{\ensuremath{\mu_{\mr{R}}}}
\newcommand{\muF}{\ensuremath{\mu_{\mr{F}}}}
\newcommand{\muQ}{\ensuremath{\mu_{\mr{Q}}}}
\newcommand{\muCKKW}{\ensuremath{\mu_{\text{CKKW}}}}
\newcommand{\done}{\ensuremath{\mr{d}}}
\newcommand{\Qcut}{\ensuremath{Q_\text{cut}}}
\newcommand{\nmax}{\ensuremath{n_\text{max}}}
\newcommand{\nmaxnlo}{\ensuremath{n_\text{max}^\text{NLO}}}
\newcommand{\mucore}{\ensuremath{\mu_\text{core}}}

\newcommand{\Bbar}{\ensuremath{\overline{\mr{B}}}}
\newcommand{\Fcal}{\ensuremath{\mc{F}}}
\newcommand{\Fbarcal}{\ensuremath{\mc{\overline{F}}}}

\newcommand{\EWvirt}{\ensuremath{\text{EW}_\text{virt}}\xspace}
\newcommand{\EWsud}{\ensuremath{\text{EW}_\text{sud}}\xspace}
\newcommand{\EWsudexp}{\ensuremath{\text{EW}_\text{sud}^\text{exp}}\xspace}
\newcommand{\NLOEWNLLEWsudexp}{NLO EW${}+{}$NLL \EWsudexp}
\newcommand{\deltaEW}{\ensuremath{\delta^\text{EW}}\xspace}
\newcommand{\deltaEWvirt}{\ensuremath{\delta_\text{virt}^\text{EW}}\xspace}
\newcommand{\deltaEWsud}{\ensuremath{\delta_\text{sud}^\text{EW}}\xspace}
\newcommand{\deltanEW}[1]{\ensuremath{\delta_{#1}^\text{EW}}\xspace}
\newcommand{\deltanBEW}[1]{\ensuremath{\delta_{#1,\mathrm{B}}^\text{EW}}\xspace}
\newcommand{\deltanSEW}[1]{\ensuremath{\delta_{#1,\mm{S}}^\text{EW}}\xspace}
\newcommand{\deltanHEW}[1]{\ensuremath{\delta_{#1,\mm{H}}^\text{EW}}\xspace}
\newcommand{\deltanEWvirt}[1]{\ensuremath{\delta_{\text{virt},#1}^\text{EW}}\xspace}
\newcommand{\deltanSEWvirt}[1]{\ensuremath{\delta_{\text{virt},#1,\mm{S}}^\text{EW}}\xspace}
\newcommand{\deltanBEWvirt}[1]{\ensuremath{\delta_{\text{virt},#1,\mathrm{B}}^\text{EW}}\xspace}
\newcommand{\deltanHEWvirt}[1]{\ensuremath{\delta_{\text{virt},#1,\mm{H}}^\text{EW}}\xspace}
\newcommand{\deltanEWsud}[1]{\ensuremath{\delta_{\text{sud},#1}^\text{EW}}\xspace}
\newcommand{\deltanSEWsud}[1]{\ensuremath{\delta_{\text{sud},#1,\mm{S}}^\text{EW}}\xspace}
\newcommand{\deltanBEWsud}[1]{\ensuremath{\delta_{\text{sud},#1,\mr{B}}^\text{EW}}\xspace}
\newcommand{\deltanHEWsud}[1]{\ensuremath{\delta_{\text{sud},#1,\mm{H}}^\text{EW}}\xspace}

\newcommand{\KEWsud}{\ensuremath{\mr{K}_\text{sud}^\text{NLL}}\xspace}
\newcommand{\KnEWsud}[1]{\ensuremath{\mr{K}_{\text{sud},#1}^\text{NLL}}\xspace}

\newcommand{\qbarbracket}{\ensuremath{\displaystyle \bar{q}^{\hspace*{-5pt}\text{\raisebox{0.75pt}{\protect\scalebox{0.5}{(}}}\hspace*{3.7pt}\text{\raisebox{0.75pt}{\protect\scalebox{0.5}{)}}}}}}
\newcommand{\shortequal}{\!\!=\!\!}

\newcommand{\mhl}{\vphantom{\int_A^B}}
\newcommand{\mhhl}{\vphantom{\frac{\pi^2}{6}}}
\newcommand{\mhhhl}{\vphantom{\frac{a}{a}}}

\newcommand{\mHl}{\vphantom{\int\limits_A^B}}

\newcommand{\nnb}{\nonumber}

\colorlet{tableoverheadcolor}{gray!37.5}
\colorlet{tableheadcolor}{gray!25}
\colorlet{tablerowcolor}{gray!12.5}

\newlength{\width}
\newlength{\height}

\def\draftdate{\relax}
\def\mda{\relax}
\def\mua{\relax}
\def\mla{\relax}
\def\draft{
\def\thtystars{******************************}
\def\sixtystars{\thtystars\thtystars}
\typeout{}
\typeout{\sixtystars**}
\typeout{* Draft mode!
         For final version remove \protect\draft\space in source file *}
\typeout{\sixtystars**}
\typeout{}
\def\draftdate{\today}
\def\mua{\marginpar[\boldmath\hfil$\uparrow$]%
                   {\boldmath$\uparrow$\hfil}\color{black}%
                    \typeout{marginpar: $\uparrow$}\ignorespaces}
\def\mda{\color{red}\marginpar[\boldmath\hfil$\downarrow$]%
                   {\boldmath$\downarrow$\hfil}%
                    \typeout{marginpar: $\downarrow$}\ignorespaces}
\def\mla{\marginpar[\boldmath\hfil$\rightarrow$]%
                   {\boldmath$\leftarrow $\hfil}%
                    \typeout{marginpar: $\leftrightarrow$}\ignorespaces}
\def\Mua{\marginpar[\boldmath\hfil$\Uparrow$]%
                   {\boldmath$\Uparrow$\hfil}\color{black}%
                    \typeout{marginpar: $\uparrow$}\ignorespaces}
\def\Mda{\color{red}\marginpar[\boldmath\hfil$\Downarrow$]%
                   {\boldmath$\Downarrow$\hfil}%
                    \typeout{marginpar: $\downarrow$}\ignorespaces}
\def\Mla{\marginpar[\boldmath\hfil\textcolor{red}{$\Rightarrow$}]%
                   {\boldmath\textcolor{red}{$\Leftarrow $}\hfil}%
                    \typeout{marginpar: $\leftrightarrow$}\ignorespaces}
\overfullrule 5pt
\oddsidemargin 15mm
\marginparwidth 29mm
}


%% file: text/appendix-ewsud.tex
\section{EW Sudakov corrections outside the strict high-energy limit}
\label{sec:app:ewsud}

Following~\cite{Denner:2000jv,Denner:2001gw},
the strict condition of the high-energy limit
that all invariants
$r_{kl}=(p_k + p_l)^2$
with external momenta $p_k$ and $p_l$
are large compared to the EW energy scale
has the consequence that a naive application
to processes with intermediate resonances,
such as $W$ and $Z$ bosons decaying into leptons,
will result in $\deltanEWsud{n} = \KnEWsud{n} = 0$.
This is because the intermediate
resonance will constrain the invariant
mass of the decay products $k$ and $l$ to be close
to its on-shell mass.
In consequence, at least this $r_{kl}$ is of the
order of the EW scale and not large compared to it.
To address this,
we follow the same strategy as is used in the YFS
resummation in processes with internal resonances
\cite{Schonherr:2008av,Kallweit:2017khh}.

We have extended the original \Sherpa implementation of
Sudakov corrections 
presented in~\cite{Bothmann:2020sxm} in the following way.
Before calculating $\KnEWsud{n}$,
we combine any lepton pair $(k,l)$  with an invariant
mass $m_{kl}$ close to the mass $M_V$ ($V=W,Z$) of a vector boson
of the same quantum numbers, based on the measure
\begin{equation}
  \Delta_{kl} =
  \frac{
    |m_{kl} - M_{V}|
  }{
    \Gamma_V
  }
  < \Deltathr\;,
\end{equation}
with $\Gamma_{V}$ the width of the vector boson.
\Deltathr\ is an arbitrary threshold parameter that we set to $10$.
If multiple possible combinations exist, \emph{e.g.}\ in $e^+e^+e^-e^-$
final states, the pair with the smallest $\Delta_{kl}$ is combined first.
The resulting amplitude now has $m \leq n$ external particles. Note that this also implies that the
phase space used to evaluate the EW Sudakov contribution changes as well, and it
now refers to $m \leq n$ (or $m \leq n+1$ for \mm{H} events) final-state particles,
with $m$ depending on the number of combined lepton pairs.
We then set
$\deltanEWsud{n}\left(\Phi_n\right) =
\KnEWsud{m}\left(\Phi_m\right)$.
The clustered phase space $\Phi_m$ is constructed by combining the four momenta
of all clustered lepton pairs ($k$, $l$), and assign it to new external vector bosons,
\emph{i.e.}\ $p_{V,kl}=p_k + p_l$.
After all possible daughter leptons are clustered,
all momenta in $\Phi_m$ are reshuffled such that all external particles are on
their mass shell, in particular $p^2_{V,kl}=m^2_V$. To this end, we use the same algorithm
as in Ref.~\cite{Bothmann:2020sxm} to ensure that all matrix elements
entering the calculation
of $\KnEWsud{n}$ have on-shell external momenta after potential $\text{SU}(2)$ rotations
of external states, see the discussion therein.

As a second provision to improve the \EWsud approximation
in phase-space regions where not all invariants are of the same large size,
we include terms proportional to
\begin{equation}
  \label{eq:neglected_dl}
  L(|r_{kl}|, s) = \frac{\alpha}{4\pi} \log^2 \left(\frac{r_{kl}}{s}\right)\;,
\end{equation}
as they appear in the split of the double logarithmic (DL) terms
in~\cite{Denner:2000jv}.
Here, $s$ is the squared partonic centre-of-mass energy.
The terms in Eq.~\eqref{eq:neglected_dl}
are neglected in~\cite{Denner:2000jv}
and in our original implementation~\cite{Bothmann:2020sxm},
as the strict high-energy limit requires $M_W^2 \ll r_{kl} \sim s$.
However, we find that in the $ZZ$ and $ZZj$ observables studied here
neglecting these terms consistently reduces the level of agreement
between \EWsud and NLO EW by up to \SI{5}{\percent},
due to the contribution of kinematic configurations
where at least for some invariants
a looser high-energy limit
$M_W^2 \ll r_{kl} \ll s$
is approached.

Hence, we have included these contributions in all \EWsud results
and made this behaviour the new default of the implementation in \Sherpa.
Note that this term is also retained in~\cite{Pagani:2021vyk}
as part of the so-called $\mathrm{SSC}^{{s \rightarrow r}_{k l}}$ contribution.


%% file: text/appendix-match.tex
\section{Matched EW corrections in multijet-merged calculations}
\label{sec:app:match}

This appendix discusses how the interplay of the EW corrections 
contained in the local $k$-factor and those explicitly effected 
onto the higher-multiplicity LO matrix elements achieves both 
the correct resummation of the multiplicity-specific EW Sudakov 
factors and the inclusive behaviour of the \EWvirt\ approximation 
of the highest multiplicity it is available for. 
Thus, it needs to be shown that the EW corrections for a) the $n=\nmaxnlo$ 
parton process are not changed when higher-multiplicity LO 
matrix elements are merged on top of it, and b) the 
$n=\nmaxnlo+l$ parton processes properly resum their 
respective Sudakov logarithms.

First, we show that for the highest multiplicity for which 
NLO EW virtual corrections are available, $n=\nmaxnlo$, the 
correction effected onto observables sensitive to it are given 
by the matched EW corrections of Eq.\ \eqref{eq:setup:deltamatchSH}. 
To this end, we examine the inclusive $n$-parton cross section 
consisting of the sum of the exclusive $n$-parton cross section 
and the inclusive $(n+1)$-parton cross section. 
Further splitting the latter into the exclusive $(n+1)$-parton 
cross section and the inclusive $(n+2)$-parton cross section
follows trivially and does not change our conclusions.
The inclusive $n$-parton cross section, integrating over both 
\emph{unresolved} and \emph{resolved} emissions (according to 
the merging scale), is then given by
\begin{equation}
  \begin{split}
    \done\sigma_{n}^{\MEPSatNLO}
    =&\;
      \done\Phi_{n}\,\Bbar_n(\Phi_{n})\,
      \left[\exp\left(\deltanEWsud{n}(\Phi_n)\mhhl\right)
            -\deltanEWsud{n}(\Phi_n)+\deltanSEWvirt{n}(\Phi_n)\right]\,
      \Theta_n(\Qcut)\\
    &\times
      \left\{
        \overline{\Delta}_n(t_c,t_n)\mHl
      \right.\\
    &\hspace*{15pt}
      \left.{}
        +\int_{t_c}^{t_n}\!\!\done\Phi_1^\prime\,
         \overline{\mr{K}}_n\left(\Phi_1^\prime\right)
        \left[
          \overline{\Delta}_n(t_{n+1}',t_n)\,
          \Theta(\Qcut-Q_{n+1}^\prime)
          \vphantom{\frac{\left(\deltanEWsud{n}\right)}{\left(\deltanEWsud{n}\right)}}
        \right.
      \right.\\
    &
      \left.
        \left.\hspace*{105pt}{}
          +\frac{\exp\left(\deltanEWsud{n+1}(\Phi_{n+1}^\prime)\mhhl\right)}
                {\exp\left(\deltanEWsud{n}(\Phi_n)\mhhl\right)}\,
           \Delta_n(t_{n+1}',t_n)\,
          \Theta(Q_{n+1}^\prime-\Qcut)
        \right]
      \right.\\
    &
      \left.\hspace*{60pt}{}\times
        \Fcal_{n+1}(t_{n+1}')
        \vphantom{\frac{\left(\deltanEWsud{n}\right)}{\left(\deltanEWsud{n}\right)}}
      \right\}
      \\
    &{}
      +\done\Phi_{n+1}\,\mr{H}_n(\Phi_{n+1})\,
       \exp\left(\deltanEWsud{n+1}(\Phi_{n+1})\mhhl\right)\,
       \Theta_n(\Qcut)\,\Fcal_{n+1}(\mu_Q^2,\Qcut)
      \;,
  \end{split}
\end{equation}
with $\Phi_{n+1}'=\Phi_n\cdot\Phi_1'$ and where we have 
explicitly extracted the relevant Sudakov factor from the 
vetoed parton shower where appropriate.
The only offending term with respect to the pure unmerged result 
is the ratio of the $n$- and $(n+1)$-parton Sudakov factor 
above the merging scale,
brought about by the local $k$-factor of Eq.\ \eqref{eq:kfacmatch}
acting on the LO $(n+1)$-parton process. 
This factor, however, is of $\order{\alpha_s}$ (contained 
in the modified splitting kernel $\overline{\mr{K}}_n$) relative to
the intended accuracy of the matched result in the $n$-parton phase space. 
Further, for $Q_{n+1}$ around the merging scale \Qcut, the ratio of Sudakov factors 
is near unity as a relatively soft $(n+1)$st emission does not 
contribute any large invariants to the existing $n$-parton ensemble. 
For $Q_{n+1}\gg\Qcut$, where the $(n+1)$st emission does contribute 
large invariants to the process, we are in the Sudakov limit where 
\deltanEWsud{n} and \deltanEWvirt{n} coincide and, thus, this ratio 
replaces the $n$-parton resummed Sudakov corrections with the more 
suitable $(n+1)$-parton resummed Sudakov corrections, a clear improvement 
over the unmerged result.
As a short aside, it is interesting to note that in case of the pure 
\EWvirt approximation of Eq.\ \eqref{eq:deltaEWvirt} or 
\eqref{eq:deltaEWvirtplus} this ratio of Sudakov factors is absent 
and the exact unmerged result for the $n$-parton process is reproduced 
by the merged calculation.

Second, we show that observables sensitive to $(n+l)$-parton processes 
with $l>0$
are exposed to the correct $(n+l)$-parton 
resummed Sudakov correction in the Sudakov regime.
We examine this explicitly for $l=1$, and the higher-multiplicity 
processes follow by analogy. 
We thus have
\begin{equation}
  \begin{split}
    \done\sigma_{n+1}^{\MEPSatNLO}
    =&\;
      \done\Phi_{n+1}\,\mr{B}_{n+1}(\Phi_{n+1})\,
      \exp\left(\deltanEWsud{n+1}(\Phi_{n+1})\mhhhl\right)\,
      \Theta_n(\Qcut)\,\Fcal(\mu_Q^2,\Qcut)\\
    &{}
      +\done\Phi_{n+1}\,\mr{D}_n(\Phi_{n+1})\,
      \exp\left(\deltanEWsud{n+1}(\Phi_{n+1})\mhhhl\right)\,
      \Theta_n(\Qcut)\mhl\\
    &\hspace*{10pt}\times
      \left\{
        \frac{\Bbar_n(\Phi_n)}{\mr{B}_n(\Phi_n)}
        \left(
          1-
          \frac{\deltanEWsud{n}(\Phi_n)-\deltanEWvirt{n}(\Phi_n)}
               {\exp\left(\deltanEWsud{n}(\Phi_n)\mhhhl\right)}
        \right)
      \right.\\
    &\hspace*{70pt}
      \left.\times
        \left[
          \frac{\exp\left(\deltanEWsud{n}(\Phi_{n})\mhhl\right)}
               {\exp\left(\deltanEWsud{n+1}(\Phi_{n+1})\mhhhl\right)}\;
          \overline{\Delta}_n(t_{n+1},t_n)\,
          \Theta(\Qcut-Q_{n+1})
        \right.
      \right.\\
    &\hspace*{85pt}
      \left.
        \left.{}
          +\Delta_n(t_{n+1},t_n)\,
          \Theta(Q_{n+1}-\Qcut)\mHl
        \right]
        \;\;-\;1\;\;\mHl
      \right\}\;
      \Fcal_{n+1}(t_{n+1})\;.
  \end{split}
\end{equation}
While the term of the first line represents the desired result, 
the second term is, in the absence of EW corrections, a correction 
of relative $\order{\alpha_s}$ ($\Bbar_n/\mr{B}_n=1+\order{\alpha_s}$), 
applied in the region where the $(n+1)$-parton matrix element can be
written as an $n$-parton matrix element times a splitting kernel, 
$\mr{D}_n=\mr{B}_n\overline{\mr{K}}_n\,\Theta(\mu_Q^2-t_{n+1})$, reproducing 
the behaviour of the unmerged result, see \cite{Hoeche:2014rya}. 
In the presence of EW corrections, however, the situation is 
slightly more involved, leading nonetheless to the same conclusion. 
First, we observe again that for small emission scales $Q_{n+1}$ the 
$(n+1)$st parton does not produce additional large invariants with respect to
those of the $n$-parton ensemble, thus 
$\deltanEWsud{n}(\Phi_n)\approx\deltanEWsud{n+1}(\Phi_{n+1})$. 
The ratio of Sudakov factors appearing in the fourth line 
for emissions below the merging scale \Qcut\ therefore vanishes in this region, 
letting the square bracket sum up to unity up to terms of 
subleading colour at relative $\order{\alpha_s}$, as in the pure QCD 
case. 
In consequence, the curly bracket produces a term of $\order{\alpha}$ 
in addition to the pure QCD case, containing the ratio of the 
difference of the \EWsud and \EWvirt approximations and the 
resummed Sudakov corrections in the underlying $n$-parton phase 
space.
This term simply replaces the $\order{\alpha}$ expansion 
of the resummed $n$-parton Sudakov correction underlying the resummed 
$(n+1)$-parton Sudakov correction in the $\mr{D}_n$ part of 
$\mr{B}_{n+1}$ of the first line ($\mr{B}_{n+1}=\mr{D}_n+\mr{H}_n$), 
\deltanEWsud{n}, with \deltanEWvirt{n}.
As both \EWvirt and \EWsud coincide in the Sudakov limit, the EW 
Sudakov logarithms of the appropriate $(n+1)$-parton phase space are 
resummed unimpededly. 
On the other hand, in the inclusive phase space, this construction 
benefits at least in part from the additional terms contained in the 
\EWvirt approximation of the underlying $n$-parton phase-space 
configuration.